\documentclass[12pt]{article}

\usepackage{amsmath}
\usepackage{amssymb}
\usepackage{epsfig}
\usepackage{graphicx}
\usepackage{cite}
\usepackage{multirow}
\usepackage{longtable}
\usepackage{lscape}
\usepackage{bbm}

\newcommand{\capdef}{}
\newcommand{\mycaption}[2][\capdef]{\renewcommand{\capdef}{#2}%
        \caption[#1]{{\footnotesize #2}}}
\makeatletter
\renewcommand{\fnum@table}{\textbf{\tablename~\thetable}}
\renewcommand{\fnum@figure}{\textbf{\figurename~\thefigure}}
\makeatother

\newcounter{myenumi}

\renewcommand{\themyenumi}{\roman{myenumi}}
{\end{list}}

\newlength{\myem}
\newcounter{mysubequation}[equation]

\makeatletter
\renewcommand{\section}{\@startsection{section}{1}{0em}{-\baselineskip}%
{\baselineskip}{\normalfont\large\bfseries}}
\renewcommand{\subsection}%
{\@startsection{subsection}{2}{0em}{-0.7\baselineskip}%
{0.7\baselineskip}{\normalfont\bfseries}}
\makeatother

\newcommand{\bi}{\begin{itemize}}
\newcommand{\ei}{\end{itemize}}

\newcommand{\be}{\begin{equation}}
\newcommand{\ee}{\end{equation}}
\newcommand{\bea}{\begin{eqnarray}}
\newcommand{\eea}{\end{eqnarray}}

\newcommand{\deltacp}{\delta}
\newcommand{\stheta}{\sin^2 2 \theta_{13}}

\newcommand{\ie}{{\it i.e.}}

\newcommand{\cf}{{\it cf.}}

\newcommand{\etc}{{\it etc.}}
\newcommand{\eq}{Eq.}

\newcommand{\fig}{Fig.}

\newcommand{\Ref}{Ref.}
\newcommand{\Refs}{Refs.}
\newcommand{\Sec}{Sec.}

\newcommand{\Tab}{Table}

\newcommand{\equ}[1]{\eq~(\ref{equ:#1})}
\newcommand{\figu}[1]{\fig~\ref{fig:#1}}

\begin{document}

\begin{titlepage}

\renewcommand{\thefootnote}{\alph{footnote}}

\vspace*{-3.cm}
\begin{flushright}
\end{flushright}


\renewcommand{\thefootnote}{\fnsymbol{footnote}}
\setcounter{footnote}{-1}

{\begin{center}
{\large\bf
Systematic Parameter Space Search \\[0.2cm]
of Extended Quark-Lepton Complementarity
} \end{center}}
\renewcommand{\thefootnote}{\alph{footnote}}

\vspace*{.8cm}
\vspace*{.3cm}
{\begin{center} {\large{\sc
 		Florian~Plentinger\footnote[1]{\makebox[1.cm]{Email:}
                florian.plentinger@physik.uni-wuerzburg.de},
 		Gerhart~Seidl\footnote[2]{\makebox[1.cm]{Email:}
                seidl@physik.uni-wuerzburg.de}, and
                Walter~Winter\footnote[3]{\makebox[1.cm]{Email:}
                winter@physik.uni-wuerzburg.de}
                }}
\end{center}}
\vspace*{0cm}
{\it
\begin{center}

       Institut f{\"u}r Theoretische Physik und Astrophysik, Universit{\"a}t W{\"u}rzburg, \\
       D-97074 W{\"u}rzburg, Germany

\end{center}}

\vspace*{1.5cm}

{\Large \bf
\begin{center} Abstract \end{center}  }

We systematically investigate the parameter space of neutrino and charged lepton mass matrices for textures
motivated by an extended quark-lepton complementarity. 
As the basic hypothesis, we postulate that all mixing angles in
$U_\ell$ and $U_\nu$ be either maximal or described by powers of a
single small quantity $\epsilon \simeq \theta_\text{C}$. All mass
hierarchies are described by this $\epsilon$ as well. In this study, we do not assume specific forms for
$U_\ell$ and $U_\nu$, such as large mixing coming from the neutrino sector only. 
We perform a systematic scan of the
$262 \, 144$ generated mixing matrices for being compatible with current experimental data, and find
a sample of $2 \, 468$ possibilities. We then analyze and classify the effective charged lepton and neutrino mass textures, where we especially focus on a subset of models getting under pressure for small $\theta_{13}$. In addition, 
we predict the mixing angle distributions from our sample of all valid textures, and study the
robustness of this prediction. We also demonstrate how our procedure can be extended to predictions of the Dirac and Majorana phases in $U_{\mathrm{PMNS}}$. For instance, we find that CP conservation in neutrino oscillations is preferred, and we can impose a lower bound on the mixing matrix
element for $0\nu\beta\beta$ decay.

\vspace*{.5cm}

\end{titlepage}

\newpage

\renewcommand{\thefootnote}{\arabic{footnote}}
\setcounter{footnote}{0}

\section{Introduction}
The Standard Model (SM) of elementary particle physics with three
light Majorana neutrinos contains 28 free
parameters. Most of them, in total 22, describe the masses and
mixings of the fermions (the remaining six parameters are the three SM
gauge couplings, the vacuum expectation value and quartic coupling of the Higgs, and the
QCD $\theta$ parameter). This large number
of parameters, especially in the fermion sector, is generally considered
as an unsatisfactory feature of the SM and one therefore seeks for models in which
the number of parameters can be minimized. One possibility to reduce
the number of parameters is to embed the SM into a Grand Unified Theory
(GUT).

By putting quarks and leptons into GUT multiplets, the masses and
mixing angles in the quark and lepton sectors become related. From this point of view, it
is thus reasonable to describe the observed hierarchical pattern of
the masses and mixing angles of quarks and charged leptons \cite{Yao:2006px} in
terms of powers of a single small expansion parameter $\epsilon$. The
expansion parameter $\epsilon$ might, for example, represent a
low-energy remnant of a flavor symmetry that has been broken at some
high scale. In fact, the CKM mixing
matrix $V_{\rm CKM}$ \cite{Cabibbo:1963yz,Kobayashi:1973fv} exhibits quark mixing angles of the orders
\begin{equation}\label{eq:CKMangles}
|V_{us}|\sim\epsilon,\quad |V_{cb}|\sim\epsilon^2,\quad
|V_{ub}|\sim\epsilon^3,
\end{equation}
where the quantity $\epsilon$ is of the order of the Cabibbo angle
$\theta_\text{C}\simeq 0.2$. Similarly, for the same value
$\epsilon\simeq \theta_\text{C}$, the mass ratios of the up-type quarks, down-type quarks, and the charged
leptons can be approximated, {\it e.g.}, by\footnote{We take here a
  fit compatible with an $SU(5)$ GUT.}
\begin{equation}\label{equ:massratios}
m_u:m_c:m_t=\epsilon^6:\epsilon^4:1,\quad
m_d:m_s:m_b=\epsilon^4:\epsilon^2:1,\quad
m_e:m_\mu:m_\tau=\epsilon^4:\epsilon^2:1,
\end{equation}
where $m_b/m_t\sim\epsilon^2$, $m_\tau/m_b\sim 1$, and $m_t\simeq
175\:\text{GeV}$. While the CKM angles and charged fermion
masses are thus strongly hierarchical, there are striking differences in
the neutrino sector. In the past few years, solar
\cite{Fukuda:2002pe,Ahmad:2002ka}, atmospheric \cite{Fukuda:1998mi},
reactor \cite{Araki:2004mb,Apollonio:2002gd}, and accelerator
\cite{Aliu:2004sq} neutrino oscillation experiments have established
with increasing precision that among the leptonic mixing angles only the reactor angle
$\theta_{13}$ is small whereas the solar angle $\theta_{12}$ and
the atmospheric angle $\theta_{23}$ are both large. Moreover, neutrino
oscillation data tells us that the neutrinos have only a mild
hierarchy. To be specific, expressing the neutrino mass ratios as in
\equ{massratios} in terms of powers of $\epsilon$, the neutrino mass
spectrum can, {\it e.g.}, be written as
\begin{equation}\label{equ:spectra}
m_1:m_2:m_3=\epsilon^2:\epsilon:1,\quad
m_1:m_2:m_3=1:1:\epsilon,\quad
m_1:m_2:m_3=1:1:1,
\end{equation}
where $m_1,m_2,$ and $m_3$ denote the 1st, 2nd, and 3rd
neutrino mass eigenvalue. In \equ{spectra}, the first equation corresponds
to a normal hierarchical, the second to an
inverted hierarchical, and the third to a degenerate neutrino mass spectrum.\footnote{More precisely, we have in
  \equ{spectra} for the inverse hierarchical case $m_2>m_1$ and
  $(m_2-m_1)/m_2\sim\epsilon^2$.}
In addition, we know from cosmological observations that the absolute neutrino mass scale is of the order $\sim
10^{-2}\dots 10^{-1}\:\text{eV}$ \cite{Tegmark:2003ud}. An attractive origin of the
smallness of neutrino masses is provided by the seesaw mechanism
\cite{typeIseesaw,typeIIseesaw}. For a summary of current
values and errors for the neutrino oscillation parameters from a global analysis, see \Tab~\ref{tab:values}.

\begin{table}
\centering
\begin{tabular}{l@{\quad}l@{\quad}c@{\quad}c@{\quad}c@{\quad}c@{\quad}}
  \hline
  Parameter & Best-fit$\pm 1\sigma$ & $1\sigma$ acc. & 2$\sigma$ range & 3$\sigma$ range & Projection $\sim$ 2016 \\
  \hline
  $\Delta m^2_{21}  \: [10^{-5} \mathrm{eV^2}]$ & $7.9\pm 0.3$          &  4\% 
     & $7.3-8.5$ & $7.1-8.9$ & - \\
  $|\Delta m^2_{31}|\: [10^{-3}\mathrm{eV^2}]$ & $2.5^{+0.20}_{-0.25}$ & 10\%
     & $2.1-3.0$ & $1.9-3.2$ & - \\
  \hline
  $\sin^2\theta_{12}$ & $0.30^{+0.02}_{-0.03}$ &  9\% & $0.26-0.36$ & $0.24-0.40$ & 4.6\% ($1\sigma$)\\
  $\sin^2\theta_{23}$ & $0.50^{+0.08}_{-0.07}$ & 16\% & $0.38-0.64$ & $0.34-0.68$ & 10\% ($1\sigma$)\\
  $\sin^2\theta_{13}$ & $-$  &$-$  & $\leq 0.025$ & $\leq 0.041$ & $\leq 0.0076$ ($3\sigma$) \\
  \hline
\end{tabular}
\mycaption{\label{tab:values} Current best-fit values with $1\sigma$ errors, relative accuracies
  at $1\sigma$, as well as $2\sigma$ and $3\sigma$ allowed ranges of
  three-flavor neutrino oscillation parameters (from a combined
  analysis of all available data \cite{Schwetz:2006dh}). In addition, we show as the last column the projected improvement of the mixing angle errors on a time scale of about ten years from now. For this projection, we assume that the current best-fit values remain unchanged, 
in particular, that $\stheta$ will not be discovered. We use the experimental bounds $\stheta \lesssim 0.03$ ($3\sigma$, year 2016 from \Ref~\cite{Huber:2006vr} based on NO$\nu$A), $\sim$10\% precision of $\sin^2 \theta_{23}$ ($1\sigma$, from \Ref~\cite{Antusch:2004yx}, based on beam experiments), 4.6\% precision of $\sin^2 \theta_{12}$ ($1\sigma$, from \Ref~\cite{Minakata:2004jt} based on SPMIN/SADO+others with an exposure of $10 \, \mathrm{GW \, kt \, yr}$, which seems to be reasonable on that time scale for a KamLAND-scale detector). These values should be interpreted with
care, because they depend on the experimental strategy.
}
\end{table}

In spite of the qualitative differences between the quark and the lepton
sector, there have recently been proposed interesting ``quark-lepton
complementarity'' (QLC) relations \cite{Smirnov:2004ju,Raidal:2004iw,Minakata:2004xt} (for an early
approach see Ref.~\cite{Petcov:1993rk}) that might point to quark-lepton unification. The QLC relations express that the solar
angle $\theta_{12}$ and the atmospheric angle $\theta_{23}$ seem to be
connected to the quark mixing angles by
\begin{equation}\label{equ:qlc}
\theta_{12}+\theta_\text{C}\approx\pi/4,\quad\theta_{23}+\theta_{cb}\approx\pi/4, 
\end{equation}
where $\theta_{cb}=\text{arcsin}\:V_{cb}$. A simple interpretation of how the QLC relations in \equ{qlc}
might arise is to assume that the small and the maximal angles in this
relation, $\theta_\text{C}$, $\theta_{cb}$, and $\pi/4$, describe the
mixing of the left-handed charged leptons and the neutrinos in flavor
basis. Consequently, the observed large leptonic mixing angles $\theta_{12}$
and $\theta_{23}$ can only arise as a result of taking the product of the charged
lepton mixing matrix $U_\ell$ and the neutrino mixing matrix
$U_\nu$ in the PMNS \cite{Pontecorvo:1957cp,Maki:1962mu} mixing matrix
$U_{\rm PMNS}=U_\ell^\dagger U_\nu$. This point of view is supported
by the fact that in explicit models,
$\theta_\text{C}$ and $\theta_{cb}$ are usually given from the outset and also maximal
leptonic mixing has meanwhile been obtained in many models (for a review
of recent developments see, {\it e.g.},
Ref.~\cite{Altarelli:2006ri}). Therefore, if the QLC
relations really arise from taking the product of $U_\ell$ and
$U_\nu$ in $U_\text{PMNS}$, it is interesting to study models with mass
matrices where, {\it e.g.}, the value of $\theta_{12}$ can be
understood in terms of $\theta_\text{C}$ and $\theta_{23}$ as a result
of combining $U_\ell$ and $U_\nu$ into $U_\text{PMNS}$.

There has been a considerable amount of work on QLC. For example,
deviations from bimaximal neutrino mixing
\cite{bimaximal} that are introduced by $\theta_\text{C}$
have been studied in
Refs.~\cite{Jezabek:1999ta,Giunti:2002pp,Frampton:2004ud}. Sum rules in the context of QLC using some assumptions on the mixing angles were presented in Refs.~\cite{Ohlsson:2005js,Antusch:2005kw}. Parameterizations of
$U_\text{PMNS}$ in terms of $\theta_\text{C}$ as an expansion
parameter were given in Refs.~\cite{Rodejohann:2003sc,Li:2005ir,Xing:2005ur,
Datta:2005ci,Everett:2005ku}. For renormalization group effects and
QLC see Ref.~\cite{Schmidt:2006rb} and references therein. Model building
realizations of QLC were discussed in
Refs.~\cite{Frampton:2004vw,Antusch:2005ca} and for the special case
$\theta_\text{C}\simeq \theta_{13}$ in Ref.~\cite{Ohlsson:2002rb}.

In this paper, we study systematically 262\,144 pairs of charged lepton
and neutrino mass matrices that lead to the
observed leptonic masses and mixing angles. Motivated by QLC, our only assumption on the mixing angles in $U_\ell$
and $U_\nu$ is that they can take any of the values
$\pi/4,\epsilon,\epsilon^2,0$, where $\epsilon$ is of the
order of the Cabibbo angle $\epsilon \simeq \theta_\text{C}$ and
represents the only small expansion parameter in our approach. Different
from earlier studies of QLC, which assume very
specific forms of $U_\ell$ and $U_\nu$
\cite{Jezabek:1999ta,Giunti:2002pp,Frampton:2004ud,Ohlsson:2005js,Antusch:2005kw},
we allow maximal ($\simeq\pi/4$) and small ($\simeq\epsilon,\epsilon^2,\dots$)
mixing angles to be generated in the
charged lepton as well as in the neutrino sector. This means, in
particular, that we do not
assume that $U_\ell$ or $U_\nu$ are necessarily
of a bimaximal mixing form but obtain this, instead, as a special
case. We consider all possible combinations of the mixing angles
$\pi/4,\epsilon,\epsilon^2,\dots$ in $U_\ell$ and $U_\nu$, and select those pairs $U_\ell$ and $U_\nu$ that
give PMNS mixing angles $\theta_{12},\theta_{13},$ and $\theta_{23}$ in agreement with observations. 
For the detailed description of our method, see \Sec~\ref{sec:method}.
We then unambiguously reconstruct directly the charged lepton and the neutrino
mass matrices from the known charged lepton and neutrino mass spectra
in Eqs.~(\ref{equ:massratios}) and (\ref{equ:spectra}).  The matrix
pairs that we obtain in this way are then presented in the flavor
basis with order of magnitude entries that are expressed as powers of $\epsilon\simeq\theta_\text{C}$, which takes the role
of a small expansion parameter for the matrices. Such approximate
forms or representations of mass matrices expanded in terms of
$\epsilon$ will be called in the following ``mass matrix textures'' or
simply ``textures''. They are presented and discussed in \Sec~\ref{sec:models}. 
We analyze the predictions for the mixing angles
from the set of mixing matrices consistent with current observations in \Sec~\ref{sec:pred}, and we study
 the robustness of these ``predictions''. Finally, we show how our framework can
be applied to phase predictions in \Sec~\ref{sec:phasesext}, such as for
 CP violation in neutrino oscillations or the
mixing matrix element in neutrinoless double beta decay.

\section{Method}
\label{sec:method}

In this section, we will introduce our method for a systematic scan
 of the textures that exhibit QLC. First, we set up the notation for
 the mass and mixing matrices. Next, we motivate and present our
 hypothesis for the charged lepton and neutrino mixing angles. We show
 how the resulting PMNS matrices are compared with
 observation using a selector and discuss how to construct from the
 selected mixing matrices the textures for the charged leptons
 and neutrinos.

\subsection{Mixing Formalism and Notation}\label{sec:notation}
In this section, we describe the formalism and notation for the leptonic mixing
parameters. Here, we follow closely Ref.~\cite{Frampton:2004ud}. A general unitary $3\times 3$ matrix $U_\text{unitary}$ can always be written as
\begin{subequations}
\begin{equation}\label{equ:unitary}
 U_\text{unitary}=
 \text{diag}\left(e^{\text{i}\varphi_{1}},e^{\text{i}\varphi_{2}},e^{\text{i}\varphi_{3}}\right)\cdot\widehat{U}\cdot\text{diag}\left(e^{\text{i}\alpha_{1}},e^{\text{i}
 \alpha_{2}}, 1 \right)
 ,
\end{equation}
where the phases $\varphi_1$, $\varphi_2$, $\varphi_3$, $\alpha_1$, and $\alpha_2$, take
their values in the interval $\left[0,2\pi\right]$ and
\begin{equation}
 \label{equ:ckm}
 \widehat{U} = \left( 
 \begin{array}{ccc}
   c_{12} c_{13} & s_{12} c_{13} & s_{13} e^{-\text{i}\widehat{\delta}} \\
   -s_{12} c_{23} - c_{12} s_{23} s_{13} e^{\text{i}\widehat{\delta}} &   c_{12} c_{23} -
 s_{12} s_{23} s_{13} e^{\text{i}\widehat{\delta}} & s_{23} c_{13} \\ 
 s_{12} s_{23} - c_{12} c_{23} s_{13}
e^{\text{i}\widehat{\delta}} & -c_{12} s_{23} - s_{12} c_{23} s_{13} e^{\text{i}\widehat{\delta}} & c_{23}
c_{13} 
 \end{array}
 \right) 
\end{equation}
\end{subequations}
is a CKM-like matrix in the standard parameterization with $s_{ij} =
 \sin\hat{\theta}_{ij}$, $c_{ij} = \cos\hat{\theta}_{ij}$, where
$\hat{\theta}_{ij}\in\{\hat{\theta}_{12},\hat{\theta}_{13},\hat{\theta}_{23}\}$ lie all in the first quadrant, {\it i.e.}
$\hat{\theta}_{ij}\in\left[0,\frac{\pi}{2}\right]$, and 
$\widehat{\delta}\in[0,2\pi]$. The matrix $\widehat{U}$ is thus described
 by 3 mixing angles $\theta_{ij}$ and one phase $\delta$, {\it i.e.}, it has 4
 parameters. The matrix $U_\text{unitary}$ has five additional phases and contains therefore in
 total 9 parameters.

Depending on whether neutrinos are Majorana or Dirac particles, the
low-energy effective Lagrangian for lepton masses takes one of
the forms
\begin{equation}\label{equ:massterms}
 \mathcal{L}_\text{M}=-(M_\ell)_{ij}e_ie^c_j-\frac{1}{2}(M_\nu^\text{Maj})_{ij}\nu_i\nu_j+\text{h.c.},\quad
\mathcal{L}_\text{D}=-(M_\ell)_{ij}e_ie^c_j-(M_\nu^\text{Dirac})_{ij}\nu_i\nu_j^c+\text{h.c.},
\end{equation}
where $e_i$ and $\nu_i$ are the left-handed charged leptons and
neutrinos that are part of the $SU(2)_L$ lepton doublets
$\ell_i=(\nu_i,e_i)^T$, while $e_i^c$ are the right-handed ($SU(2)_L$
singlet) charged leptons, and $i=1,2,3$ is the generation index. We have also extended the SM by adding to each
 generation $i$ one right-handed SM singlet neutrino $\nu_i^c$. In
 \equ{massterms}, $M_\ell$, $M_\nu^\text{Maj}$, and
 $M_\nu^\text{Dirac}$, denote the charged lepton ($M_\ell$),
 the Majorana neutrino($M_\nu^\text{Maj}$), and the Dirac
 neutrino mass matrix ($M_\nu^\text{Dirac}$), respectively. The Dirac matrices
 $M_\ell$ and $M_\nu^\text{Dirac}$ are general complex $3\times 3$ matrices,
 whereas $M_\nu^\text{Maj}$ is a complex symmetric $3\times 3$ matrix. The
 charged lepton mass matrix is diagonalized by a biunitary transformation
\begin{subequations}
\begin{equation}
 \label{equ:lepdiag}
M_\ell=U_\ell\,M_\ell^\text{diag}\,{U_\ell'}^\dagger,
\end{equation}
where $U_\ell$ and $U_\ell'$ are unitary matrices acting on the
left-handed ($U_\ell$) and right-handed ($U_\ell'$) charged leptons
$e_i$ and $e^c_i$, which span the rows and columns of $M_\ell$, respectively. In \equ{lepdiag}, the matrix
$M_\ell^\text{diag}$ is on the diagonal form
$M_\ell^\text{diag}=\text{diag}(m_e,m_\mu,m_\tau)$. Using the freedom
of re-phasing the charged lepton fields, we can, in what follows,
assume that $U_\ell$ is on a CKM-like form that is parameterized as in \equ{ckm}. We then define the 4 mixing parameters of the left-handed charged leptons,
the three mixing angles $\theta_{12}^\ell$, $\theta_{13}^\ell$,
$\theta_{23}^\ell$, and the phase $\delta^\ell$, by identifying in
\equ{ckm} $\hat{\theta}_{ij}\rightarrow\theta_{ij}^\ell$ and
$\widehat{\delta}\rightarrow\delta^\ell$. The Majorana and Dirac
neutrino mass matrices are given by
\begin{equation}\label{equ:Maj}
 M_\nu^\text{Maj}=U_\nu M_\nu^\text{diag}U_\nu^T,\quad
 M_\nu^\text{Dirac}=U_\nu\,M_\nu^\text{diag}\,{U_\nu'}^\dagger,
\end{equation}
\end{subequations}
where the unitary neutrino mixing matrix $U_\nu$ acts on the left-handed
neutrinos $\nu_i$, while $U_\nu'$ acts on the right-handed neutrinos $\nu_i^c$.
In \equ{Maj}, $M_\nu^\text{diag}$ is a diagonal matrix
$M_\nu^\text{diag}=\text{diag}(m_1,m_2,m_3)$. Since $U_\ell$ has already been
brought to a CKM-like form, we have no longer the same freedom to
remove phases in $U_\nu$ and, thus, we find that the PMNS matrix is in
general written as
\begin{subequations}
\begin{equation}\label{equ:pmnspara}
U_\text{PMNS}=U_\ell^\dagger U_\nu=U_\ell^\dagger D\widehat{U}_\nu\,K,
\end{equation}
where $\widehat{U}_\nu$ is a CKM-like matrix that is on the form as in
\equ{ckm} while
$D=\text{diag}(1,e^{\text{i}\widehat{\varphi}_1},e^{\text{i}\widehat{\varphi}_2})$
and $K=\text{diag}(e^{\text{i}\widehat{\phi}_1},e^{\text{i}\widehat{\phi}_2},1)$ are
diagonal matrices with phases in the range
$\widehat{\varphi}_1,\widehat{\varphi}_2,\widehat{\phi}_1,\widehat{\phi}_2\in[0,2\pi]$. Note that we have already removed in
$U_\text{PMNS}$ an unphysical overall phase. The CKM-like matrix
$\widehat{U}_\nu$ in \equ{pmnspara} contains four neutrino mixing
parameters, the three neutrino mixing angles
$\theta_{12}^\nu,\theta_{13}^\nu, \theta_{23}^\nu$, and a neutrino phase
$\delta^\nu$, which we define in the standard parameterization by
identifying in \equ{ckm} the neutrino mixing angles as
$\hat{\theta}_{ij}\rightarrow\theta_{ij}^\nu$ and the neutrino phase as
$\widehat{\delta}\rightarrow\delta^\nu$. The matrix in \equ{pmnspara} is written in terms
of the six phases
$\delta^\ell,\delta^\nu,\widehat{\varphi}_1,\widehat{\varphi}_2,\widehat{\phi}_1,$
and $\widehat{\phi}_2$, which lead to three physical phases. The PMNS matrix in
\equ{pmnspara} can, equivalently, also be directly written as
\begin{equation}\label{equ:standard}
 U_\text{PMNS}=U_\ell^\dagger U_\nu=\widehat{U}\cdot\text{diag}(e^{\text{i}\phi_1},e^{\text{i}\phi_2},1),
\end{equation}
\end{subequations}
where $\widehat{U}$ is a CKM-like matrix that is on the form as in \equ{ckm} and the phases $\phi_{1}$ and $\phi_2$ are Majorana
phases. The CKM-like matrix $\widehat{U}$ in \equ{standard} is
described by the solar angle $\theta_{12}$, the reactor angle
$\theta_{13}$, the atmospheric angle $\theta_{23}$, and one Dirac
CP-phase $\delta$, which we identify in the standard
parameterization of \equ{ckm} as
$\hat{\theta}_{ij}\rightarrow\theta_{ij}$ and
$\widehat{\delta}\rightarrow\delta$. The PMNS matrix has thus 3 mixing
angles and 3 phases and contains therefore 6 physical
parameters. Writing the matrix elements of $U_\text{PMNS}$ as $U_{ij}\equiv(U_\text{PMNS})_{ij}$, we read off in the standard
parameterization of \equ{ckm}, the 6 leptonic mixing
parameters as follows
\begin{subequations}\label{equ:mixingtool}
\begin{eqnarray}
\sin\theta_{13}&=&|U_{13}|,\\
\tan\theta_{12}&=&\left|\frac{U_{12}}{U_{11}}\right|,\quad\text{$\theta_{12}=\frac{\pi}{2}$ if $U_{11}=0$},\\
\tan\theta_{23}&=&\left|\frac{U_{23}}{U_{33}}\right|,\quad
 \text{$\theta_{23}=\frac{\pi}{2}$ if $U_{33}=0$},\\
\delta&=&
\left\{\begin{array}{ll}
\displaystyle{-\arg\,\left(\frac{U^*_{12}U_{13}U_{22}U^*_{23}}{s_{12}s_{13}s_{23}c_{12}c_{13}^2c_{23}}+\frac{s_{12}s_{13}s_{23}}{c_{12}c_{23}}\right)}, & \text{for $\theta_{ij}\neq0,\frac{\pi}{2}$},\\\vspace{-2mm}&\\
0 & \text{else},
\end{array}\right.\\
\phi_1&=&
\left\{\begin{array}{ll}
\arg (e^{-i\delta}U^*_{13}U_{11}), & \text{for $\theta_{ij}\neq\frac{\pi}{2}$ and $\theta_{13}\neq0$},\\\vspace{-2mm}&\\
\arg (U_{31}U^*_{33}), & \text{for $\theta_{ij}\neq\frac{\pi}{2}$ and only $\theta_{13}=0$ allowed},\\\vspace{-2mm}&\\
0 & \text{else},
\end{array}\right.\\
\phi_2&=&
\left\{\begin{array}{ll}
\arg (e^{-i\delta}U^*_{13} U_{12}), & \text{for $\theta_{ij}\neq\frac{\pi}{2}$ and $\theta_{12},\theta_{13}\neq0$},\\\vspace{-2mm}&\\
\arg (e^{i\phi_1}U^*_{11} U_{12}), & \text{for $\theta_{13},\theta_{23}=0$ and $\theta_{12}\neq0,\frac{\pi}{2}$},\\\vspace{-2mm}&\\
\arg (U_{22} U^*_{23}), & \text{for $\theta_{12}$ or $\theta_{13}=0$ but $\neq\frac{\pi}{2}$, and $\theta_{23}\neq0,\frac{\pi}{2}$,}\\\vspace{-2mm}&\\
0 & \text{else}.
\end{array}\right.
\end{eqnarray}
\end{subequations}
Note that we set the phases to $0$ if they are undefined. Since we can easily identify these cases if needed, there is no bias or constraint introduced by this choice. For example, 
for $\theta_{13}=0$, $\delta$ is undefined. In this case, $\delta$ does not affect the mixing
matrix, \ie, an arbitrary choice of $0$ does not change the physics. However, for the phase
prediction of $\delta$, the irrelevant cases can be easily eliminated by the identification of the
corresponding $\theta_{13} =0$.
Note that the above relations are valid for a general unitary matrix
$U_\text{unitary}$ as in \equ{unitary} as well. In addition, if $U_\text{PMNS}$ is already a CKM-like matrix, \ie, all unphysical phases are already removed, we simply have $\delta=\arg (U^\ast_{13})$.

\subsection{Generating the PMNS Matrices and Textures}\label{sec:generating}
Let us now introduce our method for a systematic scan
 of the textures that exhibit QLC. Instead of starting out with
 various forms for the lepton mass matrix textures and
 calculate from these the lepton masses and mixing
 angles, we follow in our procedure a reverse process: we reconstruct
 the mass matrix textures in flavor basis from the masses and mixing parameters of charged leptons and neutrinos.

Although neutrino oscillation experiments will in future allow to pin
 down the leptonic mixing angles with increasing precision, the PMNS
 matrix does not uniquely reveal the individual mixing parameters in $U_\ell$ and
$U_\nu$. Existing models and
 studies mostly suppose that the observed large solar
and atmospheric mixing angles arise mainly in the neutrino sector. It
 is, however, important to emphasize that maximal atmospheric mixing
can also equally well arise from the charged
lepton sector as proposed, {\it e.g.}, in ``lopsided'' GUT models
\cite{Albright:1998vf} (for realizations of bilarge neutrino mixing in
lopsided models see also, {\it
  e.g.},
 Refs.~\cite{Nir:1999xp,Nomura:1999ty,Albright:2001uh}). Moreover,
 the QLC relations in \equ{qlc} suggest, in particular, that both maximal
 ($\simeq\pi/4$) and small ($\simeq\theta_\text{C},\theta_{cb}$) mixing
 angles might be expected in the charged lepton and in the neutrino
 sector. Motivated by QLC, we will thus assume in our approach that the
 mixing angles $\theta_{ij}^\ell$ in $U_\ell$ and $\theta_{ij}^\nu$ in
 $U_\nu$ can a priori take any of the values in the sequence
 $\pi/4,\epsilon,\epsilon^2,\dots,$ where $\epsilon\simeq\theta_\text{C}\simeq 0.2$, and then compare the
resulting PMNS mixing angles $\theta_{ij}$ with current data. The
 choice of the angles in this sequence is a simple and
straightforward interpretation of the QLC relations in \equ{qlc} in the sense that the solar angle $\theta_{12}\simeq 33^\circ$ can
only arise as a result of taking the
product of $U_\ell$ and $U_\nu$ in $U_\text{PMNS}=U_\ell^\dagger
U_\nu$. The assumption that any of the angles $\theta_{ij}^\ell$ and
 $\theta_{ij}^\nu$ can assume any of the values
 $\pi/4,\epsilon,\epsilon^2,\dots$, generalizes the definition of QLC
 to what we call an ``extended QLC''.

Let us next specify the values of the mass parameters in our approach. Motivated by
the mass ratios of quarks and leptons in Eqs.~(\ref{equ:massratios})
and (\ref{equ:spectra}), we will approximate the diagonalized charged
lepton mass matrix $M_\ell^\text{diag}$ in
Eq.~(\ref{equ:lepdiag}) by
\begin{subequations}\label{equ:masshypo}
\begin{equation}
 M_\ell^\text{diag}=m_\tau\,\text{diag}(\epsilon^4,\epsilon^2,1),
\end{equation}
and choose for the diagonalized neutrino mass matrix
$M_\nu^\text{diag}$ in \equ{Maj} any of the cases
\begin{equation}\label{equ:neutrinomasses}
  M_\nu^\text{diag}=m_3\text\,\text{diag}(\epsilon^2,\epsilon,1),\quad
  M_\nu^\text{diag}=m_2\text\,\text{diag}(1,1,\epsilon),\quad
 M_\nu^\text{diag}=m_3\text\,\text{diag}(1,1,1),
\end{equation}
\end{subequations}
where the first, second, and third equation
corresponds to a normal, inverted, and degenerate neutrino mass
spectrum, respectively. We thus see that the quantity
$\epsilon\simeq \theta_\text{C}$ will appear in our
 approach as a small expansion parameter, which is in charge of all the
hierarchies and small effects that are exhibited by the mass spectra
and mixing angles of the fermions.

Our systematic search for textures satisfying QLC hence consists
 in the following simple three-step procedure:

{\bf First step} -- In the first step of our procedure, we consider the PMNS matrix in the parameterization $U_\text{PMNS}=U_\ell^\dagger
 U_\nu=U_\ell^\dagger D\widehat{U}_\nu\,K$ of \equ{pmnspara} and
 assume here that the entries
$\sin(\theta_{ij}^\ell)$ and $\sin(\theta_{ij}^\nu)$ in the CKM-like matrices
$U_\ell$ and $\widehat{U}_\nu$ can take all the values
\begin{equation}\label{equ:sinehypo}
s_{ij}^\ell, s_{ij}^\nu \in \{\frac{1}{\sqrt{2}}, \epsilon,
\epsilon^2,0 \},
\end{equation}
where we have defined $s_{ij}^\ell=\sin(\theta_{ij}^\ell)$ and
 $s_{ij}^\nu=\sin(\theta_{ij}^\nu)$. For definiteness, we choose in
 \equ{sinehypo} $\epsilon = 0.2$ as our standard value. Moreover, we
 will first restrict ourselves to the case of real mixing matrices
 $U_\ell$ and $U_\nu$, and assume in \equ{pmnspara} that the phases
 $\delta^\ell,\delta^\nu,\widehat{\varphi}_1,\widehat{\varphi}_2,\widehat{\phi}_1,$
 and $\widehat{\phi}_2$, are only $0$ or $\pi$. In total, this
 gives
\begin{equation}\label{equ:pairs}
4^3\times 2\times 4^3 \times 2^5 = 262\,144
\end{equation}
possible distinct pairs of mixing matrices $\{U_\ell, U_\nu\}$ ({\it
  cf.}~\equ{pmnspara}). Later,
in Secs.~\ref{sec:tuning} and \ref{sec:phases}, we will vary $\epsilon$ in the range $0.15\leq\epsilon\leq 0.25$ and
the phases $\delta^\ell$ and $\delta^\nu$ in the whole range
  $\delta^\ell,\delta^\nu\in[0,2\pi]$. In the following, we will call a ``model'' the set of twelve mixing parameters
  $\{\theta_{12}^\ell,\theta_{13}^\ell,\theta_{23}^\ell,\theta_{12}^\nu,\theta_{13}^\nu,\theta_{23}^\nu,\delta^\ell,\delta^\nu,\widehat{\varphi}_1,\widehat{\varphi}_2,\widehat{\phi}_1,\widehat{\phi}_2\}$
  that describes a pair of charged lepton and neutrino mixing matrices
  $\{U_\ell,U_\nu\}$.

{\bf Second step} -- In the second step of our method, we obtain for each of the 262\,144 matrix
pairs the corresponding PMNS matrix
$U_\text{PMNS}=U_\ell^\dagger U_\nu$, read off the PMNS mixing parameters
$\theta_{12}$, $\theta_{13}$, $\theta_{23}$, $\delta$, $\phi_1$, and $\phi_2$ using \equ{mixingtool},
and extract those pairs $\{U_\ell,U_\nu\}$, which are in agreement
with current neutrino oscillation data. In order to have an automatic
selection criterion for our candidate pairs $\{U_\ell,U_\nu\}$, we create our ``best sample'' by defining a selector
\begin{equation}
S \equiv \left( \frac{\sin^2 \theta_{12}  - 0.3}{0.3 \times \sigma_{12}} \right)^2 + \left( \frac{\sin^2 \theta_{23} -  0.5}{0.5 \times \sigma_{23}} \right )^2.
\label{equ:selector}
\end{equation}
This selector corresponds to a Gaussian $\chi^2$ with the current
best-fit values and the given relative $1\sigma$ errors $\sigma_{12} \simeq 9\%$ (for
$\sin^2\theta_{12}$) and $\sigma_{23} \simeq 16\%$ (for
$\sin^2\theta_{23}$) from \Tab~\ref{tab:values}, where we assume a
Gaussian distribution in $\sin^2 \theta_{12}$ and $\sin^2 \theta_{23}$
as an approximation.  For the best sample, we choose all models that
satisfy the selection criterion
\begin{equation}\label{equ:selectioncriterion}
S \le 11.83\quad\text{and}\quad\sin^2 \theta_{13} \le \sigma_{13}
\simeq 0.04,
\end{equation}
which corresponds to a $\Delta \chi^2$ for the $3 \sigma$ confidence
level with two degrees of freedom, and the hard cut on $\theta_{13}$
represents the current $3 \sigma$ bound according to \Tab~\ref{tab:values}. 
We do not include the numerical value of $\sin^2 \theta_{13}$ in the selector in \equ{selector}, because its current best-fit value would introduce a bias towards small $\theta_{13}$.
In some cases, we will use different values for the selection,
such as to test the experimental pressure on the model space. For example, on a time scale of about ten years from now, we choose the values from the last column in \Tab~\ref{tab:values}.
This selection process is a conservative guess/crude estimator for the
sample of models which can be still accommodated with data within the
$3 \sigma$ confidence level.  Note that we have, so far, neither used
the mass squared differences (and mass hierarchy), nor
extracted/predicted them. At this stage, in steps 1 and 2, our
procedure is independent from the mass spectra of charged leptons and
neutrinos. They become, however, important for constructing the mass
matrix textures in the next step.

{\bf Third step} -- The third step of our approach consists of considering all the pairs
$\{U_\ell,U_\nu\}$ that have been selected in the previous step 2 and
rotate for each such pair the diagonal
matrices $M_\ell^\text{diag}$ and $M_\nu^\text{diag}$ given in
Eqs.~(\ref{equ:masshypo}) back to flavor basis according to
\begin{equation}\label{equ:backrotation}
 M_\ell=U_\ell M_\ell^\text{diag},\quad
 M_\nu^\text{Maj}=U_\nu M_\nu^\text{diag}U_\nu^T,\quad
 M_\nu^\text{Dirac}=U_\nu M_\nu^\text{diag},
\end{equation}
where we have chosen, for simplicity, $U_\ell'=U_\nu'=\mathbbm{1}_3$. The Dirac
mass matrices $M_\ell$ and $M_\nu^\text{Dirac}$ can thus be viewed as
representatives of a class of mass matrices that is obtained by
introducing arbitrary rotation matrices $U_\ell'$ and $U_\nu'$ acting
on the right-handed leptons. Moreover, note that \equ{backrotation} actually describes six
different cases, depending on whether one chooses from
\equ{neutrinomasses} a normal hierarchical, inverted hierarchical, or a quasi-degenerate
neutrino mass spectrum. In the following, we will, as already
mentioned in the introduction, denote the mass
matrices $M_\ell$, $M_\nu^\text{Maj}$, and $M_\nu^\text{Dirac}$, in
\equ{backrotation} expanded in terms of $\epsilon$ as ``mass matrix textures'' or simply ``textures''.

Let us briefly summarize again our three-step procedure for a
systematic scan of real PMNS matrices and textures with QLC:
\begin{enumerate}
 \item Generate 226\,144 PMNS matrices by inserting all distinct combinations of
 $s_{ij}^\ell$, $s_{ij}^\nu$, $\delta^\ell$, $\delta^\nu$,
 $\widehat{\varphi}_1$, $\widehat{\varphi}_2$, $\widehat{\phi}_1$, and
 $\widehat{\phi}_2$, into \equ{pmnspara}. Here, $s_{ij}^\ell$ and
 $s_{ij}^\nu$ are taken from $\{\frac{1}{\sqrt{2}},\epsilon,\epsilon^2,0\}$, whereas the phases can
 assume values $\pi$ and $0$.
\item Read off the mixing parameters $\theta_{12}$,
  $\theta_{13}$, $\theta_{23}$, $\delta$, $\phi_1$, and $\phi_2$, using
  \equ{mixingtool} and select the pairs $\{U_\ell,U_\nu\}$
  where $U_\text{PMNS}=U_\ell^\dagger U_\nu$ satisfies the selection
  criterion in \equ{selectioncriterion}.
\item For each selected pair $\{U_\ell,U_\nu\}$ rotate the mass
  matrices in \equ{masshypo} back to flavor basis to obtain the
  textures $M_\ell$,
  $M_\nu^\text{Maj}$, and $M_\nu^\text{Dirac}$ ({\it
  cf.}~\equ{backrotation}). 
\end{enumerate}
We wish to emphasize that the above procedure leaves
complete freedom to let the observed leptonic mixing angles be
generated by contributions from both the charged lepton and/or neutrino sector in the
most general way. Another
advantage of our method, that we will exploit later, is that it allows to scan very quickly the parameter space
of complex phases by varying the parameters directly in the PMNS
matrix instead of varying the highly redundant phases in the mass matrices. Our procedure to construct mass matrix textures in this way is extremely simple and efficient. In particular,
to derive the textures, there is no need to perform any
diagonalization in the whole process.

\section{Scanning the Parameter Space for Individual Models}\label{sec:models}
In this section, we apply our procedure outlined in Sec.~\ref{sec:generating} for extracting the mass
matrix textures from the best sample obtained with the selector in
\equ{selector}, and make it explicit for several examples. Thereby, as
explained above in step 3 of our procedure, we start with the diagonal
forms of the charged lepton and normal hierarchical neutrino mass
matrices in Eqs.~(\ref{equ:masshypo}) and obtain the mass matrices by
rotating back to flavor basis following \equ{backrotation}, where we
set for convenience $U_\ell'$ and $U_\nu'$ equal to the unit
matrix. In the following, we make first a subjective choice and focus
in Sec.~\ref{sec:current} on several textures which are of special interest like (i) the combination of a
CKM-like matrix for $U_\ell$ with a bimaximal mixing matrix $U_\nu$,
(ii) a texture with very small $\theta_{13}$, and (iii) the sets which
represent the best approximation to the current best-fit values, \ie,
those for which the selector $S$ in Eq.~(\ref{equ:selector}) takes the
lowest value. Later, in Sec.~\ref{sec:in10years}, we present a complete list of the textures that exhibit a small $\theta_{13}$, as well as survive increased experimental pressure from the other parameters.

\subsection{Several Examples of Models Compatible with Current Bounds}\label{sec:current}

Here, we present a subjective selection of models compatible with current bounds.
In addition, we describe how we identify the leading order terms in the textures.

\subsubsection*{Matrices with CKM-like Plus Bimaximal Mixing}
Let us first consider the case where $U_\ell$ is CKM-like and $U_\nu$
is on a bimaximal mixing form. In the parameterization of
\equ{pmnspara}, this can be realized in our
procedure by taking for the charged lepton and
neutrino mixing parameters
\begin{subequations}
\begin{eqnarray}
 (s^\ell_{12}, s^\ell_{13}, s^\ell_{23},
  \delta^\ell)&=&(\epsilon,0,\epsilon^2,\pi)\\
(s^\nu_{12}, s^\nu_{13}, s^\nu_{23},\delta^\nu, \widehat{\varphi}_1, \widehat{\varphi}_2, \widehat{\phi}_1, \widehat{\phi}_2)&=&(\frac{1}{\sqrt{2}},\epsilon,\frac{1}{\sqrt{2}},0,0,0,\pi,\pi),
\end{eqnarray}
which lead to the PMNS mixing angles
\begin{equation}\label{equ:PMNSanglesCKM+Unu}
 (\theta_{12}, \theta_{13}, \theta_{23})=(36.5^\circ,3.6^\circ,43.8^\circ).
\end{equation}
\end{subequations}
The corresponding charged lepton, neutrino, and PMNS mixing matrices
are
\begin{subequations}
\begin{equation}\label{equ:CKM+Unu}
U_\ell=
\left(
\begin{array}{ccc}
 1-\frac{\epsilon ^2}{2} & \epsilon  & 0 \\
 -\epsilon  & 1-\frac{\epsilon ^2}{2} & \epsilon ^2 \\
 0 & -\epsilon ^2 & 1
\end{array}
\right),
\quad
U_\nu=
\left(
\begin{array}{ccc}
 -\frac{1}{\sqrt{2}}+\frac{\epsilon ^2}{2 \sqrt{2}} & -\frac{1}{\sqrt{2}}+\frac{\epsilon ^2}{2 \sqrt{2}} & \epsilon  \\
 \frac{1}{2}+\frac{\epsilon }{2} &-\frac{1}{2} +\frac{\epsilon }{2} & \frac{1}{\sqrt{2}}-\frac{\epsilon ^2}{2 \sqrt{2}} \\
 -\frac{1}{2}+\frac{\epsilon }{2} & \frac{1}{2}+\frac{\epsilon }{2} & \frac{1}{\sqrt{2}}-\frac{\epsilon ^2}{2 \sqrt{2}}
\end{array}
\right),
\end{equation}
\begin{equation}
U_\text{PMNS}=\left(
\begin{array}{rrr}
 -0.7+0.5\,\epsilon+1.2\,\epsilon ^2 &
   -0.7-0.5\,\epsilon+1.2\,\epsilon ^2 &
  1.7 \,\epsilon  \\
0.5+1.2\,\epsilon-0.8\,\epsilon ^2 & -0.5+1.2 \,\epsilon+0.8\,\epsilon^2 & 0.7-\,\epsilon ^2 \\
 -0.5+0.5\,\epsilon-0.5\,\epsilon ^2 & 0.5+0.5\,\epsilon+0.5\,\epsilon ^2 &
 0.7-1.1\,\epsilon ^2
\end{array}
\right).
\end{equation}
The CKM matrix and the bimaximal mixing matrix are, on
the other hand, given by
\begin{equation}\label{equ:CKM+Ubimax}
V_\text{CKM}=
\left(
\begin{array}{ccc}
 1-\frac{\lambda ^2}{2} & \lambda  & 0 \\
 -\lambda  & 1-\frac{\lambda ^2}{2} & A\,\lambda ^2 \\
 0 & -A\,\lambda ^2 & 1
\end{array}
\right)
\quad
\text{and}
\quad
U_\text{bimax}=
\left(
\begin{array}{ccc}
 \frac{1}{\sqrt{2}} & -\frac{1}{\sqrt{2}} & 0  \\
 \frac{1}{2} &\frac{1}{2}& -\frac{1}{\sqrt{2}} \\
 \frac{1}{2}& \frac{1}{2}& \frac{1}{\sqrt{2}}
\end{array}
\right),
\end{equation}
\end{subequations}
where we have used for $V_\text{CKM}$ the Wolfenstein parameterization
to second order in $\lambda=0.22$. The PMNS mixing angles in
$U_\text{PMNS}=V_\text{CKM}^\dagger U_\text{bimax}$ are $(\theta_{12}, \theta_{13}, \theta_{23})_\text{CKM+bimax}=\left(36^\circ,9^\circ,45^\circ\right)$.
Comparison of Eqs.~(\ref{equ:CKM+Unu}) and
(\ref{equ:CKM+Ubimax}) yields that $U_\ell$ reproduces exactly the CKM matrix in
the Wolfenstein parameterization, whereas the entries of $U_\nu$ differ from
those in $U_\text{bimax}$ by terms of the orders $\epsilon$ or
$\epsilon^2$. These terms lead to the deviations between the numerical
values of $(\theta_{12}, \theta_{13},
\theta_{23})_\text{CKM+bimax}$
and $(\theta_{12},\theta_{13},\theta_{23})$ in
\equ{PMNSanglesCKM+Unu}. Note that choosing the phases of $U_\nu$
such that all signs of the entries in
$U_\text{bimax}$ are reproduced but without changing $U_\ell$,
changes the PMNS
mixing angles in \equ{PMNSanglesCKM+Unu} to
$(\theta_{12},\theta_{13},\theta_{23})\rightarrow\left(33.8^\circ,10.0^\circ,43.4^\circ\right)$.

Now, we rotate the mass matrices $M_\ell^\text{diag}=m_\tau\,\text{diag}(\epsilon^4,\epsilon^2,1)$ and
 $M_\nu^\text{diag}=m_3\,\text{diag}(\epsilon^2,\epsilon,1)$ back to
 flavor basis following \equ{backrotation}, where $U_\ell$
 and $U_\nu$ are parameterized in terms of powers of $\epsilon$ as in
\equ{CKM+Unu}. In flavor basis, the resulting mass matrix
 textures of charged leptons and neutrinos are then given by
\begin{subequations}\label{equ:texturesCKM+Unu}
\begin{equation}
M_\ell=m_\tau\left(
\begin{array}{ccc}
 0 & 0 & 0 \\
 0 & \epsilon ^2 & \epsilon ^2 \\
 0 & 0 & 1
\end{array}
\right)
\rightarrow
\left(
\begin{array}{ccc}
 0 & 0 & 0 \\
 0 & \epsilon ^2 & \epsilon ^2 \\
 0 & 0 & 1
\end{array}
\right),
\end{equation}
\begin{equation}
M_\nu^\text{Maj}=m_3
\left(
\begin{array}{ccc}
 \frac{\epsilon }{2}+\frac{3 \epsilon ^2}{2} & \frac{3 \epsilon }{2 \sqrt{2}}-\frac{\epsilon ^2}{\sqrt{2}} &
   \frac{\epsilon }{2 \sqrt{2}} \\
 \frac{3 \epsilon }{2 \sqrt{2}}-\frac{\epsilon ^2}{\sqrt{2}} & \frac{1}{2}+\frac{\epsilon
   }{4}-\frac{3 \epsilon ^2}{4} & \frac{1}{2}-\frac{\epsilon }{4}-\frac{3 \epsilon ^2}{4} \\
 \frac{\epsilon }{2 \sqrt{2}} & \frac{1}{2}-\frac{\epsilon }{4}-\frac{3 \epsilon ^2}{4} & \frac{1}{2}+\frac{\epsilon }{4}+\frac{\epsilon
   ^2}{4}
\end{array}
\right)
\rightarrow
\left(
\begin{array}{ccc}
 \epsilon  & \epsilon  & \epsilon  \\
 \epsilon  & 1 & 1 \\
 \epsilon  & 1 & 1
\end{array}
\right),
\end{equation}
and
\begin{equation}
M_\nu^\text{Dirac}=m_3
\left(
\begin{array}{ccc}
 -\frac{\epsilon ^2}{\sqrt{2}} & -\frac{\epsilon }{\sqrt{2}} & \epsilon  \\
 \frac{\epsilon ^2}{2} & -\frac{\epsilon }{2}+\frac{\epsilon ^2}{2} & \frac{1}{\sqrt{2}}-\frac{\epsilon ^2}{2 \sqrt{2}}
   \\
 -\frac{\epsilon ^2}{2} & \frac{\epsilon }{2}+\frac{\epsilon ^2}{2} & \frac{1}{\sqrt{2}}-\frac{\epsilon ^2}{2
   \sqrt{2}}
\end{array}
\right)
\rightarrow
\left(
\begin{array}{ccc}
 \epsilon ^2 & \epsilon  & \epsilon  \\
 \epsilon ^2 & \epsilon  & 1 \\
 \epsilon ^2 & \epsilon  & 1
\end{array}
\right),
\end{equation}
\end{subequations}
where we have, for simplicity, assumed the special case of a normal hierarchical neutrino
mass spectrum. Starting, instead with
$M_\nu^\text{diag}=m_3\,\text{diag}(1,1,\epsilon)$ or
$M_\nu^\text{diag}=m_3\,\text{diag}(1,1,1)$, we obtain the
corresponding textures of the neutrinos for the case of inverse hierarchical
or degenerate neutrino masses. In Eqs.~(\ref{equ:texturesCKM+Unu}), ``$\rightarrow$'' symbolizes, up to an overall mass scale, the identification of the leading
order terms in the expansion in $\epsilon$ that contribute to the mass matrix
elements in $M_\ell$, $M_\nu^\text{Maj}$, and
$M_\nu^\text{Dirac}$. This identification is possible for all the mass
matrices which are currently valid at the 3$\sigma$ level, since it
turns out that the contributions to a mass matrix element belonging to
different orders in $\epsilon$ can always be clearly separated from each other.
In other words, for the given expansions of
the mass matrix elements in powers of $\epsilon$, the zeroth
order term in $\epsilon$ (if non-zero), is always larger than the
higher order terms, and the first order term in $\epsilon$ (if non-zero) is
always larger than the second order term. In the thus identified leading order
contributions to the mass matrices, we can then further approximate
and set the order unity coefficients equal to one. This leads us finally to a rough texture with matrix elements $1$, $\epsilon$, $\epsilon^2$, and 0. These
entries have thus to be understood as order of magnitude entries. Note that such textures become 't Hooft natural \cite{tHooft:1979} when they
 arise from a spontaneously broken flavor symmetry
 \cite{Froggatt:1978nt} (see also Ref.~\cite{Leurer:1992wg}). An origin of the small number $\epsilon$ is provided in terms of an
 anomalous $U(1)$ symmetry
 \cite{Green:1984sg}, which has been employed
 for generating textures in various models
 \cite{Ibanez:1994ig} (for an anomaly-free approach see, {\it e.g.}, Ref.~\cite{Enkhbat:2005xb}).

In the rest of the text, we will always
proceed exactly the same way and determine the textures for charged
leptons and neutrinos as it was done above in arriving at
Eqs.~(\ref{equ:texturesCKM+Unu}).

\subsubsection*{Matrices with Very Small $\boldsymbol{\theta_{13}}$ -- Near-CKM Plus Bimaximal Mixing}

We find a class of 64 models with very small $\theta_{13} \simeq 0.2^\circ$. All these models
have in common that $\theta_{12}$ is slightly off the current best-fit value, which means that they
will be excluded in ten years by our assumptions. Note, however, that experimentally $\theta_{13}$ and
$\theta_{12}$ are two different degrees of freedom, which means that there could well be much stronger
pressure coming from $\theta_{13}$ than from $\theta_{12}$. In this case, this class of models
may be surviving very long. For one typical representative combining near-CKM-type mixing in $U_\ell$ with with near-bimaximal mixing in $U_\nu$, the charged lepton and neutrino mixing parameters are
\begin{subequations}
\begin{eqnarray}
(s^\ell_{12}, s^\ell_{13}, s^\ell_{23}, \delta^\ell) & = & (\epsilon,\epsilon,0,\pi)\,,\\
(s^\nu_{12}, s^\nu_{13}, s^\nu_{23},\delta^\nu, \widehat{\varphi}_1, \widehat{\varphi}_2, \widehat{\phi}_1, \widehat{\phi}_2) & = & (\frac{1}{\sqrt{2}},0,\frac{1}{\sqrt{2}},0,0,0,\pi,\pi),
\end{eqnarray}
\end{subequations}
which lead to the PMNS mixing angles
\begin{equation}
(\theta_{12},\theta_{13},\theta_{23})=(28.7^\circ,0.2^\circ,46.1^\circ).
\end{equation}
The charged lepton, neutrino, and PMNS mixing matrices are
\begin{subequations}
\begin{equation}
U_\ell=
\left(
\begin{array}{ccc}
1-\epsilon ^2 & \epsilon & -\epsilon  \\
 -\epsilon & 1-\frac{\epsilon^2 }{2} & 0 \\
 \epsilon& \epsilon^{2} & 1-\frac{\epsilon ^2}{2}
\end{array}
\right),
\quad
U_\nu=
\left(
\begin{array}{ccc}
 -\frac{1}{\sqrt{2}} &  -\frac{1}{\sqrt{2}} & 0 \\
 \frac{1}{2} & -\frac{1}{2} &  \frac{1}{\sqrt{2}} \\
-\frac{1}{2} & \frac{1}{2} &  \frac{1}{\sqrt{2}}
\end{array}
\right),
\end{equation}
\begin{equation}
U_\text{PMNS}=
\left(
\begin{array}{rrr}
 - \frac{1}{\sqrt{2}}- \epsilon+\frac{\epsilon ^2}{\sqrt{2}}   & \
- \frac{1}{\sqrt{2}}+ \epsilon+\frac{\epsilon ^2}{\sqrt{2}}   & 0 \\
 \frac{1}{2}- \frac{\epsilon }{\sqrt{2}}-\frac{3 \epsilon ^2}{4}  & \
- \frac{1}{2}- \frac{\epsilon }{\sqrt{2}}+\frac{3 \epsilon ^2}{4} & \
 \frac{1}{\sqrt{2}}+ \frac{\epsilon ^2}{2 \sqrt{2}}\\
 - \frac{1}{2}+ \frac{\epsilon }{\sqrt{2}}+\frac{\epsilon ^2}{4}  &  \frac{1}{2}+ \frac{\epsilon }{\sqrt{2}} - \
\frac{\epsilon ^2}{4}  & \
\frac{1}{\sqrt{2}} - \frac{\epsilon ^2}{2 \sqrt{2}}
\end{array}
\right).
\end{equation}
Here, $U_\ell$ is on a bimaximal mixing form but $U_\nu$ is not
exactly the CKM-matrix (mainly due to the entry $\sim\epsilon$ in the
1-3 and 3-1 element). For the textures of charged leptons and
neutrinos we thus find
\begin{equation}
M_\ell=m_\tau
\left(
\begin{array}{ccc}
 0 & 0 & -\epsilon  \\
 0 & \epsilon ^2 & 0 \\
 0 & 0 & 1-\frac{\epsilon ^2}{2}
\end{array}
\right)
\rightarrow
\left(
\begin{array}{ccc}
 0 & 0 & \epsilon  \\
 0 & \epsilon ^2 & 0 \\
 0 & 0 & 1
\end{array}
\right),
\end{equation}
\begin{equation}
M_\nu^\text{Maj}=m_3
\left(
\begin{array}{ccc}
 \frac{\epsilon }{2}+\frac{\epsilon ^2}{2} & \frac{\epsilon }{2 \
\sqrt{2}} - \frac{\epsilon ^2}{2 \sqrt{2}} & - \frac{\epsilon }{2 \sqrt{2}}+\frac{\epsilon ^2}{2 \
\sqrt{2}}  \\
 \frac{\epsilon }{2 \sqrt{2}} - \frac{\epsilon ^2}{2 \sqrt{2}} & \frac{1}{2}+ \frac{\epsilon }{4} +\frac{\
\epsilon ^2}{4}   & \frac{1}{2}- \frac{\epsilon }{4}- \
\frac{\epsilon ^2}{4} \\
 - \frac{\epsilon }{2 \sqrt{2}}+\frac{\epsilon ^2}{2 \sqrt{2}}  & \frac{1}{2}  - \frac{\epsilon }{4} - \
\frac{\epsilon ^2}{4}& \frac{1}{2} + \frac{\epsilon }{4} + \
\frac{\epsilon ^2}{4}
\end{array}
\right)
\rightarrow
\left(
\begin{array}{ccc}
 \epsilon & \epsilon  & \epsilon \\
 \epsilon  & 1  & 1  \\
 \epsilon & 1  & 1
\end{array}
\right),
\end{equation}
and
\begin{equation}
M_\nu^\text{Dirac}=m_3
\left(
\begin{array}{ccc}
 -\frac{\epsilon ^2}{\sqrt{2}} & - \frac{\epsilon }{\sqrt{2}} & 0 \\
 \frac{\epsilon ^2}{2} & - \frac{\epsilon }{2} & \frac{1}{\sqrt{2}} \\
 -\frac{\epsilon ^2}{2} & \frac{\epsilon }{2} & \frac{1}{\sqrt{2}}
\end{array}
\right)
\rightarrow
\left(
\begin{array}{ccc}
 \epsilon ^2 & \epsilon  & 0 \\
  \epsilon ^2 & \epsilon  & 1  \\
 \epsilon ^2 & \epsilon & 1
\end{array}
\right),
\end{equation}
\end{subequations}
which is similar to the example given before.

\subsubsection*{Best-Fit Matrices}\label{sec:bestfitmatrices}

Let us now consider the two best examples of models which provide
perfect fits to current data. These two examples minimize the selector
in Eq.~(\ref{equ:selector}), which takes for these models
the smallest possible value $S=0.12$. We denote
these textures as ``best-fit matrices'' since they provide a ``perfect
fit'' to current data. Both examples exhibit similar
PMNS mixing angles with a $\theta_{13}$ that is
in the reach of next-generation neutrino oscillation experiments.
For example, $\theta_{13}$ is large enough to be discovered by Double Chooz, which means
that for large $\theta_{13}$, these representatives would be the perfect candidates.

{\bf Example (a)} -- Our first example for the best-fit matrices has the charged lepton and
neutrino mixing parameters
\begin{subequations}
\begin{eqnarray}
(s^\ell_{12}, s^\ell_{13}, s^\ell_{23}, \delta^\ell) & = & (\frac{1}{\sqrt{2}},\epsilon,\frac{1}{\sqrt{2}},\pi)\,,\\
(s^\nu_{12}, s^\nu_{13}, s^\nu_{23},\delta^\nu, \widehat{\varphi}_1, \widehat{\varphi}_2, \widehat{\phi}_1, \widehat{\phi}_2) & = & (\epsilon^2,\frac{1}{\sqrt{2}},\epsilon,0,\pi,\pi,\pi,\pi),
\end{eqnarray}
\end{subequations}
which lead to the PMNS mixing angles
\begin{equation}
(\theta_{12},\theta_{13},\theta_{23})=(33.4^\circ,7.5^\circ,43.5^\circ).
\end{equation}
The charged lepton, neutrino, and PMNS mixing matrices are
\begin{subequations}
\begin{equation}
U_\ell=
\left(
\begin{array}{ccc}
 \frac{1}{\sqrt{2}}-\frac{\epsilon ^2}{2 \sqrt{2}} & \frac{1}{\sqrt{2}}-\frac{\epsilon ^2}{2 \sqrt{2}} & -\epsilon  \\
 -\frac{1}{2}+\frac{\epsilon }{2} & \frac{1}{2}+\frac{\epsilon }{2} & \frac{1}{\sqrt{2}}-\frac{\epsilon ^2}{2 \sqrt{2}} \\
 \frac{1}{2}+\frac{\epsilon }{2} & -\frac{1}{2}+\frac{\epsilon }{2} & \frac{1}{\sqrt{2}}-\frac{\epsilon ^2}{2 \sqrt{2}}
\end{array}
\right),
\quad
U_\nu=
\left(
\begin{array}{ccc}
 -\frac{1}{\sqrt{2}} & -\frac{\epsilon ^2}{\sqrt{2}} & \frac{1}{\sqrt{2}} \\
 -\frac{\epsilon }{\sqrt{2}}-\epsilon ^2 & 1-\frac{\epsilon ^2}{2} & -\frac{\epsilon }{\sqrt{2}} \\
 -\frac{1}{\sqrt{2}}+\frac{\epsilon ^2}{2 \sqrt{2}} & -\epsilon-\frac{\epsilon ^2}{\sqrt{2}}  & -\frac{1}{\sqrt{2}}+\frac{\epsilon ^2}{2\sqrt{2}}
\end{array}
\right),
\end{equation}
\begin{equation}
U_\text{PMNS}=
\left(
\begin{array}{rrr}
 -0.9+0.6 \,\epsilon ^2 & -0.5-1.1 \,\epsilon ^2 & 0.1-0.4 \,\epsilon ^2 \\
-0.1 -0.7\,\epsilon-0.8 \,\epsilon ^2 &   0.5+\epsilon -0.9 \,\epsilon ^2& 0.9-0.7\,\epsilon-0.8 \,\epsilon ^2 \\
 -0.5+0.2 \,\epsilon-0.2 \,\epsilon ^2  & 0.7-0.7\,\epsilon-1.2 \,\epsilon ^2 &
   -0.5-1.2 \,\epsilon+0.5\,\epsilon ^2 
\end{array}
\right).
\end{equation}
Note that, in this case, $U_\ell$ is on a bimaximal mixing form but
$U_\nu$ is neither the CKM-matrix nor CKM-like. For the textures of
charged leptons and neutrinos we find
\begin{equation}
M_\ell=m_\tau
\left(
\begin{array}{ccc}
 0 & \frac{\epsilon ^2}{\sqrt{2}} & -\epsilon  \\
 0 & \frac{\epsilon ^2}{2} & \frac{1}{\sqrt{2}}-\frac{\epsilon ^2}{2 \sqrt{2}} \\
 0 & -\frac{\epsilon ^2}{2} & \frac{1}{\sqrt{2}}-\frac{\epsilon ^2}{2 \sqrt{2}}
\end{array}
\right)
\rightarrow
\left(
\begin{array}{ccc}
 0 & \epsilon ^2 & \epsilon  \\
 0 & \epsilon ^2 & 1 \\
 0 & \epsilon ^2 & 1
\end{array}
\right),
\end{equation}
\begin{equation}
M_\nu^\text{Maj}=m_3
\left(
\begin{array}{ccc}
 \frac{1}{2}+\frac{\epsilon ^2}{2} & -\frac{\epsilon }{2} & -\frac{1}{2}+\frac{3 \epsilon ^2}{4} \\
 -\frac{\epsilon }{2} & \epsilon+\frac{\epsilon ^2}{2}  & \frac{\epsilon }{2}-\epsilon ^2 \\
 -\frac{1}{2}+\frac{3 \epsilon ^2}{4} & \frac{\epsilon }{2}-\epsilon ^2 & \frac{1}{2}
\end{array}
\right)
\rightarrow
\left(
\begin{array}{ccc}
 1 & \epsilon  & 1 \\
 \epsilon  & \epsilon  & \epsilon  \\
 1 & \epsilon  & 1
\end{array}
\right),
\end{equation}
and
\begin{equation}
M_\nu^\text{Dirac}=m_3
\left(
\begin{array}{ccc}
 -\frac{\epsilon ^2}{\sqrt{2}} & 0 & \frac{1}{\sqrt{2}} \\
 0 & \epsilon  & -\frac{\epsilon }{\sqrt{2}} \\
 -\frac{\epsilon ^2}{\sqrt{2}} & -\epsilon ^2 & -\frac{1}{\sqrt{2}}+\frac{\epsilon ^2}{2 \sqrt{2}}
\end{array}
\right)
\rightarrow
\left(
\begin{array}{ccc}
 \epsilon ^2 & 0 & 1 \\
 0 & \epsilon  & \epsilon  \\
 \epsilon ^2 & \epsilon ^2 & 1
\end{array}
\right).
\end{equation}
\end{subequations}
We thus see that $M_\ell$ is on a typical lopsided form with a strictly
hierarchical structure within each row. On the other hand,
$M_\nu^\text{Maj}$ and $M_\nu^\text{Dirac}$ have nearly degenerate large
entries in the first and third row/column with small entries in the
second row/column. 

{\bf Example (b)} -- Our second example for the best-fit matrices has the charged lepton and neutrino mixing parameters
\begin{subequations}
\begin{eqnarray}
(s^\ell_{12}, s^\ell_{13}, s^\ell_{23}, \delta^\ell) & = & (\frac{1}{\sqrt{2}},\epsilon,\epsilon,0),\\
(s^\nu_{12}, s^\nu_{13}, s^\nu_{23},\delta^\nu, \widehat{\varphi}_1, \widehat{\varphi}_2, \widehat{\phi}_1, \widehat{\phi}_2) & = & (\epsilon^2,\frac{1}{\sqrt{2}},\frac{1}{\sqrt{2}},\pi,\pi,\pi,\pi,\pi),
\end{eqnarray}
\end{subequations}
which lead to the PMNS mixing angles
\begin{equation}
(\theta_{12},\theta_{13},\theta_{23})=(33.4^\circ,7.5^\circ,43.5^\circ).
\end{equation}
Thus, the charged lepton, neutrino, and PMNS mixing matrices are
\begin{subequations}
\begin{equation}
U_\ell=
\left(
\begin{array}{ccc}
 \frac{1}{\sqrt{2}}-\frac{\epsilon ^2}{2 \sqrt{2}} & \frac{1}{\sqrt{2}}-\frac{\epsilon ^2}{2 \sqrt{2}} & \epsilon  \\
 -\frac{1}{\sqrt{2}}-\frac{\epsilon ^2}{2 \sqrt{2}} & \frac{1}{\sqrt{2}}-\frac{3 \epsilon ^2}{2 \sqrt{2}} & \epsilon  \\
 0 & -\sqrt{2} \epsilon  & 1-\epsilon ^2
\end{array}
\right),
\quad
U_\nu=
\left(
\begin{array}{ccc}
 -\frac{1}{\sqrt{2}} & -\frac{\epsilon ^2}{\sqrt{2}} & -\frac{1}{\sqrt{2}} \\
 \frac{1}{2}-\frac{\epsilon ^2}{\sqrt{2}} & \frac{1}{\sqrt{2}}+\frac{\epsilon ^2}{2}& -\frac{1}{2} \\
 \frac{1}{2}+\frac{\epsilon ^2}{\sqrt{2}} & -\frac{1}{\sqrt{2}}+\frac{\epsilon ^2}{2} & -\frac{1}{2}
\end{array}
\right),
\end{equation}
\begin{equation}
U_\text{PMNS}=
\left(
\begin{array}{rrr}
-0.9+0.6 \,\epsilon ^2 & -0.5-1.1 \,\epsilon ^2 & -0.1+0.4 \,\epsilon ^2 \\
-0.1-0.7\,\epsilon-0.8 \,\epsilon ^2 & 0.5+\epsilon -0.9 \,\epsilon ^2 & -0.9+0.7\,\epsilon+0.8\,\epsilon ^2 \\
 0.5-0.2 \,\epsilon +0.2 \,\epsilon ^2 & -0.7+0.7\,\epsilon+1.2 \,\epsilon ^2 &
   -0.5-1.2 \,\epsilon +0.5\,\epsilon ^2
\end{array}
\right).
\end{equation}
\end{subequations}
As a consequence, we find for the textures of charged leptons and
neutrinos
\begin{subequations}
\begin{equation}
M_\ell=m_\tau
\left(
\begin{array}{ccc}
 0 & \frac{\epsilon ^2}{\sqrt{2}} & \epsilon  \\
 0 & \frac{\epsilon ^2}{\sqrt{2}} & \epsilon  \\
 0 & 0 & 1-\epsilon ^2
\end{array}
\right)
\rightarrow
\left(
\begin{array}{ccc}
 0 & \epsilon ^2 & \epsilon  \\
 0 & \epsilon ^2 & \epsilon  \\
 0 & 0 & 1
\end{array}
\right),
\end{equation}
\begin{equation}
M_\nu^\text{Maj}=m_3
\left(
\begin{array}{ccc}
 \frac{1}{2}+\frac{\epsilon ^2}{2} & \frac{1}{2 \sqrt{2}}-\frac{\epsilon ^2}{2 \sqrt{2}} & \frac{1}{2
   \sqrt{2}}-\frac{\epsilon ^2}{2 \sqrt{2}} \\
 \frac{1}{2 \sqrt{2}}-\frac{\epsilon ^2}{2 \sqrt{2}} & \frac{1}{4}+\frac{\epsilon }{2}+\frac{\epsilon ^2}{4} &
   \frac{1}{4}-\frac{\epsilon }{2}+\frac{\epsilon ^2}{4} \\
 \frac{1}{2 \sqrt{2}}-\frac{\epsilon ^2}{2 \sqrt{2}} & \frac{1}{4}-\frac{\epsilon }{2}+\frac{\epsilon ^2}{4} &
   \frac{1}{4}+\frac{\epsilon }{2}+\frac{\epsilon ^2}{4}
\end{array}
\right)
\rightarrow
\left(
\begin{array}{ccc}
1 & 1 & 1 \\
1 & 1 & 1 \\
1 & 1 & 1
\end{array}
\right),
\end{equation}
and
\begin{equation}
M_\nu^\text{Dirac}=m_3
\left(
\begin{array}{ccc}
 -\frac{\epsilon ^2}{\sqrt{2}} & 0 & -\frac{1}{\sqrt{2}} \\
 \frac{\epsilon ^2}{2} & \frac{\epsilon }{\sqrt{2}} & -\frac{1}{2} \\
 \frac{\epsilon ^2}{2} & -\frac{\epsilon }{\sqrt{2}} & -\frac{1}{2}
\end{array}
\right)
\rightarrow
\left(
\begin{array}{ccc}
 \epsilon ^2 & 0 & 1 \\
 \epsilon ^2 & \epsilon  & 1 \\
 \epsilon ^2 & \epsilon  & 1
\end{array}
\right).
\end{equation}
\end{subequations}
In this case, $M_\ell$ is no longer of the usual lopsided form but has
only entries of similar (small) orders in the first and second
row. The neutrino mass matrices, however, have entries in
the last column that differ only by order unity factors, which leads,
after neglecting these factors, for $M_\nu^\text{Maj}$ to a ``democratic'' or ``anarchic'' texture. 

\clearpage
\subsection{Textures with Small $\boldsymbol{\theta_{13}}$ Surviving Increased Experimental Pressure}\label{sec:in10years}
Let us now concentrate on models with a small $\theta_{13}$
that would survive a projected increased experimental pressure in ten
years, \ie, pressure from a stronger $\theta_{13}$ bound as well as from the other parameters as estimated in
the last column of \Tab~\ref{tab:values}. The reason for considering textures with small $\theta_{13}$ is
two-fold. First, from a theoretical point of view, an understanding of
textures with small $\theta_{13}$ seems to require interesting 
model-building assumptions, {\it e.g.}, discrete non-Abelian flavor
symmetries \cite{Babu:2002dz}. Second, from an experimental perspective, a general survey shows
that most of the existing GUT model-building approaches predict a comparatively
large $\theta_{13}$, which is close to the current upper
bound \cite{Albright:2006cw}. These predictions from model-building can soon be tested in next-generation experiments such as Double Chooz, \ie, the measurement of $\theta_{13}$ exerts a strong
experimental pressure on currently allowed models.

In \Tab~\ref{tab:textures}, we have summarized all possible real
textures (only leading order entries) that would survive a projection of the experimental bounds on
the leptonic mixing angles in ten years. In order to extract these, we have used the
extrapolated errors from the last column in \Tab~\ref{tab:values} in \equ{selector} and \equ{selectioncriterion}
by assuming that the current best-fit values hold. Of course, we could as well extract the
textures for any other set of best-fit values with the same procedure, but this would exceed
the scope of this paper (see also \Ref~\cite{psw-web}).
We have listed the textures
$M_\ell$, $M_\nu^\text{Maj}$, and $M_\nu^\text{Dirac}$, for normal
hierarchical, inverted hierarchical, and degenerate neutrino
masses. The textures in
\Tab~\ref{tab:textures} have been
selected from the 264\,144 initial models following the procedure in
Sec.~\ref{sec:generating}, which assumes that the charged lepton and
neutrino mixing matrices $U_\ell$ and $U_\nu$ are real. The textures in \Tab~\ref{tab:textures} therefore
represent all the models which survive the projected experimental bounds
in ten years for the CP conserving case. The number of such models is
less than $\sim 0.01\%$ of the initial sample.

In the notation and parameterization of
Sec.~\ref{sec:generating}, \Tab~\ref{tab:textures} shows the individual mixing parameters
$(s_{12}^\ell,s_{13}^\ell,s_{23}^\ell)$,
$(s_{12}^\nu,s_{13}^\nu,s_{23}^\nu)$, and $(\delta^\ell,\delta^\nu,\widehat{\varphi}_1,\widehat{\varphi}_2)$, of $U_\ell$
and $U_\nu$, as well as the PMNS mixing angles
$(\theta_{12},\theta_{13},\theta_{23})$. Notice that we have dropped the phases
$\widehat{\phi}_1$ and $\widehat{\phi}_2$, since they can be absorbed
into the definition of the Majorana phases $\phi_1$ and
$\phi_2$. In the CP conserving case considered here, however, the
Majorana phases are either 0 or $\pi$, and thus play no role for the
structure of the textures (or the PMNS mixing angles). Moreover,
\Tab~\ref{tab:textures} contains for degenerate neutrino masses only the
Dirac neutrino mass matrix $M_\nu^\text{Dirac}$, since in the
CP conserving case $M_\nu^\text{Maj}$ would only be proportional to
the unit matrix.

For the textures \#6, \#11, and \#12, we have given two sets of PMNS
mixing angles, which correspond to different choices of the phases $0$
and $\pi$ in $U_\ell$ and $U_\nu$, resulting in a variation of the
mixing parameters. This is not the case for all other textures in the
list. Moreover, even though different phase combinations change the
PMNS mixing angles, the textures of the mass matrices are, for normal
hierarchical neutrino masses, invariant under such
permutations of the phases. For an inverted hierarchical
or degenerate neutrino mass spectrum, however, a
different choice of phase combinations can manifest itself in a
slightly different mass matrix texture, even in the CP conserving case.

\begin{landscape}
\begin{footnotesize}
\begin{longtable}{|@{\hspace{0.5mm}}c@{\hspace{0.5mm}}|@{\hspace{0.5mm}}c@{\hspace{0.5mm}}||@{\hspace{0.5mm}}c@{\hspace{0.5mm}}|@{\hspace{0.5mm}}c@{\hspace{0.5mm}}||c@{\hspace{0.5mm}}|@{\hspace{0.5mm}}c@{\hspace{0.5mm}}||c@{\hspace{0.5mm}}||@{\hspace{0.5mm}}c@{\hspace{0.5mm}}|@{\hspace{0.5mm}}c@{\hspace{0.5mm}}|}\hline

\multirow{4}{.3cm}{\#}&&\multicolumn{2}{c||}{}&\multicolumn{2}{c||}{}&&
\multirow{3}{2.3cm}{\begin{minipage}{2.3cm}\vspace{.25cm}\centering
$(s^\ell_{12}, s^\ell_{13}, s^\ell_{23})$\vspace{.1cm}\\ $(s^\nu_{12}, s^\nu_{13}, s^\nu_{23})$\vspace{.1cm}\\ $(\delta^\ell,\delta^\nu, \widehat{\varphi}_1, \widehat{\varphi}_2)$ \end{minipage}}&\multirow{4}{2cm}{$(\theta_{12}, \theta_{13}, \theta_{23})$} \\

&&\multicolumn{2}{c||}{\begin{normalsize}Normal Hierarchy\end{normalsize}}& \multicolumn{2}{c||}{\begin{normalsize}Inverted Hierarchy\end{normalsize}}&\begin{normalsize}Degenerate\end{normalsize}& & \\

&&\multicolumn{2}{c||}{}& \multicolumn{2}{c||}{}& & &\\
\vspace{-2mm}
 & \begin{normalsize}$M_\ell$\end{normalsize} & \begin{normalsize}$M_\nu^\text{Maj}$\end{normalsize} & \begin{normalsize}$M_\nu^\text{Dirac}$\end{normalsize}& \begin{normalsize}$M_\nu^\text{Maj}$\end{normalsize} & \begin{normalsize}$M_\nu^\text{Dirac}$\end{normalsize}&\begin{normalsize}$M_\nu^\text{Dirac}$\end{normalsize} &  &  \\

&&&&&&&&\\
\hline\endhead&&&&&&&&\vspace{-1mm}\\

1 &  $\left(
\begin{array}{lll}
 0 & 0 & \epsilon  \\
 0 & \epsilon ^2 & 0 \\
 0 & 0 & 1
\end{array}
\right)$  & 
 $\left(
\begin{array}{lll}
 \epsilon & \epsilon  & \epsilon  \\
 \epsilon  & 1 & 1 \\
 \epsilon  & 1 & 1
\end{array}
\right)$  & 
 $\left(
\begin{array}{lll}
 \epsilon ^2 & \epsilon  & \epsilon  \\
 \epsilon ^2 & \epsilon  & 1 \\
 \epsilon ^2 & \epsilon  & 1
\end{array}
\right)$  & 
 \hspace{-2mm}$\left(
\begin{array}{ccc}
 1 & \epsilon  & \epsilon  \\
 \epsilon  & 1 & 1 \\
 \epsilon  & 1 & 1
\end{array}
\right)$  &  $\left(
\begin{array}{ccc}
 1 & 1 & \epsilon ^2 \\
 1 & 1 & \epsilon  \\
 1 & 1 & \epsilon 
\end{array}
\right)$  
& \hspace{-2mm}$\left(
\begin{array}{ccc}
 1 & 1 & \epsilon \\
 1 & 1 & 1  \\
 1 & 1 & 1 
\end{array}
\right)$ &
\begin{minipage}{2.3cm}\begin{center}
 $(\epsilon^2,\, \epsilon,\, 0)$ \vspace{.1cm}\\
 $(\frac{1}{\sqrt{2}},\, \epsilon,\,\frac{1}{\sqrt{2}})$ \vspace{.1cm}\\
 $(\xi,\,\pi,\,0,\,\xi+\pi)$
\end{center}\end{minipage} & 
$\left(35.2^\circ, 4.9^\circ, 43.8^\circ\right)$
 \vspace{-1mm}\\*&&&&&&&&\\*\hline

&&&&&&&&\\*\vspace{-1mm}2 &  $\left(
\begin{array}{lll}
 0 & 0 & \epsilon  \\
 0 & \epsilon ^2 & \epsilon ^2 \\
 0 & 0 & 1
\end{array}
\right)$  & 
 $\left(
\begin{array}{lll}
 \epsilon & \epsilon  & \epsilon  \\
 \epsilon  & 1 & 1 \\
 \epsilon  & 1 & 1
\end{array}
\right)$  & 
 $\left(
\begin{array}{lll}
 \epsilon ^2 & \epsilon  & \epsilon  \\
 \epsilon ^2 & \epsilon  & 1 \\
 \epsilon ^2 & \epsilon  & 1
\end{array}
\right)$  & 
 \hspace{-2mm}$\left(
\begin{array}{ccc}
 1 & \epsilon  & \epsilon  \\
 \epsilon  & 1 & 1 \\
 \epsilon  & 1 & 1
\end{array}
\right)$  &  $\left(
\begin{array}{ccc}
 1 & 1 & \epsilon ^2 \\
 1 & 1 & \epsilon  \\
 1 & 1 & \epsilon 
\end{array}
\right)$  
& \hspace{-2mm}$\left(
\begin{array}{ccc}
 1 & 1 & \epsilon \\
 1 & 1 & 1  \\
 1 & 1 & 1 
\end{array}
\right)$ &
\begin{minipage}{2.3cm}\begin{center}
 $(\epsilon^2,\, \epsilon,\, \epsilon^2)$ \vspace{.1cm}\\
 $(\frac{1}{\sqrt{2}},\, \epsilon,\, \frac{1}{\sqrt{2}}$ \vspace{.1cm}\\
 $(\pi,\,\pi,\,0,\,0)$
\end{center}\end{minipage} & 
$\left(35.5^\circ, 4.5^\circ, 41.6^\circ\right)$
\vspace{-1mm} \\*&&&&&&&&\\*\hline

&&&&&&&&\\*\vspace{-1mm}3 &  $\left(
\begin{array}{lll}
 0 & 0 & \epsilon ^2 \\
 0 & \epsilon ^2 & 0 \\
 0 & 0 & 1
\end{array}
\right)$  & 
 $\left(
\begin{array}{lll}
 \epsilon & \epsilon  & \epsilon  \\
 \epsilon  & 1 & 1 \\
 \epsilon  & 1 & 1
\end{array}
\right)$  & 
 $\left(
\begin{array}{lll}
 \epsilon ^2 & \epsilon  & \epsilon  \\
 \epsilon ^2 & \epsilon  & 1 \\
 \epsilon ^2 & \epsilon  & 1
\end{array}
\right)$  
& \hspace{-3.5mm} $\left(
\begin{array}{ccc}
 1 & \epsilon  & \epsilon  \\
 \epsilon  & 1 & 1 \\
 \epsilon  & 1 & 1
\end{array}
\right)$  &  $\left(
\begin{array}{ccc}
 1 & 1 & \epsilon ^2 \\
 1 & 1 & \epsilon  \\
 1 & 1 & \epsilon 
\end{array}
\right)$ 
&\hspace{-2mm}$\left(
\begin{array}{ccc}
 1 & 1 & \epsilon \\
 1 & 1 & 1  \\
 1 & 1 & 1 
\end{array}
\right)$ & 
\begin{minipage}{2.3cm}\begin{center}
 $(\epsilon,\, \epsilon^2,\, 0)$ \vspace{.1cm}\\
 $(\frac{1}{\sqrt{2}},\, \epsilon,\, \frac{1}{\sqrt{2}})$ \vspace{.1cm}\\
 $(\xi,\,0,\,0,\,\xi+\pi)$
\end{center}\end{minipage} & 
$\left(35.2^\circ, 4.8^\circ, 46.6^\circ\right)$
 \vspace{-1mm}\\*&&&&&&&&\\*\hline

&&&&&&&&\\*\vspace{-1mm}4 &  $\left(
\begin{array}{lll}
 0 & 0 & \epsilon ^2 \\
 0 & \epsilon ^2 & \epsilon ^2 \\
 0 & 0 & 1
\end{array}
\right)$  & 
 $\left(
\begin{array}{lll}
 \epsilon & \epsilon  & \epsilon  \\
 \epsilon  & 1 & 1 \\
 \epsilon  & 1 & 1
\end{array}
\right)$  & 
 $\left(
\begin{array}{lll}
 \epsilon ^2 & \epsilon  & \epsilon  \\
 \epsilon ^2 & \epsilon  & 1 \\
 \epsilon ^2 & \epsilon  & 1
\end{array}
\right)$  & \hspace{-3mm}
 $\left(
\begin{array}{ccc}
 1 & \epsilon  & \epsilon  \\
 \epsilon  & 1 & 1 \\
 \epsilon  & 1 & 1
\end{array}
\right)$  &  $\left(
\begin{array}{ccc}
 1 & 1 & \epsilon ^2 \\
 1 & 1 & \epsilon  \\
 1 & 1 & \epsilon 
\end{array}
\right)$ 
&\hspace{-3mm} $\left(
\begin{array}{ccc}
 1 & 1 & \epsilon \\
 1 & 1 & 1  \\
 1 & 1 & 1 
\end{array}
\right)$ &
\begin{minipage}{2.3cm}\begin{center}
 $(\epsilon,\, \epsilon^2,\, \epsilon^2)$ \vspace{.1cm}\\
 $(\frac{1}{\sqrt{2}},\, \epsilon,\, \frac{1}{\sqrt{2}})$ \vspace{.1cm}\\
 $(0,\,0,\,0,\,\pi)$
\end{center}\end{minipage} & 
$\left(35.5^\circ, 4.5^\circ, 48.9^\circ\right)$
 \vspace{-1mm}\\*&&&&&&&&\\*\hline

&&&&&&&&\\*\vspace{-1mm}5 &  $\left(
\begin{array}{lll}
 0 & 0 & \epsilon \\
 0 & \epsilon ^2 & 0 \\
 0 & 0 & 1
\end{array}
\right)$  & 
 $\left(
\begin{array}{lll}
 \epsilon  & \epsilon  & \epsilon \\
 \epsilon  & 1 & 1 \\
 \epsilon & 1 & 1
\end{array}
\right)$  & 
 $\left(
\begin{array}{lll}
 \epsilon ^2 & \epsilon  & \epsilon ^2 \\
 \epsilon ^2 & \epsilon  & 1 \\
 \epsilon ^2 & \epsilon  & 1
\end{array}
\right)$  
& \hspace{-2.7mm} $\left(
\begin{array}{ccc}
 1 & \epsilon ^2 & \epsilon ^2 \\
 \epsilon ^2 & 1 & 1 \\
 \epsilon ^2 & 1 & 1
\end{array}
\right)$  &  $\left(
\begin{array}{ccc}
 1 & 1 & 0 \\
 1 & 1 & \epsilon  \\
 1 & 1 & \epsilon 
\end{array}
\right)$ 
&\hspace{-3mm} $\left(
\begin{array}{ccc}
 1 & 1 & \epsilon^2 \\
 1 & 1 & 1  \\
 1 & 1 & 1 
\end{array}
\right)$ & 
\begin{minipage}{2.3cm}\begin{center}
 $(\epsilon^2,\, \epsilon,\, 0)$ \vspace{.1cm}\\
 $(\frac{1}{\sqrt{2}},\, \epsilon^2,\, \frac{1}{\sqrt{2}})$ \vspace{.1cm}\\
 $(\xi,\,\pi,\,0,\,\xi+\pi)$
\end{center}\end{minipage} & 
$\left(35.2^\circ, 3.7^\circ, 45.1^\circ\right)$
 \vspace{-1mm}\\*&&&&&&&&\\*\hline

&&&&&&&&\\*\vspace{-1mm}6 &  $\left(
\begin{array}{lll}
 0 & 0 & \epsilon \\
 0 & \epsilon ^2 & \epsilon ^2 \\
 0 & 0 & 1
\end{array}
\right)$  & 
 $\left(
\begin{array}{lll}
 \epsilon  & \epsilon  & \epsilon \\
 \epsilon  & 1 & 1 \\
 \epsilon & 1 & 1
\end{array}
\right)$  & 
 $\left(
\begin{array}{lll}
 \epsilon ^2 & \epsilon  & \epsilon ^2 \\
 \epsilon ^2 & \epsilon  & 1 \\
 \epsilon ^2 & \epsilon  & 1
\end{array}
\right)$  
& \hspace{-2.7mm} $\left(
\begin{array}{ccc}
 1 & \epsilon ^2 & \epsilon ^2 \\
 \epsilon ^2 & 1 & 1 \\
 \epsilon ^2 & 1 & 1
\end{array}
\right)$  &  $\left(
\begin{array}{ccc}
 1 & 1 & 0 \\
 1 & 1 & \epsilon  \\
 1 & 1 & \epsilon 
\end{array}
\right)$ 
& \hspace{-1.8mm}$\left(
\begin{array}{ccc}
 1 & 1 & \epsilon^2 \\
 1 & 1 & 1  \\
 1 & 1 & 1 
\end{array}
\right)$ & 
\begin{minipage}{2.3cm}\begin{center}
 $(\epsilon^2,\, \epsilon,\, \epsilon^2)$ \vspace{.1cm}\\
 $(\frac{1}{\sqrt{2}},\, \epsilon^2,\, \frac{1}{\sqrt{2}})$ \vspace{.1cm}\\
 $(0,\,\pi,\,0,\,\pi)$ \vspace{.1cm}\\
 $(\pi,\,\pi,\,0,\,0)$
\end{center}\end{minipage} & 
\begin{minipage}{2.8cm}\begin{center}
\vspace{1.2cm}
$\left(35.0^\circ, 3.8^\circ, 47.7^\circ\right)$\vspace{.1cm}\\
$\left(35.5^\circ, 4.6^\circ, 43.1^\circ\right)$
\end{center}\end{minipage}
 \vspace{-1mm}\\*&&&&&&&&\\*\hline
\newpage
&&&&&&&&\\*\vspace{-1mm}7 &  $\left(
\begin{array}{lll}
 0 & 0 & \epsilon ^2 \\
 0 & \epsilon ^2 & 0 \\
 0 & 0 & 1
\end{array}
\right)$  & 
 $\left(
\begin{array}{lll}
 \epsilon  & \epsilon & \epsilon  \\
 \epsilon & 1 & 1 \\
 \epsilon  & 1 & 1
\end{array}
\right)$  & 
 $\left(
\begin{array}{lll}
 \epsilon ^2 & \epsilon  & \epsilon ^2 \\
 \epsilon ^2 & \epsilon  & 1 \\
 \epsilon ^2 & \epsilon  & 1
\end{array}
\right)$ 
& \hspace{-3mm} $\left(
\begin{array}{ccc}
 1 & \epsilon ^2 & \epsilon ^2 \\
 \epsilon ^2 & 1 & 1 \\
 \epsilon ^2 & 1 & 1
\end{array}
\right)$  &  $\left(
\begin{array}{ccc}
 1 & 1 & 0 \\
 1 & 1 & \epsilon  \\
 1 & 1 & \epsilon 
\end{array}
\right)$  
&\hspace{-3mm} $\left(
\begin{array}{ccc}
 1 & 1 & \epsilon^2 \\
 1 & 1 & 1  \\
 1 & 1 & 1 
\end{array}
\right)$ & 
\begin{minipage}{2.3cm}\begin{center}
 $(\epsilon,\, \epsilon^2,\, 0)$ \vspace{.1cm}\\
 $(\frac{1}{\sqrt{2}},\, \epsilon^2,\, \frac{1}{\sqrt{2}})$ \vspace{.1cm}\\
 $(\xi,\,0,\,0,\,\xi+\pi)$ 
\end{center}\end{minipage} & 
$\left(35.2^\circ, 3.7^\circ, 45.1^\circ\right)$
\vspace{-1mm} \\*&&&&&&&&\\*\hline

&&&&&&&&\\*\vspace{-1mm}8 &  $\left(
\begin{array}{lll}
 0 & 0 & \epsilon ^2 \\
 0 & \epsilon ^2 & \epsilon ^2 \\
 0 & 0 & 1
\end{array}
\right)$  & 
 $\left(
\begin{array}{lll}
 \epsilon  & \epsilon & \epsilon  \\
 \epsilon & 1 & 1 \\
 \epsilon  & 1 & 1
\end{array}
\right)$  & 
 $\left(
\begin{array}{lll}
 \epsilon ^2 & \epsilon  & \epsilon ^2 \\
 \epsilon ^2 & \epsilon  & 1 \\
 \epsilon ^2 & \epsilon  & 1
\end{array}
\right)$  
&  \hspace{-1.8mm}$\left(
\begin{array}{ccc}
 1 & \epsilon ^2 & \epsilon ^2 \\
 \epsilon ^2 & 1 & 1 \\
 \epsilon ^2 & 1 & 1
\end{array}
\right)$  &  $\left(
\begin{array}{ccc}
 1 & 1 & 0 \\
 1 & 1 & \epsilon  \\
 1 & 1 & \epsilon 
\end{array}
\right)$ 
& \hspace{-1.8mm}$\left(
\begin{array}{ccc}
 1 & 1 & \epsilon^2 \\
 1 & 1 & 1  \\
 1 & 1 & 1 
\end{array}
\right)$ & 
\begin{minipage}{2.3cm}\begin{center}
 $(\epsilon,\, \epsilon^2,\, \epsilon^2)$ \vspace{.1cm}\\
 $(\frac{1}{\sqrt{2}},\, \epsilon^2,\, \frac{1}{\sqrt{2}})$ \vspace{.1cm}\\
 $(\pi,\,0,\,0,\,0)$ \vspace{.1cm}\\
 $(0,\,0,\,0,\,\pi)$
\end{center}\end{minipage} & 
\begin{minipage}{2.8cm}\begin{center}
\vspace{1.2cm}
$\left(35.0^\circ, 3.9^\circ, 42.8^\circ\right)$\vspace{.1cm}\\
$\left(35.5^\circ, 4.7^\circ, 47.3^\circ\right)$
\end{center}\end{minipage}
 \vspace{-1mm}\\*&&&&&&&&\\*\hline\hline

&&&&&&&&\\*\vspace{-1mm}9 &  $\left(
\begin{array}{lll}
 0 & 0 & \epsilon  \\
 0 & \epsilon ^2 & 1 \\
 0 & \epsilon ^2 & 1
\end{array}
\right)$  & 
 $\left(
\begin{array}{lll}
 \epsilon & \epsilon  & \epsilon  \\
 \epsilon  & \epsilon  & \epsilon ^2 \\
 \epsilon  & \epsilon ^2 & 1
\end{array}
\right)$  & 
 $\left(
\begin{array}{lll}
 \epsilon ^2 & \epsilon  & \epsilon  \\
 \epsilon ^2 & \epsilon  & \epsilon ^2 \\
 0 & \epsilon ^2 & 1
\end{array}
\right)$  
& \hspace{-3mm} $\left(
\begin{array}{ccc}
 1 & 0 & \epsilon  \\
 0 & 1 & \epsilon ^2 \\
 \epsilon  & \epsilon ^2 & \epsilon 
\end{array}
\right)$  &  $\left(
\begin{array}{ccc}
 1 & 1 & \epsilon ^2 \\
 1 & 1 & 0 \\
 \epsilon  & \epsilon  & \epsilon 
\end{array}
\right)$ 
&\hspace{-3mm} $\left(
\begin{array}{ccc}
 1 & 1 & \epsilon \\
 1 & 1 & \epsilon^2 \\
 \epsilon  & \epsilon  & 1
\end{array}
\right)$ & 
\begin{minipage}{2.3cm}\begin{center}
 $(\epsilon^2,\, \epsilon,\, \frac{1}{\sqrt{2}})$ \vspace{.1cm}\\
 $(\frac{1}{\sqrt{2}},\, \epsilon,\, \epsilon^2)$ \vspace{.1cm}\\
 $(0,\,0,\,0,\,0)$ 
\end{center}\end{minipage} & 
$\left(35.5^\circ, 4.5^\circ, 41.6^\circ\right)$
\vspace{-1mm} \\*&&&&&&&&\\*\hline

&&&&&&&&\\*\vspace{-1mm}10 &  $\left(
\begin{array}{lll}
 0 & 0 & \epsilon ^2 \\
 0 & \epsilon ^2 & 1 \\
 0 & \epsilon ^2 & 1
\end{array}
\right)$  & 
 $\left(
\begin{array}{lll}
 \epsilon & \epsilon  & \epsilon  \\
 \epsilon  & \epsilon  & \epsilon ^2 \\
 \epsilon  & \epsilon ^2 & 1
\end{array}
\right)$  & 
 $\left(
\begin{array}{lll}
 \epsilon ^2 & \epsilon  & \epsilon  \\
 \epsilon ^2 & \epsilon  & \epsilon ^2 \\
 0 & \epsilon ^2 & 1
\end{array}
\right)$  
& \hspace{-2mm}$\left(
\begin{array}{ccc}
 1 & 0 & \epsilon  \\
 0 & 1 & \epsilon ^2 \\
 \epsilon  & \epsilon ^2 & \epsilon 
\end{array}
\right)$  &  $\left(
\begin{array}{ccc}
 1 & 1 & \epsilon ^2 \\
 1 & 1 & 0 \\
 \epsilon  & \epsilon  & \epsilon 
\end{array}
\right)$ 
&\hspace{-3mm} $\left(
\begin{array}{ccc}
 1 & 1 & \epsilon \\
 1 & 1 & \epsilon^2 \\
 \epsilon  & \epsilon  & 1
\end{array}
\right)$ & 
\begin{minipage}{2.3cm}\begin{center}
 $(\epsilon,\, \epsilon^2,\, \frac{1}{\sqrt{2}})$ \vspace{.1cm}\\
 $(\frac{1}{\sqrt{2}},\, \epsilon,\, \epsilon^2)$ \vspace{.1cm}\\
 $(0,\,0,\,0,\,\pi)$ 
\end{center}\end{minipage} & 
$\left(35.5^\circ, 4.5^\circ, 48.9^\circ\right)$
 \vspace{-1mm}\\*&&&&&&&&\\*\hline

&&&&&&&&\\*\vspace{-1mm}11a &  $\left(
\begin{array}{lll}
 0 & 0 & \epsilon  \\
 0 & \epsilon ^2 & 1 \\
 0 & \epsilon ^2 & 1
\end{array}
\right)$  & 
 $\left(
\begin{array}{lll}
 \epsilon  & \epsilon  & \epsilon ^2 \\
 \epsilon  & \epsilon  & \epsilon ^2 \\
 \epsilon ^2 & \epsilon ^2 & 1
\end{array}
\right)$  & 
 $\left(
\begin{array}{lll}
 \epsilon ^2 & \epsilon  & \epsilon ^2 \\
 \epsilon ^2 & \epsilon  & \epsilon ^2 \\
 0 & 0 & 1
\end{array}
\right)$  
& \hspace{-2.8mm} $\left(
\begin{array}{ccc}
 1 & 0 & \epsilon ^2 \\
 0 & 1 & \epsilon ^2 \\
 \epsilon ^2 & \epsilon ^2 & \epsilon 
\end{array}
\right)$  &  $\left(
\begin{array}{ccc}
 1 & 1 & 0 \\
 1 & 1 & 0 \\
 0 & \epsilon ^2 & \epsilon 
\end{array}
\right)$ 
& \hspace{-2mm}$\left(
\begin{array}{ccc}
 1 & 1 & \epsilon ^2 \\
 1 & 1 & \epsilon ^2 \\
 0 & \epsilon ^2 & 1
\end{array}
\right)$ & 
\begin{minipage}{2.3cm}\begin{center}
 $(\epsilon^2,\, \epsilon,\, \frac{1}{\sqrt{2}})$ \vspace{.1cm}\\
 $(\frac{1}{\sqrt{2}},\, \epsilon^2,\, \epsilon^2)$ \vspace{.1cm}\\
 $(0,\,\pi,\,0,\,\pi)$
\end{center}\end{minipage} & 
$\left(35.5^\circ, 4.7^\circ, 47.3^\circ\right)$
\vspace{-1mm}\\*&&&&&&&&\\*\hline

&&&&&&&&\\*\vspace{-1mm}11b &  $\left(
\begin{array}{lll}
 0 & 0 & \epsilon  \\
 0 & \epsilon ^2 & 1 \\
 0 & \epsilon ^2 & 1
\end{array}
\right)$  & 
 $\left(
\begin{array}{lll}
 \epsilon  & \epsilon  & \epsilon ^2 \\
 \epsilon  & \epsilon  & \epsilon ^2 \\
 \epsilon ^2 & \epsilon ^2 & 1
\end{array}
\right)$  & 
 $\left(
\begin{array}{lll}
 \epsilon ^2 & \epsilon  & \epsilon ^2 \\
 \epsilon ^2 & \epsilon  & \epsilon ^2 \\
 0 & 0 & 1
\end{array}
\right)$  
&\hspace{-3mm} $\left(
\begin{array}{ccc}
 1 & 0 & \epsilon ^2 \\
 0 & 1 & \epsilon ^2 \\
 \epsilon ^2 & \epsilon ^2 & \epsilon 
\end{array}
\right)$ & $\left(
\begin{array}{ccc}
 1 & 1 & 0 \\
 1 & 1 & 0 \\
 \epsilon ^2 & 0 & \epsilon 
\end{array}
\right)$ &\hspace{-3mm}
$\left(
\begin{array}{ccc}
 1 & 1 & \epsilon ^2 \\
 1 & 1 & \epsilon ^2 \\
 \epsilon ^2 & 0 & 1
\end{array}
\right)$ & 
\begin{minipage}{2.3cm}\begin{center}
 $(\epsilon^2,\, \epsilon,\, \frac{1}{\sqrt{2}})$ \vspace{.1cm}\\
 $(\frac{1}{\sqrt{2}},\, \epsilon^2,\, \epsilon^2)$ \vspace{.1cm}\\
 $(0,\,0,\,0,\,0)$
\end{center}\end{minipage} & 
$\left(35.0^\circ, 3.9^\circ, 42.8^\circ\right)$
\vspace{-1mm}\\*&&&&&&&&\\*\hline
\newpage
&&&&&&&&\\*\vspace{-1mm}12a &  $\left(
\begin{array}{lll}
 0 & 0 & \epsilon ^2 \\
 0 & \epsilon ^2 & 1 \\
 0 & \epsilon ^2 & 1
\end{array}
\right)$  & 
 $\left(
\begin{array}{lll}
 \epsilon  & \epsilon  & \epsilon ^2 \\
 \epsilon  & \epsilon  & \epsilon ^2 \\
 \epsilon ^2 & \epsilon ^2 & 1
\end{array}
\right)$  & 
 $\left(
\begin{array}{lll}
 \epsilon ^2 & \epsilon  & \epsilon ^2 \\
 \epsilon ^2 & \epsilon  & \epsilon ^2 \\
 0 & 0 & 1
\end{array}
\right)$  
& \hspace{-3mm} $\left(
\begin{array}{ccc}
 1 & 0 & \epsilon ^2 \\
 0 & 1 & \epsilon ^2 \\
 \epsilon ^2 & \epsilon ^2 & \epsilon 
\end{array}
\right)$  &  $\left(
\begin{array}{ccc}
 1 & 1 & 0 \\
 1 & 1 & 0 \\
 0 & \epsilon ^2 & \epsilon 
\end{array}
\right)$ 
& \hspace{-3mm} $\left(
\begin{array}{ccc}
 1 & 1 & \epsilon ^2 \\
 1 & 1 & \epsilon ^2 \\
 0 & \epsilon ^2 & 1
\end{array}
\right)$ & 
\begin{minipage}{2.3cm}\begin{center}
 $(\epsilon,\, \epsilon^2,\, \frac{1}{\sqrt{2}})$ \vspace{.1cm}\\
 $(\frac{1}{\sqrt{2}},\, \epsilon^2,\, \epsilon^2)$ \vspace{.1cm}\\
 $(0,\,\pi,\,0,\,0)$ 
\end{center}\end{minipage} & 
$\left(35.5^\circ, 4.6^\circ, 43.1^\circ\right)$
 \vspace{-1mm}\\*&&&&&&&&\\*\hline

&&&&&&&&\\*\vspace{-1mm}12b &  $\left(
\begin{array}{lll}
 0 & 0 & \epsilon ^2 \\
 0 & \epsilon ^2 & 1 \\
 0 & \epsilon ^2 & 1
\end{array}
\right)$  & 
 $\left(
\begin{array}{lll}
 \epsilon  & \epsilon  & \epsilon ^2 \\
 \epsilon  & \epsilon  & \epsilon ^2 \\
 \epsilon ^2 & \epsilon ^2 & 1
\end{array}
\right)$  & 
 $\left(
\begin{array}{lll}
 \epsilon ^2 & \epsilon  & \epsilon ^2 \\
 \epsilon ^2 & \epsilon  & \epsilon ^2 \\
 0 & 0 & 1
\end{array}
\right)$  
&\hspace{-2mm}$\left(
\begin{array}{ccc}
 1 & 0 & \epsilon ^2 \\
 0 & 1 & \epsilon ^2 \\
 \epsilon ^2 & \epsilon ^2 & \epsilon 
\end{array}
\right)$ & $\left(
\begin{array}{ccc}
 1 & 1 & 0 \\
 1 & 1 & 0 \\
 \epsilon ^2 & 0 & \epsilon 
\end{array}
\right)$ &\hspace{-2.8mm}
$\left(
\begin{array}{ccc}
 1 & 1 & \epsilon ^2 \\
 1 & 1 & \epsilon ^2 \\
 \epsilon ^2 & 0 & 1
\end{array}
\right)$ & 
\begin{minipage}{2.3cm}\begin{center}
 $(\epsilon,\, \epsilon^2,\, \frac{1}{\sqrt{2}})$ \vspace{.1cm}\\
 $(\frac{1}{\sqrt{2}},\, \epsilon^2,\, \epsilon^2)$ \vspace{.1cm}\\
 $(0,\,0,\,0,\,\pi)$
\end{center}\end{minipage} & 
$\left(35.0^\circ, 3.8^\circ, 47.7^\circ\right)$
\vspace{-1mm}\\*&&&&&&&&\\*\hline

&&&&&&&&\\*\vspace{-1mm}13 &  $\left(\begin{array}{lll}
 0 & 0 & \epsilon  \\
 0 & \epsilon ^2 & 1 \\
 0 & \epsilon ^2 & 1
\end{array}\right)$  & 
 $\left(
\begin{array}{lll}
 \epsilon & \epsilon  & \epsilon  \\
 \epsilon  &\epsilon  & \epsilon ^2 \\
 \epsilon  & \epsilon ^2 & 1
\end{array}\right)$  & 
 $\left(
\begin{array}{lll}
 \epsilon ^2 & \epsilon  & \epsilon  \\
 \epsilon ^2 & \epsilon  & 0 \\
 0 & \epsilon ^2 & 1
\end{array}\right)$  
& \hspace{-3mm} $\left(
\begin{array}{ccc}
 1 & 0 & \epsilon  \\
 0 & 1 & 0 \\
 \epsilon  & 0 & \epsilon 
\end{array}
\right)$  &  $\left(
\begin{array}{ccc}
 1 & 1 & \epsilon ^2 \\
 1 & 1 & 0 \\
 \epsilon  & \epsilon  & \epsilon 
\end{array}
\right)$ 
& \hspace{-2mm}$\left(
\begin{array}{ccc}
 1 & 1 & \epsilon \\
 1 & 1 & 0 \\
 \epsilon  & \epsilon  & 1
\end{array}
\right)$ & 
\begin{minipage}{2.3cm}\begin{center}
 $(\epsilon^2,\, \epsilon,\, \frac{1}{\sqrt{2}})$ \vspace{.1cm}\\
 $(\frac{1}{\sqrt{2}},\, \epsilon,\,0)$ \vspace{.1cm}\\
 $(0,\,\xi,\,0,\,\xi)$
\end{center}\end{minipage} & $\left(35.2^\circ, 4.9^\circ, 43.8^\circ\right)$
\vspace{-1mm} \\*&&&&&&&&\\*\hline

&&&&&&&&\\*\vspace{-1mm}14 &  $\left(
\begin{array}{lll}
 0 & 0 & \epsilon ^2 \\
 0 & \epsilon ^2 & 1 \\
 0 & \epsilon ^2 & 1
\end{array}
\right)$  & 
 $\left(
\begin{array}{lll}
 \epsilon & \epsilon  & \epsilon  \\
 \epsilon  & \epsilon  & \epsilon ^2 \\
 \epsilon  & \epsilon ^2 & 1
\end{array}
\right)$  & 
 $\left(
\begin{array}{lll}
 \epsilon ^2 & \epsilon  & \epsilon  \\
 \epsilon ^2 & \epsilon  & 0 \\
 0 & \epsilon ^2 & 1
\end{array}
\right)$  
& \hspace{-2.2mm} $\left(
\begin{array}{ccc}
 1 & 0 & \epsilon  \\
 0 & 1 & 0 \\
 \epsilon  & 0 & \epsilon 
\end{array}
\right)$  &  $\left(
\begin{array}{ccc}
 1 & 1 & \epsilon ^2 \\
 1 & 1 & 0 \\
 \epsilon  & \epsilon  & \epsilon 
\end{array}
\right)$ 
& \hspace{-2mm}$\left(
\begin{array}{ccc}
 1 & 1 & \epsilon \\
 1 & 1 & 0 \\
 \epsilon  & \epsilon  & 1
\end{array}
\right)$ & 
\begin{minipage}{2.3cm}\begin{center}
 $(\epsilon,\, \epsilon^2,\, \frac{1}{\sqrt{2}})$ \vspace{.1cm}\\
 $(\frac{1}{\sqrt{2}},\, \epsilon,\, 0)$ \vspace{.1cm}\\
 $(0,\,\xi,\,0,\,\xi+\pi)$
\end{center}\end{minipage} & 
$\left(35.2^\circ, 4.8^\circ, 46.6^\circ\right)$
 \vspace{-1mm}\\*&&&&&&&&\\*\hline

&&&&&&&&\\*\vspace{-1mm}15 &  $\left(
\begin{array}{lll}
 0 & 0 & \epsilon  \\
 0 & \epsilon ^2 & 1 \\
 0 & \epsilon ^2 & 1
\end{array}
\right)$  & 
 $\left(
\begin{array}{lll}
 \epsilon  & \epsilon  & \epsilon ^2 \\
 \epsilon  & \epsilon  & 0 \\
 \epsilon ^2 & 0 & 1
\end{array}
\right)$  & 
 $\left(
\begin{array}{lll}
 \epsilon ^2 & \epsilon  & \epsilon ^2 \\
 \epsilon ^2 & \epsilon  & 0 \\
 0 & 0 & 1
\end{array}
\right)$  
& \hspace{-3mm} $\left(
\begin{array}{ccc}
 1 & 0 & \epsilon ^2 \\
 0 & 1 & 0 \\
 \epsilon ^2 & 0 & \epsilon 
\end{array}
\right)$  &  $\left(
\begin{array}{ccc}
 1 & 1 & 0 \\
 1 & 1 & 0 \\
 \epsilon ^2 & \epsilon ^2 & \epsilon 
\end{array}
\right)$ 
& \hspace{-1.8mm}$\left(
\begin{array}{ccc}
 1 & 1 & \epsilon ^2 \\
 1 & 1 & 0 \\
 \epsilon ^2 & \epsilon ^2 & 1
\end{array}
\right)$ & 
\begin{minipage}{2.3cm}\begin{center}
 $(\epsilon^2,\, \epsilon,\, \frac{1}{\sqrt{2}})$ \vspace{.1cm}\\
 $(\frac{1}{\sqrt{2}},\, \epsilon^2,\, 0)$ \vspace{.1cm}\\
 $(0,\,\xi,\,0,\,\xi)$
\end{center}\end{minipage} & 
$\left(35.2^\circ, 4.2^\circ, 45.4^\circ\right)$
\vspace{-1mm}\\*&&&&&&&&\\*\hline

&&&&&&&&\\*\vspace{-1mm}16 &  $\left(
\begin{array}{lll}
 0 & 0 & \epsilon ^2 \\
 0 & \epsilon ^2 & 1 \\
 0 & \epsilon ^2 & 1
\end{array}
\right)$  & 
 $\left(
\begin{array}{lll}
 \epsilon  & \epsilon  & \epsilon ^2 \\
 \epsilon  & \epsilon  & 0 \\
 \epsilon ^2 & 0 & 1
\end{array}
\right)$  & 
 $\left(
\begin{array}{lll}
 \epsilon ^2 & \epsilon  & \epsilon ^2 \\
 \epsilon ^2 & \epsilon  & 0 \\
 0 & 0 & 1
\end{array}
\right)$  
& \hspace{-3mm} $\left(
\begin{array}{ccc}
 1 & 0 & \epsilon ^2 \\
 0 & 1 & 0 \\
 \epsilon ^2 & 0 & \epsilon 
\end{array}
\right)$  &  $\left(
\begin{array}{ccc}
 1 & 1 & 0 \\
 1 & 1 & 0 \\
 \epsilon ^2 & \epsilon ^2 & \epsilon 
\end{array}
\right)$ 
&\hspace{-3mm} $\left(
\begin{array}{ccc}
 1 & 1 & \epsilon ^2 \\
 1 & 1 & 0 \\
 \epsilon ^2 & \epsilon ^2 & 1
\end{array}
\right)$ & 
\begin{minipage}{2.3cm}\begin{center}
 $(\epsilon,\, \epsilon^2,\, \frac{1}{\sqrt{2}})$ \vspace{.1cm}\\
 $(\frac{1}{\sqrt{2}},\, \epsilon^2,\, 0)$ \vspace{.1cm}\\
 $(0,\, \xi,\,0,\,\xi+\pi)$
\end{center}\end{minipage} & 
$\left(35.2^\circ, 3.7^\circ, 45.1^\circ\right)$
\vspace{-1mm} \\*&&&&&&&&\\*
\hline\newpage\hline\hline
&&&&&&&&\\*\vspace{-1mm}17 &  $\left(
\begin{array}{lll}
 0 & \epsilon ^2 & \epsilon  \\
 0 & \epsilon ^2 & 1 \\
 0 & \epsilon ^2 & 1
\end{array}
\right)$  & 
 $\left(
\begin{array}{lll}
 1 & \epsilon ^2 & 1 \\
 \epsilon ^2 & \epsilon  & \epsilon ^2 \\
 1 & \epsilon ^2 & 1
\end{array}
\right)$  & 
 $\left(
\begin{array}{lll}
 \epsilon ^2 & \epsilon ^2 & 1 \\
 0 & \epsilon  & 0 \\
 \epsilon ^2 & \epsilon ^2 & 1
\end{array}
\right)$  
&  \hspace{-2mm}$\left(
\begin{array}{ccc}
 1 & 0 & 1 \\
 0 & 1 & 0 \\
 1 & 0 & 1
\end{array}
\right)$  &  $\left(
\begin{array}{ccc}
 1 & \epsilon  & \epsilon  \\
 \epsilon  & 1 & 0 \\
 1 & \epsilon  & \epsilon 
\end{array}
\right)$ 
&\hspace{-3mm} $\left(
\begin{array}{ccc}
 1 & \epsilon  & 1  \\
 \epsilon  & 1 & 0 \\
 1 & \epsilon  & 1
\end{array}
\right)$ & 
\begin{minipage}{2.3cm}\begin{center}
 $(\frac{1}{\sqrt{2}},\, \epsilon,\, \frac{1}{\sqrt{2}})$ \vspace{.1cm}\\
 $(\epsilon,\, \frac{1}{\sqrt{2}},\, 0)$ \vspace{.1cm}\\
 $(\pi,\, \xi,\,\pi,\,\xi+\pi)$
\end{center}\end{minipage} & 
$\left(35.2^\circ, 3.8^\circ, 50.8^\circ\right)$
 \vspace{-1mm}\\*&&&&&&&&\\*\hline

&&&&&&&&\\*\vspace{-1mm}18 &  $\left(
\begin{array}{lll}
 0 & \epsilon ^2 & \epsilon  \\
 0 & \epsilon ^2 & 1 \\
 0 & \epsilon ^2 & 1
\end{array}
\right)$  & 
 $\left(
\begin{array}{lll}
 1 & \epsilon ^2 & 1 \\
 \epsilon ^2 & \epsilon  & \epsilon ^2 \\
 1 & \epsilon ^2 & 1
\end{array}
\right)$  & 
 $\left(
\begin{array}{lll}
 \epsilon ^2 & \epsilon ^2 & 1 \\
 0 & \epsilon  & \epsilon ^2 \\
 \epsilon ^2 & \epsilon ^2 & 1
\end{array}
\right)$ 
&\hspace{-3mm} $\left(
\begin{array}{ccc}
 1 & \epsilon ^2 & 1 \\
 \epsilon ^2 & 1 & \epsilon ^2 \\
 1 & \epsilon ^2 & 1
\end{array}
\right)$  &  $\left(
\begin{array}{ccc}
 1 & \epsilon  & \epsilon  \\
 \epsilon  & 1 & 0 \\
 1 & \epsilon  & \epsilon 
\end{array}
\right)$  
&\hspace{-3mm} $\left(
\begin{array}{ccc}
 1 & \epsilon  & 1  \\
 \epsilon  & 1 & \epsilon^2 \\
 1 & \epsilon  & 1
\end{array}
\right)$ & 
\begin{minipage}{2.3cm}\begin{center}
 $(\frac{1}{\sqrt{2}},\, \epsilon,\, \frac{1}{\sqrt{2}})$ \vspace{.1cm}\\
 $(\epsilon,\, \frac{1}{\sqrt{2}},\, \epsilon^2)$ \vspace{.1cm}\\
 $(\pi,\, \pi,\,\pi,\,0)$
\end{center}\end{minipage} & 
$\left(33.6^\circ, 3.1^\circ, 52.2^\circ\right)$
 \vspace{-1mm}\\*&&&&&&&&\\*\hline\hline

&&&&&&&&\\*\vspace{-1mm}19 &  $\left(
\begin{array}{lll}
 0 & \epsilon ^2 & \epsilon  \\
 0 & \epsilon ^2 & 0 \\
 0 & 0 & 1
\end{array}
\right)$  & 
 $\left(
\begin{array}{lll}
 1 & 1 & 1 \\
 1 & 1 & 1 \\
 1 & 1 & 1
\end{array}
\right)$  & 
 $\left(
\begin{array}{lll}
 \epsilon ^2 & \epsilon ^2 & 1 \\
 \epsilon ^2 & \epsilon  & 1 \\
 \epsilon ^2 & \epsilon  & 1
\end{array}
\right)$  
& \hspace{-3mm} $\left(
\begin{array}{ccc}
 1 & 1 & 1 \\
 1 & 1 & 1 \\
 1 & 1 & 1
\end{array}
\right)$  &  $\left(
\begin{array}{ccc}
 1 & \epsilon  & \epsilon  \\
 1 & 1 & \epsilon  \\
 1 & 1 & \epsilon 
\end{array}
\right)$ 
& \hspace{-2mm}$\left(
\begin{array}{ccc}
 1 & \epsilon  & 1  \\
 1 & 1 & 1 \\
 1 & 1 & 1
\end{array}
\right)$ & 
\begin{minipage}{2.3cm}\begin{center}
 $(\frac{1}{\sqrt{2}},\, \epsilon,\, 0)$ \vspace{.1cm}\\
 $(\epsilon,\, \frac{1}{\sqrt{2}},\, \frac{1}{\sqrt{2}})$ \vspace{.1cm}\\
 $(\xi,\, \pi,\,\pi,\,\xi+\pi)$
\end{center}\end{minipage} & 
$\left(35.2^\circ, 3.8^\circ, 50.8^\circ\right)$
\vspace{-1mm}\\*&&&&&&&&\\*\hline

\vspace{-1mm}&&&&&&&&\\*\vspace{-1mm}20 &  $\left(
\begin{array}{lll}
 0 & \epsilon ^2 & \epsilon  \\
 0 & \epsilon ^2 & \epsilon ^2 \\
 0 & 0 & 1
\end{array}
\right)$  & 
 $\left(
\begin{array}{lll}
 1 & 1 & 1 \\
 1 & 1 & 1 \\
 1 & 1 & 1
\end{array}
\right)$  & 
 $\left(
\begin{array}{lll}
 \epsilon ^2 & \epsilon ^2 & 1 \\
 \epsilon ^2 & \epsilon  & 1 \\
 \epsilon ^2 & \epsilon  & 1
\end{array}
\right)$  
& \hspace{-3mm} $\left(
\begin{array}{ccc}
 1 & 1 & 1 \\
 1 & 1 & 1 \\
 1 & 1 & 1
\end{array}
\right)$  &  $\left(
\begin{array}{ccc}
 1 & \epsilon  & \epsilon  \\
 1 & 1 & \epsilon  \\
 1 & 1 & \epsilon 
\end{array}
\right)$ 
& \hspace{-2mm}$\left(
\begin{array}{ccc}
 1 & \epsilon  & 1 \\
 1 & 1 & 1 \\
 1 & 1 & 1
\end{array}
\right)$ & 
\begin{minipage}{2.3cm}\begin{center}
 $(\frac{1}{\sqrt{2}},\, \epsilon,\, \epsilon^2)$ \vspace{.1cm}\\
 $(\epsilon,\, \frac{1}{\sqrt{2}},\, \frac{1}{\sqrt{2}})$ \vspace{.1cm}\\
 $(\pi,\, \pi,\,\pi,\,0)$
\end{center}\end{minipage} & $\left(33.6^\circ, 3.1^\circ, 52.2^\circ\right)$
\vspace{-1mm}\\*&&&&&&&&\\*\hline
\multicolumn{6}{c}{}\vspace{-.15cm}\\
\caption{\label{tab:textures} Mass matrix textures surviving the
  projected increased experimental pressure in ten years (see
  \Tab~\ref{tab:values}). For the mass matrices of charged leptons and
  Dirac neutrinos (for normal hierarchical, inverted hierarchical, and
  degenerate neutrinos, respectively), we use the basis in which the
  mixing angles of the right-handed fields are zero. Note that in the
  case of degenerate neutrinos, the Majorana mass matrix textures are
  simply the unit matrix and they are therefore neglected. The corresponding mixing angles and phases of $U_\ell, U_\nu$ (with $\xi\in\{0,\,\pi\}$) as well as the mixing angles of $U_\text{PMNS}$ are also shown. Note that in some cases, a phase variation in $U_\ell$ and $U_\nu$, without changing their mixing angles, can lead to different forms of $U_\text{PMNS}$. In these cases, we relate the causal phases of  $U_\ell$ and $U_\nu$ to their corresponding PMNS mixing angles.}
\end{longtable}
\end{footnotesize}
\end{landscape}

The textures in \Tab~\ref{tab:textures} can be roughly divided into
four classes:
\begin{enumerate}
 \item ``Semi-anarchic'' textures (\#1--8): $\theta_{12}^\nu$ and $\theta_{23}^\nu$ are maximal, all other mixing
 angles are small. Here, $M_\ell$ has a strongly hierarchical
 structure, whereas $M_\nu^\text{Maj}$
 and $M_\nu^\text{Dirac}$ have several leading order entries.
 \item Lopsided textures (\#9--16): $\theta_{23}^\ell$ and
 $\theta_{12}^\nu$ are maximal, all other mixing angles are
 small. $M_\ell$ is on a lopsided (highly asymmetric) form.
 \item ``Diamond'' textures (\# 17--18): $\theta_{12}^\ell$,
 $\theta_{23}^\ell$, and $\theta_{13}^\nu$, are maximal, all other
 mixing angles are small. $M_\ell$ is on a lopsided form. In the
 neutrino mass matrix, leading order entries are situated in the
 corners.
 \item Anarchic textures (\#19--20): $\theta_{12}^\ell$,
 $\theta_{13}^\nu$, and $\theta_{23}^\nu$, are maximal, all other
 mixing angles are small. $M_\ell$ has a strongly hierarchical
 structure, whereas $M_\nu^\text{Maj}$ and
 $M_\nu^\text{Dirac}$ have many leading order entries.
\end{enumerate}
We have called the matrices \#17 and \#18 ``diamond'' textures,
because of the diamond-like structure of the small entries in
$M_\nu^\text{Maj}$. Most of the textures in \Tab~\ref{tab:textures}
fall either into the first or the second class: 8
textures are semi-anarchic and 8 are lopsided. The remaining four
textures are equally distributed between the diamond and the anarchic
forms. For the semi-anarchic and lopsided textures, the charged lepton
and neutrino mixing matrices exhibit each a single maximal mixing
angle. The diamond and anarchic textures, on the other hand,
have in $U_\ell$ (diamond) or $U_\nu$ (anarchic) an additional maximal
mixing angle such that the total number of maximal mixing
angles in the charged lepton and neutrino sectors is three.

The diamond textures (\#17 and \#18) are
interesting, since they can be viewed as representing a generalization
of the lopsided models with two maximal mixing angles in the charged
lepton sector. They also have (like the anarchic examples) the unusual
feature that $\theta_{13}^\nu$ is maximal. Let us therefore briefly
consider these mass matrices in some more detail. For the diamond textures, the charged
lepton, neutrino, and PMNS mixing matrices are

\begin{equation}
U_\ell=
\left(
\begin{array}{ccc}
 \frac{1}{\sqrt{2}}-\frac{\epsilon ^2}{2 \sqrt{2}} & \frac{1}{\sqrt{2}}-\frac{\epsilon ^2}{2 \sqrt{2}} & -\epsilon  \\
 \frac{\epsilon }{2}-\frac{1}{2} & \frac{\epsilon }{2}+\frac{1}{2} & \frac{1}{\sqrt{2}}-\frac{\epsilon ^2}{2 \sqrt{2}} \\
 \frac{\epsilon }{2}+\frac{1}{2} & \frac{\epsilon }{2}-\frac{1}{2} & \frac{1}{\sqrt{2}}-\frac{\epsilon ^2}{2 \sqrt{2}}
\end{array}
\right),
\quad
U_\nu=
\left(
\begin{array}{ccc}
 \frac{1}{\sqrt{2}}-\frac{\epsilon ^2}{2 \sqrt{2}} & \frac{\epsilon }{\sqrt{2}} & \frac{1}{\sqrt{2}} \\
 \epsilon  & \frac{\epsilon ^2}{2}-1 & 0 \\
 \frac{1}{\sqrt{2}}-\frac{\epsilon ^2}{2 \sqrt{2}} & \frac{\epsilon }{\sqrt{2}} & -\frac{1}{\sqrt{2}}
\end{array}
\right),
\end{equation}
\begin{equation}
U_\text{PMNS}=
\left(
\begin{array}{rrr}
0.9-0.1\,\epsilon-0.2\,\epsilon^2 & 0.5+0.4\,\epsilon+0.1\,\epsilon^2 & 0.1-0.4\,\epsilon-0.3\,\epsilon^2 \\
0.1+0.9\,\epsilon+0.2\,\epsilon^2 & -0.5-0.4\,\epsilon+0.6\,\epsilon^2 & 0.9-0.4\,\epsilon-0.3\,\epsilon^2 \\
0.5-0.5\,\epsilon^2 & -0.7+0.5\,\epsilon & -0.5-0.7\,\epsilon+0.3\,\epsilon^2
\end{array}
\right).
\end{equation}
We thus obtain for the expansion of the diamond textures in
powers of $\epsilon$ the forms
\begin{subequations}\label{equ:diamond}
\begin{equation}
M_\ell=
m_\tau\left(
\begin{array}{ccc}
 0 & \frac{\epsilon ^2}{\sqrt{2}} & -\epsilon  \\
 0 & \frac{\epsilon ^2}{2} & \frac{1}{\sqrt{2}}-\frac{\epsilon ^2}{2 \sqrt{2}} \\
 0 & -\frac{\epsilon ^2}{2} & \frac{1}{\sqrt{2}}-\frac{\epsilon ^2}{2 \sqrt{2}}
\end{array}
\right),
\end{equation}
and
\begin{equation}
M_\nu^\text{Maj}=
m_3\left(
\begin{array}{ccc}
 \frac{1}{2}+\frac{\epsilon ^2}{2} & -\frac{\epsilon ^2}{\sqrt{2}} & -\frac{1}{2}+\frac{\epsilon ^2}{2} \\
 -\frac{\epsilon ^2}{\sqrt{2}} & \epsilon  & -\frac{\epsilon ^2}{\sqrt{2}} \\
 -\frac{1}{2}+\frac{\epsilon ^2}{2} & -\frac{\epsilon ^2}{\sqrt{2}} & \frac{1}{2}+\frac{\epsilon ^2}{2}
\end{array}
\right),\quad
M_\nu^\text{Dirac}=
m_3\,\left(
\begin{array}{ccc}
 -\frac{\epsilon ^2}{\sqrt{2}} & -\frac{\epsilon ^2}{\sqrt{2}} & -\frac{1}{\sqrt{2}} \\
 0 & \epsilon  & 0 \\
 -\frac{\epsilon ^2}{\sqrt{2}} & -\frac{\epsilon ^2}{\sqrt{2}} & \frac{1}{\sqrt{2}}
\end{array}
\right).
\end{equation}
\end{subequations}
Keeping in the expansion in $\epsilon$ only the leading order
coefficients and approximating these by one, we arrive from
Eqs.~(\ref{equ:diamond}) at the textures \#17 and \#18 in \Tab~\ref{tab:textures}. The best-fit matrices of Sec.~\ref{sec:bestfitmatrices} that provide a
perfect fit to current data are not contained in \Tab~\ref{tab:textures}, since they
exhibit a reactor angle $\theta_{13}=7.5^\circ$ that would be ruled
out by the projected confidence levels in 10 years. The diamond textures \#17
and \#18 can, however, be considered as a crude approximation of the
best-fit matrices.

It is interesting to compare the results in \Tab~\ref{tab:textures} also
with tri-bimaximal mixing \cite{tribimaximal}, which has the PMNS
angles $(\theta_{12}, \theta_{13},
\theta_{23})_\text{tribi}=\left(35^\circ,0^\circ,45^\circ\right)$. In
our procedure, neither $U_\ell$ or $U_\nu$ are on a tri-bimaximal
form, since our parameter space does not contain
$s_{12}^\text{tribi}=1/\sqrt{3}$. Nevertheless, we can associate some
textures in \Tab~\ref{tab:textures} with perturbations to
tri-bimaximal mixing. For instance, the textures \#5, \#7, \#15,
and \#16, could be viewed as describing (small) deviations from tri-bimaximal
mixing (deviations from tri-bimaximal mixing have been considered
previously, {\it e.g.}, in Ref.~\cite{Plentinger:2005kx} and references therein). Different from the
literature, however, none of the examples for tri-bimaximal-like mixing in
\Tab~\ref{tab:textures} is associated with a charged lepton mixing
matrix matrix $U_\ell$ that is close to unity. The tri-bimaximal-like
mixing emerges in these cases, instead, from large mixings in both the charged lepton and the neutrino sector.

Up to corrections of the order $\sim \epsilon^3$, all the data models
(in the sense of Sec.~\ref{sec:generating}) for the textures \#1--16
in \Tab~\ref{tab:textures} exhibit values for $\theta_{12}$ that
follow the relation
\begin{equation}\label{equ:sumt12}
\theta_{12}+\frac{\epsilon}{\sqrt{2}}+\frac{\epsilon^2}{\sqrt{2}}=\frac{\pi}{4}.
\end{equation}
This corresponds to the usual QLC relationship in \equ{qlc}. The
models for the diamond and anarchic textures \#17--20 satisfy, instead, the new sum rule
\begin{equation}\label{equ:sumt12new}
\theta_{12}+\frac{3}{5+2\sqrt{2}}\,\epsilon =\arctan(2-\sqrt{2}).
\end{equation}
This new relation is an outcome of our {\it extended} QLC
approach. The models for the textures \#1--16 exhibit
\begin{eqnarray}\label{equ:sumt13t23}
\theta_{13}=\mathcal{O}(\epsilon) & \text{and} & \theta_{23}=\frac{\pi}{4}+\mathcal{O}(\epsilon^2),
\end{eqnarray}
whereas the models for the diamond and anarchic textures \#17--20 reveal
\begin{subequations}\label{equ:sumt13t23new}
\begin{eqnarray}
\theta_{13}=\arcsin(\frac{1}{4}(2-\sqrt{2}))-\frac{1}{\sqrt{5+2\sqrt{2}}}\,\epsilon
 ~,\label{equ:theta13new}
 \\ \theta_{23}=\arctan(1+\frac{1}{\sqrt{2}})+\frac{1}{17}(2-11\sqrt{2})\,\epsilon~.\label{equ:theta23new}
\end{eqnarray}
\end{subequations}
In Eq.~(\ref{equ:sumt13t23}), $\theta_{13}$ is proportional to
 $\epsilon$, but it is nevertheless small due to a small
 coefficient multiplying $\epsilon$. In contrast to this, in
 \equ{theta13new}, a small $\theta_{13}$ arises from $\sim\epsilon$
 corrections to a small $\mathcal{O}(1)\approx8^\circ$ term. A close
 to maximal $\theta_{23}$ results in \equ{theta23new} from a $\sim
 \epsilon$ correction to
 $\arctan\left(1+{1}/{\sqrt{2}}\right)\approx60^\circ$. We would like
 to emphasize again that the new sum rules in Eqs.~(\ref{equ:sumt12new}) and (\ref{equ:sumt13t23new}) are,
 together with the known relations of Eqs.~(\ref{equ:sumt12}) and
 (\ref{equ:sumt13t23}), consequences of the extended QLC approach.

\section{Distribution of Observables: Mixing Angles}
\label{sec:pred}

In this section, we investigate the ensemble of models found in our best sample. 
This includes all models as defined by our selector (\cf~\equ{selector}) for the currently
allowed parameter ranges. 
We will be mainly interested in distributions of observables from this ensemble
of models on a statistical basis rather than individual models. Note that there will be
no predictions for the mass hierarchy, as well as there will be no dependence of the
predictions on the hierarchy.  The reason for this is the independence of the selector and
the model building process on the mass hierarchy (only the neutrino mass textures depend on
the hierarchy, which we do not consider in this section; this means that step~3 in \Sec~\ref{sec:generating}
will not be necessary). For most of this section, we
assume real mass matrices, but we will demonstrate the dependence on the Dirac-like phases $\delta^\ell$ and
$\delta^\nu$.
In all of the figures shown in this section, a ``valid'' model corresponds to a model
generated by our procedure which is consistent with current bounds. Therefore, the interpretation 
of the figures is different from so-called ``scatter plots'' showing the parameter space density for certain assumptions on the input variable distributions. Our figures represent discrete predictions 
for specific models, where each point/model can be connected with a specific texture.

Note that the interpretation of the distributions of observables as predictions has to be 
done with great care, since in nature only one model is actually implemented. In addition,
one has to use a measure in theory space, which we have done by our discrete choices of mixing angles
in $U_\ell$ and $U_\nu$. However, one can use excluded regions in the parameter space (for the
given assumptions) for strong conclusions. In addition, it is useful to study the 
model parameter space as a whole in order to obtain hints on how experiments can test this
parameter space most efficiently (\cf~\Sec~\ref{sec:exppresure}). 

\subsection{Distribution of Mixing Angles for Fixed $\boldsymbol{\epsilon}$}

\begin{figure}[p]
\begin{center}
\includegraphics[width=0.9\textwidth]{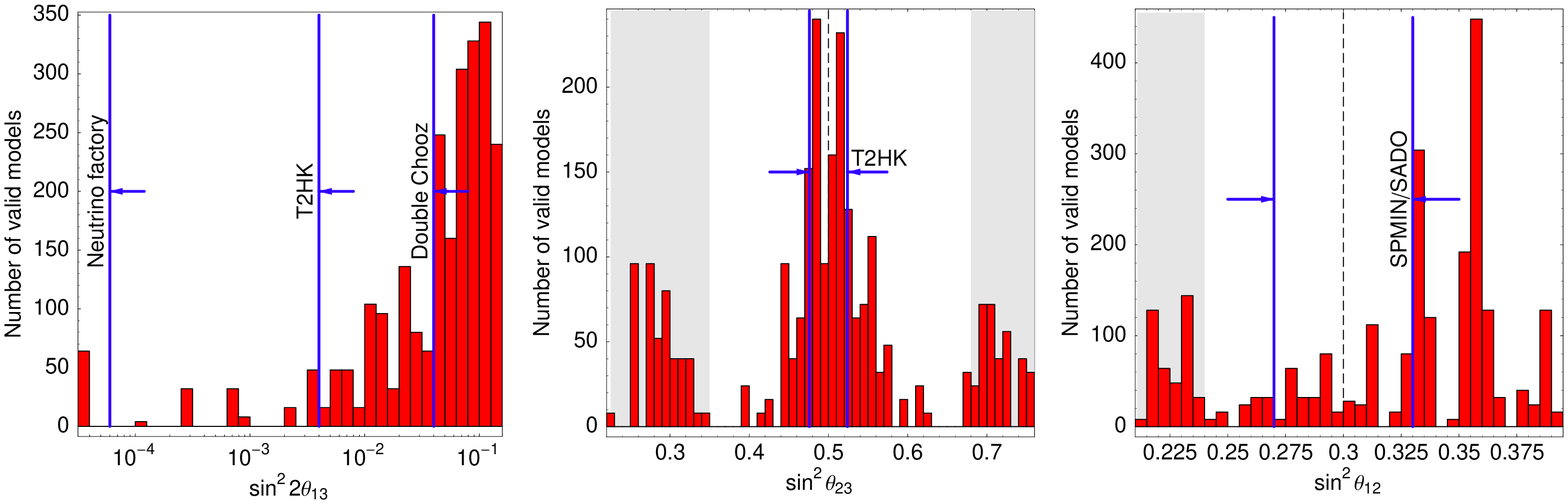}
\end{center}
\mycaption{\label{fig:histeps02} Distributions of $\stheta$ (left), $\sin^2 \theta_{23}$ (middle), and $\sin^2 \theta_{12}$ (right), of the models of our
best sample for $\epsilon=0.2$ fixed. The bars show the number of selected models per bin, \ie, per specific parameter range. The gray-shaded regions mark the current $3 \sigma$-excluded regions (\cf~\Tab~\ref{tab:values}).}
\end{figure}
\begin{figure}[p]
\begin{center}
\includegraphics[width=0.9\textwidth]{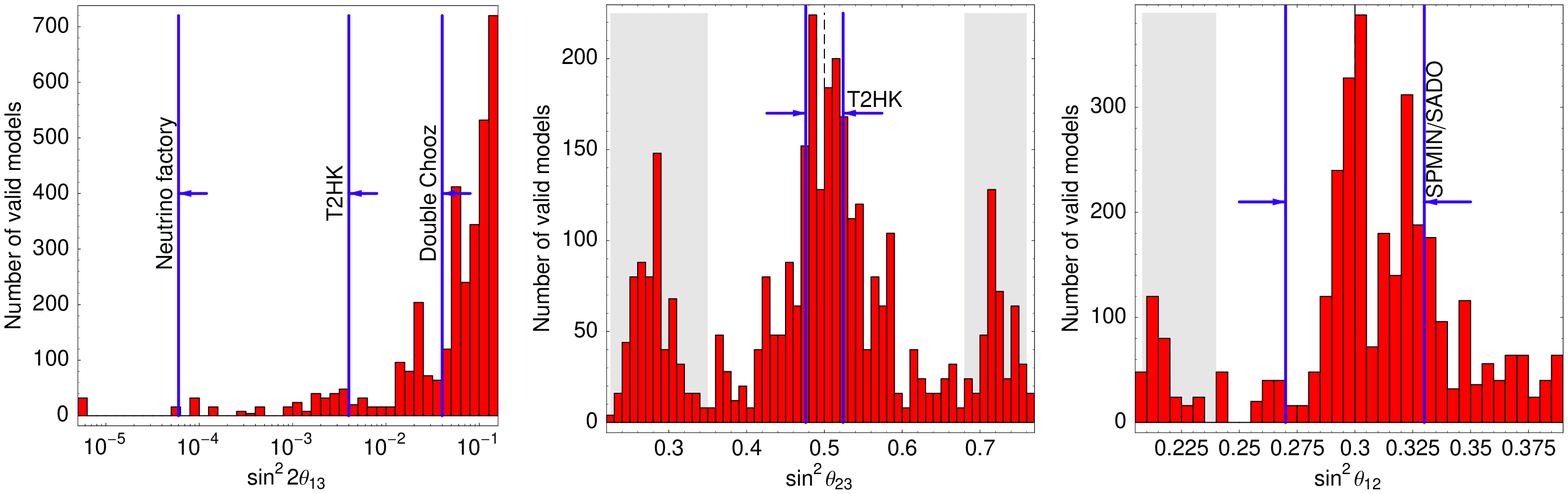}
\end{center}
\mycaption{\label{fig:histepsmin} Distributions of $\stheta$ (left), $\sin^2 \theta_{23}$ (middle), and $\sin^2 \theta_{12}$ (right), of the models of our best sample for the optimal $\epsilon$ (see main text).}
\end{figure}
\begin{figure}[p]
\begin{center}
\includegraphics[width=0.9\textwidth]{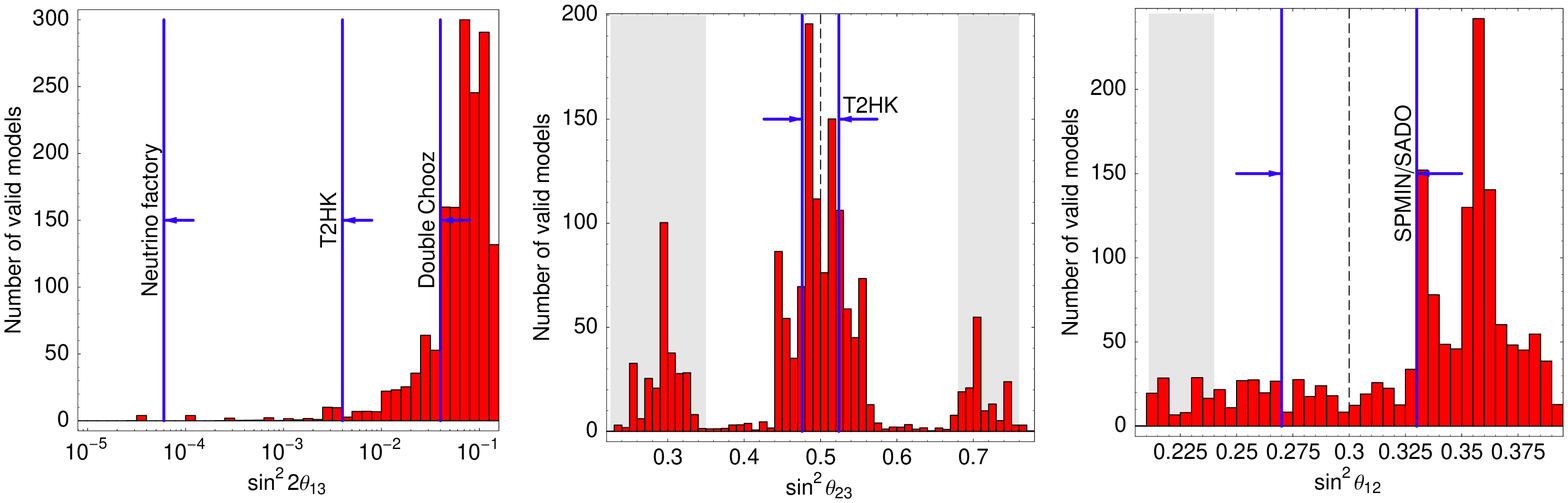}
\end{center}
\mycaption{\label{fig:histdcp} Distributions of $\stheta$ (left), $\sin^2 \theta_{23}$ (middle), and $\sin^2 \theta_{12}$ (right), of the models of our best sample averaged over the Dirac phases $\delta^\ell$ and $\delta^\nu$ using uniform distributions (see main text).}
\end{figure}

Let us first of all fix $\epsilon$ to $0.2$, as we have done before. In this case, our best sample
contains all allowed models for this fixed $\epsilon$, which are $2 \, 468$ out of the initial $262 \, 144$
models. We show the distributions of mixing angles for these models in \figu{histeps02} in terms of histograms, where the gray-shaded regions mark the current $3 \sigma$-excluded regions.
The vertical lines/arrows mark the $3 \sigma$ exclusion potential of selected future experiments.\footnote{The sensitivities are, for Double Chooz, taken from \Ref~\cite{Huber:2006vr}, for T2HK, taken from \Ref~\cite{Huber:2005jk}, for the neutrino factory,
taken from \Ref~\cite{Huber:2003ak} for two baselines, and for SPMIN/SADO (including solar data), taken from \Refs~\cite{Minakata:2004jt,Bandyopadhyay:2004cp} for a luminosity of $20 \, \mathrm{GW \, kt \, yr}$.}
For $\stheta$, the distribution is peaking at rather large values, \ie, most models are within the range of Double Chooz or the T2K and NO$\nu$A superbeams. T2HK could then exclude most of the rest, and a neutrino factory
almost all of the rest. We will discuss the number of allowed models as a function of the expected precisions in greater detail later. In fact, the smallest $\stheta$ in all of the best sample models is $3.3 \cdot 10^{-5}$, \ie, there is not a single model with $\stheta \equiv 0$. This value is not far below the reach of a neutrino factory.\footnote{Extending the allowed values for $s_{ij}^\ell$ and $s_{ij}^\nu$ by $\pm \arctan 1/\sqrt{2}$ would produce models with $\stheta \equiv 0$. However, this class of models
is not included in our initial hypothesis.} 

For $\sin^2 \theta_{23}$, the selected models peak around maximal mixing and at $\sin^2 \theta_{23} = 0.5 \pm \epsilon$, as one may expect by considering $\epsilon$ as a perturbation. This means that the value of $\epsilon$ may determine the deviation from maximal mixing for a large class of models. Interestingly, the peaks around $0.3$ and $0.7$ are already under strong experimental pressure right now, which means that an exclusion may come shortly. However, choosing a somewhat smaller $\epsilon$ may change this argument, as we will discuss in the next subsection. Note that there are gaps between the outer peaks and the maximal mixing peak with no models at all, which makes these two classes very distinguishable. We have also tested the predictions for $\theta_{23} > \pi/4$ versus $\theta_{23} < \pi/4$, but we have not found a substantial deviation from a 50:50 distribution. 

As far as $\sin^2 \theta_{12}$ is concerned, the distribution is rather flat in the currently allowed
region. In fact, the main peak around $\sin^2 \theta_{12} \simeq 0.36$ is off the current best-fit value
(but well within the currently allowed range),
which means that future precision measurements (such as by a large reactor experiment SADO/SPMIN) could
exert strong pressure on the models. Except from these parameters, we have also tested the predictions for
$\deltacp$. Since we assume real matrices in this part, only $0$ and $\pi$ can be generated 
(CP conservation). We have not found any substantial deviation from an equal distribution.

It is now interesting to compare our distributions with the literature, where we have chosen two specific
examples for $\stheta$ predictions. In \Ref~\cite{deGouvea:2003xe}, random mixing matrices and their
predictions for $\sin^2 \theta_{13}$ were investigated (``anarchy''). Our distribution for $\stheta$ follows
the general anarchy trend, which is not surprising for a statistical ensemble of models. Note, however, that our initial assumptions are qualitatively very different and we have a class of discrete models here. For example, we observe a number of discrete possible models for very small $\stheta$. One can also see this
excess on a linear scale in $\stheta$: The more or less uniform distribution in $\stheta$ (creating a distribution in $\mathrm{log}(\stheta)$ peaking at large values) has a tendency to small values of $\stheta$.
Another study, which reviews existing models in the literature by the model class and their predictions for $\stheta$, is \Ref~\cite{Albright:2006cw}. Although our model is motivated by GUTs, it does not at all show the $\stheta$ distribution obtained for GUT models in the literature (which peak around $\stheta \simeq 0.04$). In particular, we obtain an excess of very large and very small $\stheta$ models. We find this result very interesting, because our approach does not imply a particular matter of taste for the result, whereas in the literature, many models might have been biased with respect to the outcome. 

\subsection{Tuning $\boldsymbol{\epsilon}$}\label{sec:tuning}

So far, we have assumed $\epsilon \simeq \theta_\text{C} \simeq 0.2$ fixed. However, it may well be that a 
slightly different choice for $\epsilon$ will shift specific models in our selection range, \ie, that there
will be more models allowed. From neutrino physics only, one can derive $\epsilon$ using specific assumptions. For example, for a normal hierarchy of neutrino masses using $m_1:m_2:m_3=\epsilon^2:\epsilon:1$, we obtain from the best-fit values and allowed ranges in \Tab~\ref{tab:values} the range $0.15 \lesssim \epsilon \lesssim 0.22$ ($3\sigma$) with a best-fit value $\epsilon \simeq 0.18$. Of course, these values and ranges are based on more assumptions that we have used before, and one may find counter-arguments from the quark sector (such as using the Cabibbo angle instead). Therefore, in order to test the effect of different $\epsilon$'s, we vary $\epsilon$ in a symmetric range around $0.2$ between $0.15$ and $0.25$. Then we choose, for each model, the $\epsilon$ which minimizes our selector value. 

First of all, we observe that the number of valid models according to our selector 
increases from $2 \, 468$ to $3 \, 316$. We show the distribution of these models in \figu{histepsmin},
which is similar to \figu{histeps02} and should be compared with that. 
We find that for $\sin^2 \theta_{23}$ (middle panel), the gaps between
$\sin^2 \theta_{23} \simeq 0.5 \pm 0.2$ and $\sin^2 \theta_{23} \simeq
0.5$ (maximal mixing) are getting filled by some models. This
observation supports the hypothesis that the positions of the outer
peaks are indeed determined by the value of $\epsilon$. For $\sin^2
\theta_{12}$ (right panel), the main peak is now at the current
best-fit value $\sin^2 \theta_{12} \simeq 0.3$. This means that the
relatively small error on $\sin^2 \theta_{12}$ exerts pressure on the
choice of $\epsilon$. However, this also implies that an adjustment of
$\epsilon$ can, for most models, circumvent the fact that for fixed
$\epsilon$ most models do not hit the currently allowed value of
$\sin^2 \theta_{12}$ for future precisions of $\sin^2 \theta_{12}$
(\cf~\figu{histeps02}, right panel). As the last indicator, we observe
that the $\stheta$ distribution is now even more peaking at larger
values of $\stheta$ (\cf~\figu{histepsmin}, left panel), with most models being very close to the current bound. However, a small number of models with small $\stheta$ still survives. In summary, we find that adjusting $\epsilon$ does not change the qualitative conclusions from the previous section, but it allows for even more allowed models with a tendency of better $\sin^2 \theta_{12}$ fits and larger predicted values for $\stheta$.

\subsection{Impact of Non-Vanishing Dirac-Like Phases}
\label{sec:phases}

We now discuss the impact of Dirac-like CP phases $\delta^\ell$ and $\delta^\nu$ in $U_\ell$ and $U_\nu$,
respectively. So far, we have assumed that all phases be $0$ or $\pi$ in \equ{pmnspara}, and we have obtained the CP conserving values $0$ and $\pi$ for $\deltacp$ in $U_{\mathrm{PMNS}}$ without significant preference for either one. 
Let us now still fix $\widehat{\varphi}_1$, $\widehat{\varphi}_2$, $\widehat{\phi}_1$, and $\widehat{\phi}_2$, in \equ{pmnspara} to their CP conserving values $0$ or $\pi$, and average over all possible Dirac-like phase values of $\delta^\ell$ and $\delta^\nu$. Note that this implies that there will be non-trivial values for both the Dirac and Majorana phases in $U_{\mathrm{PMNS}}$, which can be interpreted for predictions as well. Since these predictions are a bit more model-dependent than most of this section, we discuss them in the next section. 

As far as the procedure is concerned, we now vary $\delta^\ell$ and $\delta^\nu$ from $0$ to $2 \pi$ in 32 equi-distant steps (excluding $2 \pi$), \ie, we generate $32 \times 32= 1\, 024$ models instead of four before (two values $0$ and $\pi$ times two phases). This implies that we simulate a uniform distribution in these phases. Then we choose the allowed models according to our selector, calculate the parameter predictions, and, in order to compare with previous results, divide the number of valid models in each bin by $1 \, 024/4=256$ (averaging). Indeed, using this averaging process, we find a number of $1 \, 570$ valid models (instead of $2\, 468$ for the Dirac-like phases fixed at the CP conserving values). Though the order of magnitude is the same, this implies that about $64\%$ of all phases are as good as $0$ and $\pi$ (on average), whereas one may naively expect about $50\%$.

We show the effects of the Dirac phase averaging in \figu{histdcp}, which should be compared with \figu{histeps02}. The qualitative result is, again, very similar to \figu{histeps02}. For $\stheta$ (left panel), the distribution becomes smoother because of the averaging process. Similar to the $\epsilon$ adjustment, larger values of $\stheta$ become more pronounced. For $\sin^2 \theta_{23}$ (middle panel), we find some models in the gaps between $0.5$ and $0.5 \pm 0.2$, as one may expect from a smearing of a subset of models. In addition, some peaks at around $0.5 \pm 0.2$ become relatively more pronounced, because there is a subset of models which is independent of $\delta^\ell$ and $\delta^\nu$ (for $\theta_{13}^\ell=0$ and/or $\theta_{13}^\nu=0$).
For $\sin^2 \theta_{12}$, we observe that this smearing fills the gaps which are present for fixed $\epsilon$, but the qualitative picture remains unchanged. Therefore, we do not find significant deviations from our earlier distributions. However, there is a slight excess of models favoring $\theta_{23} < \pi/4$ versus $\theta_{23} > \pi/4$, as it is already obvious from \figu{histdcp}, middle panel. As the most interesting part, we now also obtain non-trivial predictions for $\deltacp$ and the Majorana phases in $U_{\mathrm{PMNS}}$. We discuss these in the next section.

\subsection{Increase of the Experimental Pressure}
\label{sec:exppresure}

\begin{figure}[t!]
\begin{center}
\includegraphics[width=0.8\textwidth]{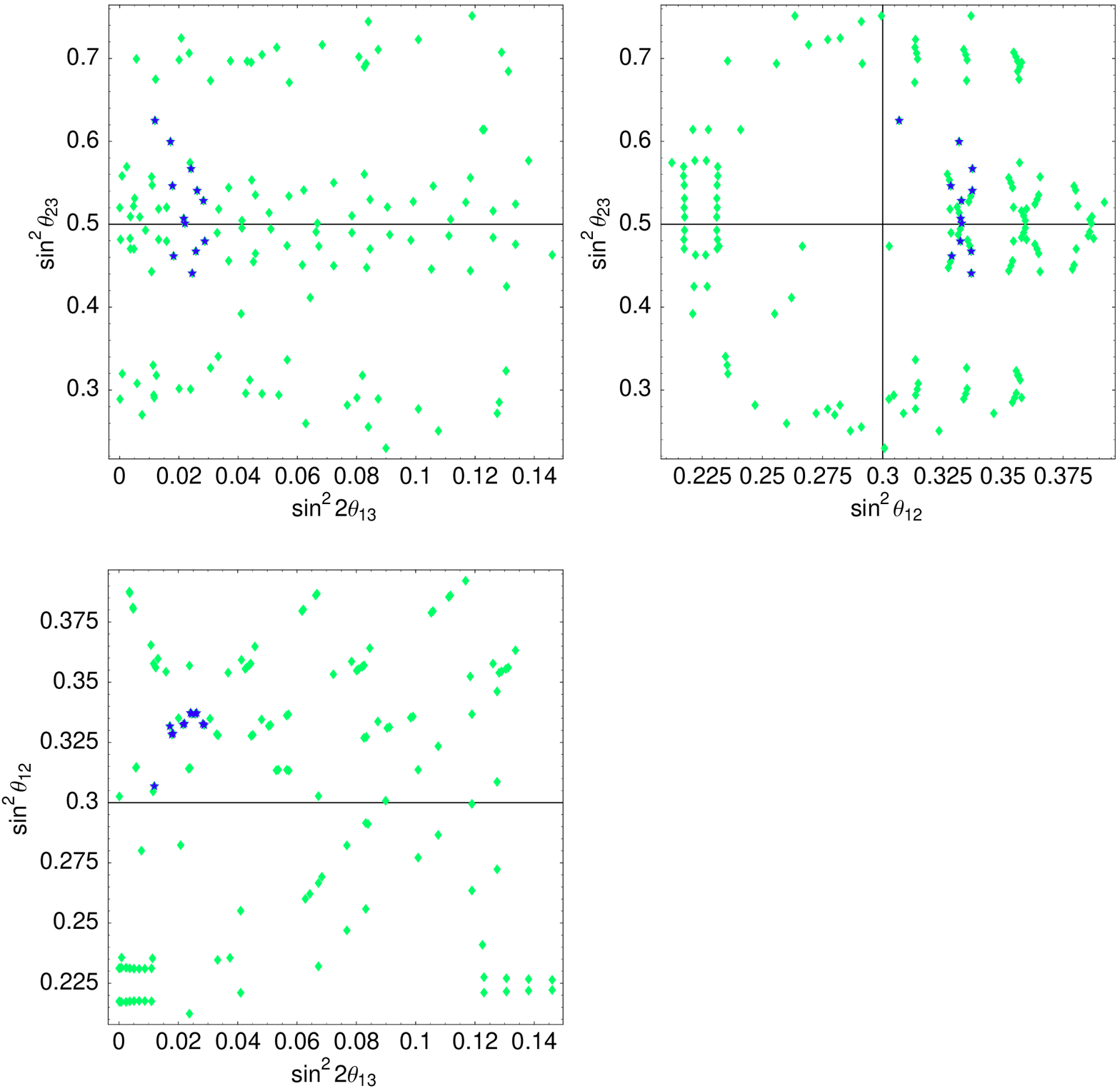}
\end{center}
\mycaption{\label{fig:pic2016} Distribution of models for fixed $\epsilon=0.2$ for the best sample based
on current experimental bounds (all marks), as well as remaining models for the bounds expected in ten years from now (dark/blue stars only); \cf~last column of \Tab~\ref{tab:values} for this projection. The distributions are given in the $\stheta$-$\sin^2 \theta_{23}$ (upper left), $\sin^2 \theta_{12}$-$\sin^2 \theta_{23}$ (upper right), and $\stheta$-$\sin^2 \theta_{12}$ (lower left) planes, where the best-fit values are marked by lines. Each point corresponds to one or more models predicting these parameter values.
}
\end{figure}

Here, we analyze future improving experimental constraints and how they affect the selection of models,
where we restrict this discussion to $\epsilon=0.2$ fixed.
First of all, we show in \figu{pic2016} the distribution of models in the $\stheta$-$\sin^2 \theta_{23}$ (upper left), $\sin^2 \theta_{12}$-$\sin^2 \theta_{23}$ (upper right), and $\stheta$-$\sin^2 \theta_{12}$ (lower left) planes, where each point corresponds to one or more models with a specific parameter combination. The figure shows all the models from our best sample based on current experimental bounds, whereas the dark stars mark the models remaining after increasing the experimental pressure
on a time scale of ten years. The corresponding scenario ``In ten years'' is defined in \Tab~\ref{tab:values}, where we assume that $\theta_{13}$ will not be discovered. 
Obviously, the improving $\stheta$ bound restricts the model space very strongly (left panels), where many models close to maximal mixing will survive (upper left panel). However, only very few models close to the $\sin^2 \theta_{12}$ best-fit value survive (lower left panel). In particular, none of the remaining models can reproduce both the solar and atmospheric best-fit values closely (upper right panel). Note that the models with extremely small $\stheta$ will be eliminated as well, mainly because of the pressure from $\sin^2 \theta_{12}$ (\cf~lower left panel).
Therefore, we expect that the strongest experimental pressure on this model space will come from $\stheta$ and $\sin^2 \theta_{12}$, which are two different degrees of freedom since they will be driven by different classes of experiments (such as beams and short-baseline reactor experiments versus solar and long-baseline reactor experiments).
Note that this discussion assumes that $\stheta$ will not be discovered, whereas a $\stheta$ discovery would, depending on the new best-fit value, select a different class of models.

\begin{figure}[t!]
\begin{center}
\includegraphics[width=\textwidth]{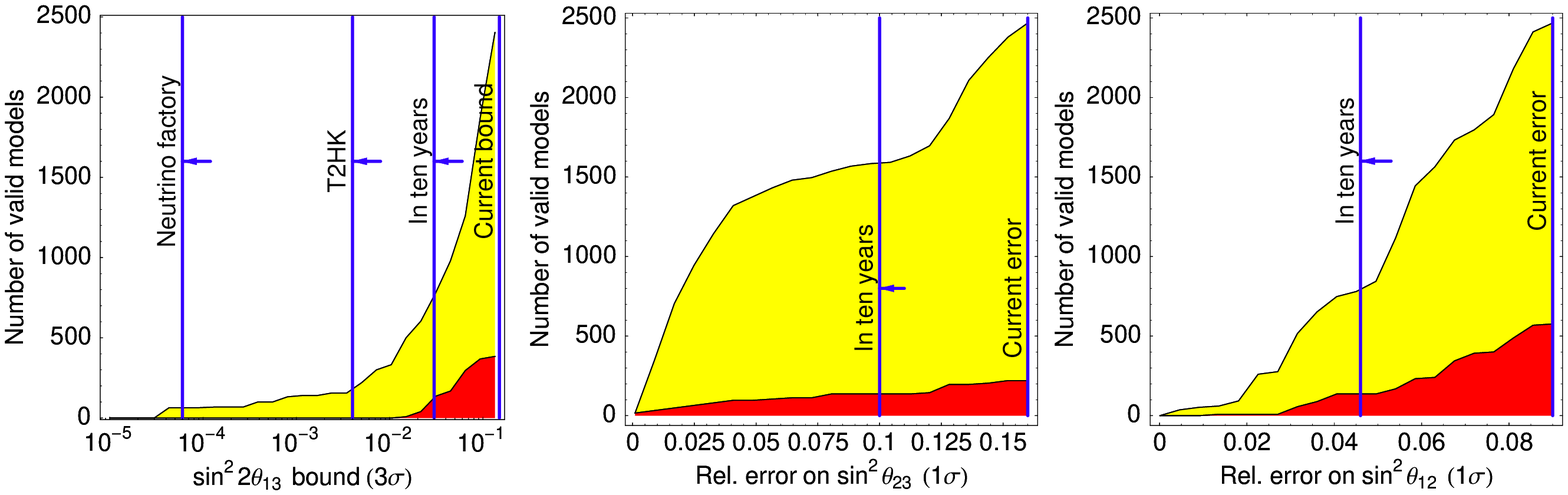}
\end{center}
\mycaption{\label{fig:exppress} Number of valid models as function of the $\stheta$ bound (left), error on $\sin^2 \theta_{23}$ (middle), and error on $\sin^2 \theta_{12}$ (right). The right edges of the plots correspond to the current errors, the left edges to exactly known parameters. The light (yellow) curves refer to the current errors for the other two degrees of freedom. The dark (red) curves assume the other errors to be as in the scenario ``In ten years'' (\cf~\Tab~\ref{tab:values}). The other indicated experiments marked by the vertical lines  are referred to in the caption of \figu{histeps02}.
}
\end{figure}

In order to demonstrate the continuous dependence on experimental
constraints, we show in \figu{exppress} the number of valid models as
a function of the $\stheta$ bound (left), error on $\sin^2 \theta_{23}$ (middle), and error on $\sin^2 \theta_{12}$ (right). The right edges of the plots correspond to the current errors, the left edges to exactly known parameters. The light (yellow) curves fix the errors for the other two degrees of freedom to their current values, whereas the dark (red) curves assume the other errors from the scenario ``In ten years'' defined in the caption of \figu{pic2016}. 
Comparing the light/yellow curves only, the strongest experimental
pressure will be exerted by $\stheta$ and $\sin^2 \theta_{12}$, which
may each independently reduce the number of models to one third on a
timescale of ten years. This means that $\stheta$ experiments affect
the valid models as much as solar and potential long-baseline reactor
experiments (SPMIN/SADO), and already one of these experiment classes
already acts as a strong model discriminator. The reason is the
generic prediction of the value of $\sin^2 \theta_{12}$ from the
concepts of maximal mixing combined with $\epsilon$ deviations,
whereas maximal mixing is used as an initial hypothesis for our
models. In order to estimate the combined potential of different
experimental degrees of freedom, we show as the dark/red curves the
number of valid models for smaller errors on the parameters not shown (approximately on a time scale of ten years). Obviously, $\stheta$ is a necessary discriminator to exclude all possible models very quickly, whereas no further increase in the $\stheta$ bound would make a further model discrimination very hard. 

\section{Distribution of Observables: Dirac and Majorana PMNS Phases}
\label{sec:phasesext}

We now demonstrate that the results in \Sec~\ref{sec:phases} can be used for non-trivial predictions of the
Dirac and Majorana phases in $U_{\mathrm{PMNS}}$. We have decoupled
this discussion from the previous sections, because the results in
this chapter will be more dependent on the assumptions used. However,
we believe that already the procedure employed in this section
warrants some attention, because it opens a new way of model building
predictions. The actual dependence on the assumptions, which need to
be chosen according to the underlying theory, deserves further studies. Note that, as in the last section,
the ``predictions'' can be used to study the impact of experiments on the parameter space of models,
but one has to be careful to interpret them as predictions for the actual theory.

In \Sec~\ref{sec:phases}, we have assumed that $\widehat{\varphi}_1$,
$\widehat{\varphi}_2$, $\widehat{\phi}_1$, and $\widehat{\phi}_2$, in
\equ{pmnspara} be $0$ or $\pi$, but the Dirac-like phases
$\delta^\ell$ and $\delta^\nu$ be uniformly distributed. We have then
generated all possible models for different phases and selected our
best sample which has been compatible with current data. These phases
generate non-trivial Dirac ($\deltacp$) and Majorana ($\phi_1$,
$\phi_2$) phases in the product $U_{\mathrm{PMNS}} = U_\ell^\dagger
U_\nu$, which can be predicted from our best sample. One can now argue
if our set of assumptions is plausible (such as the set of values used
for the mixing angles), if one should include variations of the other
phases, if the uniform distributions for the Dirac-like phases make
sense, \etc~Therefore, one has to be careful with the interpretation
of the results in this section. It is, however, the purpose of this section to demonstrate how our procedure can provide non-trivial predictions for phases under certain assumptions. 

\begin{figure}[t!]
\begin{center}
\includegraphics[angle=+90,width=\textwidth]{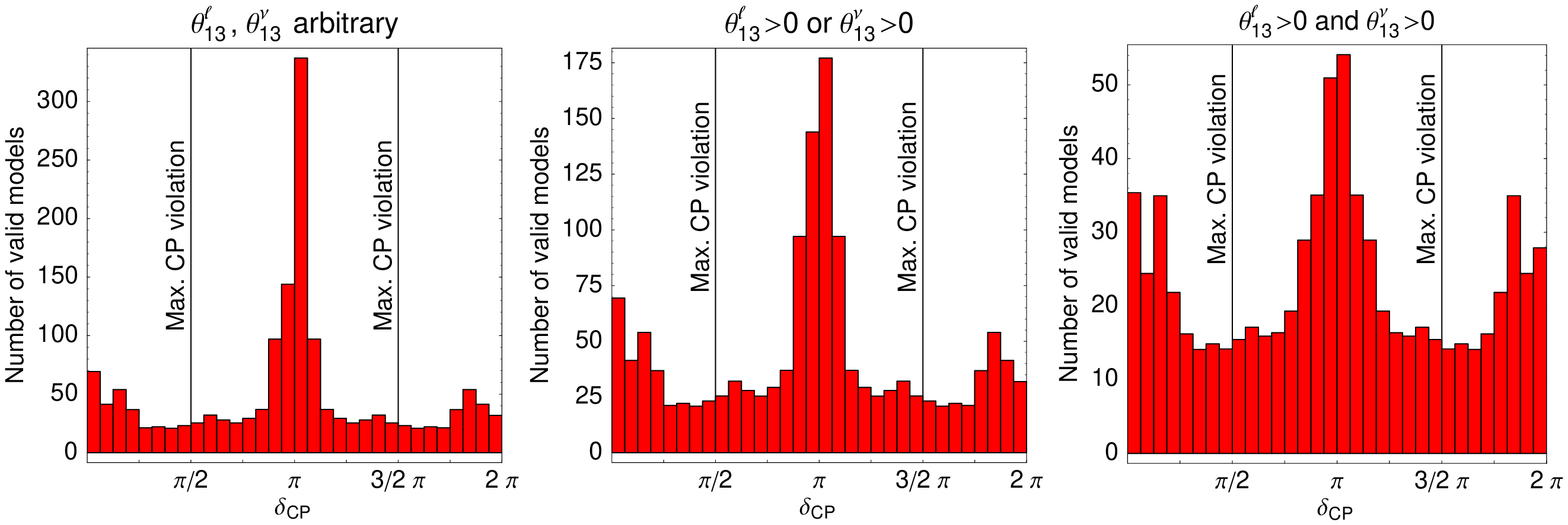}
\end{center}
\mycaption{\label{fig:dcppred} Distribution of the Dirac phase $\deltacp$ for our best sample for all models of that sample (left), only models with at least one $\theta_{13}^\ell>0$ or $\theta_{13}^\nu>0$ (middle), and models with both $\theta_{13}^\ell>0$ and $\theta_{13}^\nu>0$. There are no particular assumptions for $\theta_{13}$ (in $U_{\mathrm{PMNS}}$).
}
\end{figure}

First of all, we show in \figu{dcppred} the distribution for $\deltacp$ (in $U_{\mathrm{PMNS}}$) for our best sample. As before, we have generated 32 $\times$ 32 models for all possible pairs of Dirac-like phases $\delta^\ell$ and $\delta^\nu$, have chosen our best sample, and then have normalized the histograms by $4/1024=1/256$ (because we replace four choices of the phases by 1024). In the left plot, we show the distribution for all models in the best sample. Obviously, there is a strong peak at $\deltacp \simeq \pi$, which predominantly comes from models with $\theta_{13}^\ell=\theta_{13}^\nu=0$. In this case, $U_\ell$ and $U_\nu$ are real matrices, and the product has to be real as well, which only allows the phases $0$ and $\pi$ for $\deltacp$. There is no smearing of these models coming from the phase averaging, because $\delta^\ell$ and $\delta^\nu$ are undefined. This means that the averaging over these phases leaves the peak untouched and smears out many of the other models, leading to a relative enhancement at the CP conserving choice. The strong preference of $\pi$ compared to $0$ comes from the selector from a non-trivial restriction of the parameter space in all three mixing angles, it is not present before the selection process.

\begin{figure}[t!]
\begin{center}
\includegraphics[width=10cm]{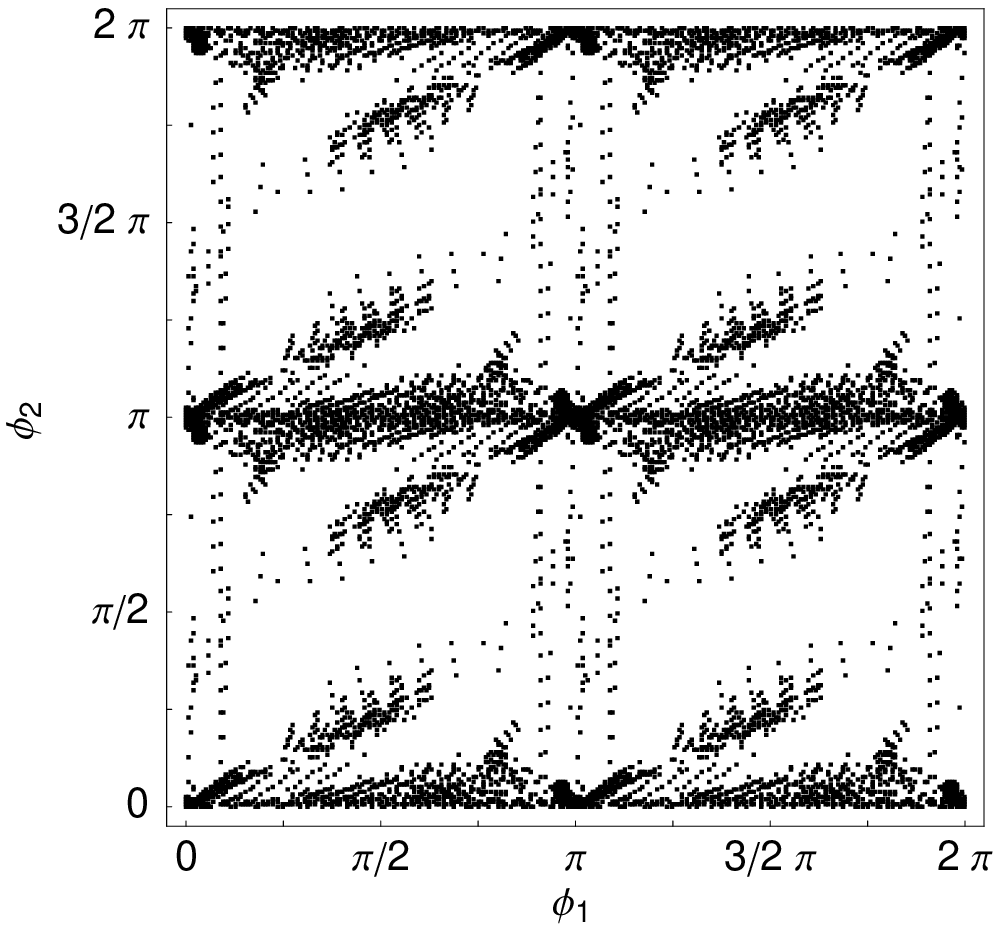}
\end{center}
\mycaption{\label{fig:majphases} Distribution of the Majorana phases $\phi_1$ and $\phi_2$ from our best sample. Each point
corresponds to one or more models predicting the corresponding phases.
}
\end{figure}

In some sense, the overall preference of CP conservation is connected with our choice of $0$ as one of the values for the mixing angles $\theta_{13}^\ell$ and $\theta_{13}^\nu$, which means that it is model-dependent. One can now argue that it does not make sense to consider the cases with $\theta_{13}^{\ell}=0$ or $\theta_{13}^{\nu}=0$ because in this case
the Dirac-like phases are not physical and therefore introduce a bias
(these models have a stronger relative weight because the phases are not varied).
 Therefore, we show in \figu{dcppred}, middle and right, only the models with at least one ``physical'' (=defined) Dirac-like phase (middle) and two physical Dirac-like phases (right) in  $U_\ell$ and $U_\nu$. Obviously, there is a shift in the preference of $\deltacp$. However, all of the plots in \figu{dcppred} have in common that maximal CP violation is disfavored. The reason for the peaks close to CP conservation is the generation of the PMNS Dirac phase
in $(U_{\mathrm{PMNS}})_{e3}$ as a combination of the Dirac-like
phases from  $U_\ell$ and $U_\nu$ (times some factors) plus a constant
term acting CP conserving. If this constant term is larger than the
phase-dependent factors, there will be CP conservation disturbed by
some $\epsilon$-size (or $\epsilon^2$-size) contribution coming from
the mixing angles. If this constant term is smaller than the
phase-dependent factors, the uniform phase distributions will
translate into uniform distributions for the PMNS Dirac phase.
Note that we have still kept all the other phases 
$\widehat{\varphi}_1$, $\widehat{\varphi}_2$, $\widehat{\phi}_1$, and $\widehat{\phi}_2$, fixed at their CP conserving values -- which is a model dependent assumption (and a constraint from computation power). However, note that $\widehat{\phi}_1$ and $\widehat{\phi}_2$ only affect the Majorana phases in $U_{\mathrm{PMNS}}$. 

Our approach does not only predict the PMNS Dirac phase, but also a non-trivial set of Majorana phases. 
We show in \figu{majphases} this distribution for our best sample. Obviously, there is some clustering close to the CP conserving values, but, in principle, the whole parameter space is covered. In addition, note that the parameter space $[\pi, 2 \pi]$ is redundant for $0\nu\beta\beta$ decay, because only the square of the phase enters. 

\begin{figure}[t!]
\begin{center}
\includegraphics[width=\textwidth]{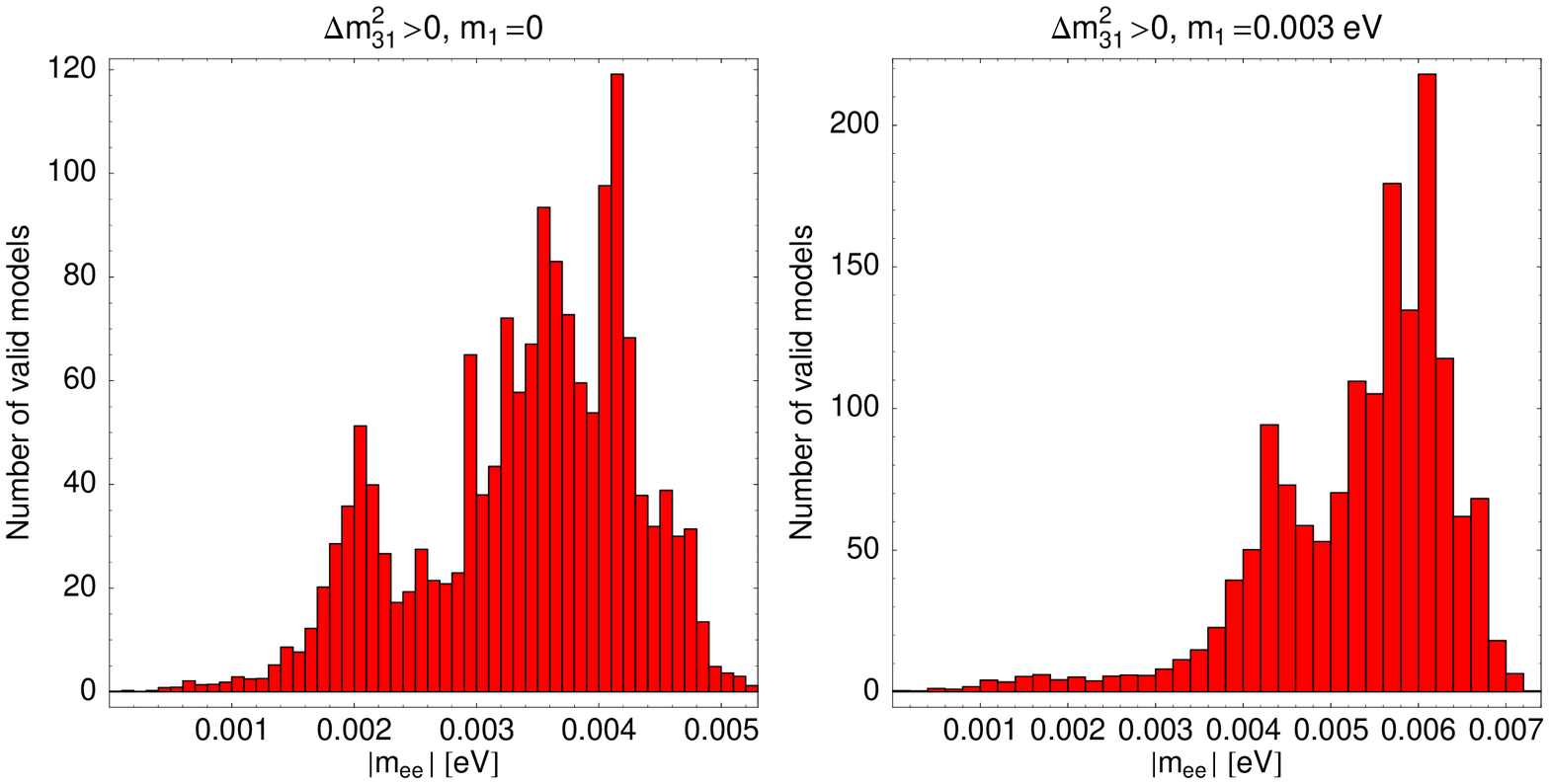}
\end{center}
\mycaption{\label{fig:meedistr} Distribution of $|m_{ee}|$ from our best sample for a normal mass hierarchy
and $m_1$ as chosen in the plot labels.
}
\end{figure}

The predictions for these phases for one specific model become very interesting in combination with the mixing angles for that model, because this set of parameters can be used for a direct prediction of the mass matrix element for $0\nu\beta\beta$ decay for Majorana neutrino masses. 
We follow the calculations in \Ref~\cite{Lindner:2005kr} (for a recent review, see also \Ref~\cite{Petcov:2004wz} and references therein), where the absolute value of the mass matrix element 
$m_{ee}$, which is proportional to the rate of the $0\nu\beta\beta$ decay, is given by\footnote{Note the change in notation for the phases: We read off $\phi_1$ and $\phi_2$ in a way  which is related to $\alpha$ and $\beta$ in \Ref~\cite{Lindner:2005kr} by $\alpha=\phi_2-\phi_1$, $\beta=-\phi_1-\delta$.} 
\be
\left| m_{ee} \right| \equiv \left| \sum U_{e i}^2 \, m_i \right| 
\mbox{ with } 
m_{ee} = |m_{ee}^{(1)}| + |m_{ee}^{(2)}| \, e^{2 i (\phi_2-\phi_1)} + 
|m_{ee}^{(3)}| \, e^{-2 i(\phi_1+\delta)} 
\label{equ:mee}
\ee
and
\begin{eqnarray}
|m_{ee}^{(1)}| &=& m_1 \, |U_{e1}|^{2} = m_1 \, c_{12}^{2} \, c_{13}^{2} ~,
\nonumber\\
|m_{ee}^{(2)}| &=& m_2 \, |U_{e2}|^{2} = m_2 \, s_{12}^{2} \, c_{13}^{2} ~,\\
|m_{ee}^{(3)}| &=& m_3 \, |U_{e3}|^{2} = m_3 \, s_{13}^{2}~.\nonumber
\end{eqnarray}
Given the mass hierarchy and using the mass squared differences from \Tab~\ref{tab:values}, one can
use these equations to compute $\left| m_{ee} \right|$ as a function of the lightest neutrino mass $m$ ($m_1$ for the normal hierarchy, and $m_3$ for the inverted hierarchy). Each of our models predicts the combination $(\phi_1,\phi_2,\theta_{13},\theta_{12},\delta)$ relevant for $0\nu\beta\beta$ decay, which results in a particular prediction of $\left| m_{ee} \right|$ as function of $m$. We show this prediction for the normal mass hierarchy for two different values of $m=m_1$ in \figu{meedistr}. The left plot corresponds to a vanishing $m_1$, the right plot to the ``chimney'', where $\left| m_{ee} \right|$  may even vanish.
From these distributions one can read off what may be obvious: Making
the $0\nu\beta\beta$ decay rate vanish could mean fine-tuning, because it requires a specific
combination of phases and mixing angles (\cf~\fig~1 in \Ref~\cite{Lindner:2005kr} for illustration). Therefore, none of our models predicts a vanishing $0\nu\beta\beta$ decay rate.

\begin{figure}[t!]
\begin{center}
\includegraphics[width=\textwidth]{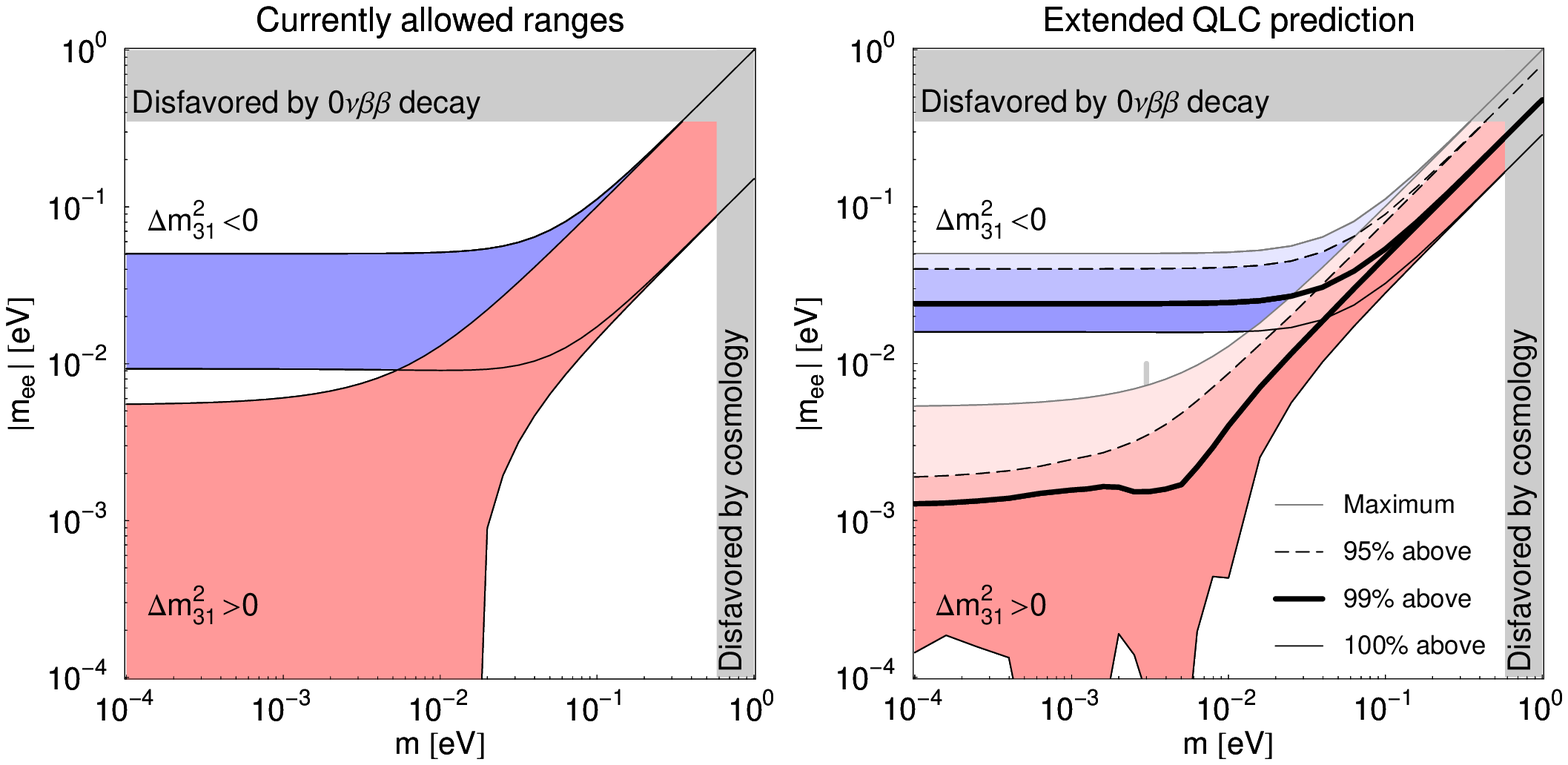}
\end{center}
\mycaption{\label{fig:0nbb} General theory allowed regions (left) and distribution of our best sample
prediction (right)  for $|m_{ee}|$ as a function of the lightest neutrino mass $m$. In the left plot, we show the currently allowed regions for $|m_{ee}|$, where the mixing angles are varied in the current $3 \sigma$ ranges from \Tab~\ref{tab:values} (computed with the formulas given in \Ref~\cite{Lindner:2005kr}).
Note that fixing $\theta_{13}$ would result in the appearance of the ``chimney''. 
In the right plot, the bands correspond to all possible models from our best sample. The different intermediate curves limit the regions where 99\% and 95\% of the models can be found above. The vertical line corresponds to the choice of \figu{meedistr}, right (for the normal hierarchy). Note that in the right panel, we use both the  mixing angle and Majorana phase predictions for each model, \ie, we use the combination of all available predictions for that model (and not only the Majorana phases).
In both panels, the limit from cosmology is obtained from a combined SDSS and WMAP analysis~\cite{Tegmark:2003ud}, and the the limit from $0\nu\beta\beta$-decay is obtained by the Heidelberg-Moscow collaboration~\cite{Klapdor-Kleingrothaus:2000sn} (with an uncertainty coming from the calculations of the nuclear matrix elements). 
We fix the mass squared differences in both panels to their best-fit values in \Tab~\ref{tab:values}.
}
\end{figure}

We show in \figu{0nbb} the theory allowed regions (left) and our model
predictions (right) as function of the lightest neutrino mass. The
general theory prediction is obtained by varying the mixing angles
(and phases) within their current $3 \sigma$ allowed ranges. Because of the relatively weak $\stheta$ bound the chimney is not explicitly visible in the left plot, and a vanishing $0\nu\beta\beta$ decay rate is possible for the normal hierarchy. In the right panel of \figu{0nbb}, we compute for each $m$ the minimum and maximum of $|m_{ee}|$ from all models in the best sample. In addition, we show the curves where 99\% and 95\% of all models are above. As the most interesting result, we find that 99\% of all normal mass hierarchy models are above $\sim 0.001 \, \mathrm{eV}$ independent of $m_1$. In addition, comparing the two panels of \figu{0nbb}, we find that the actually predicted ranges will be much more narrow than the corresponding theoretically allowed ranges. Again, this result depends on the assumptions used for the phase generation, in particular, the selection of values for the mixing angles, the uniform distributions of the Dirac-like phases $\delta^\ell$ and $\delta^\nu$, and the selection of CP conserving values for the other phases. However, the obvious observation that one needs to fine-tune the parameters in order to make $0\nu\beta\beta$ decay vanish, seems to be quite general. Therefore, we expect that in most concrete realizations,
there will be non-vanishing $0\nu\beta\beta$ decay  for Majorana neutrino masses.

\section{Summary and Conclusions}
\label{sec:summary}

As a minimal unified approach to the fermion mass and mixing parameters in the Standard Model, 
it is attractive to assume that all deviations from symmetries or zeros, such as 
deviations from maximal atmospheric neutrino mixing, the small mixing angle $\theta_{13}$ in $U_{\mathrm{PMNS}}$, the Cabibbo angle $\theta_\text{C}$ in $V_{\mathrm{CKM}}$, and the mass hierarchies,
may be described by powers of a {\em single} small quantity $\epsilon \simeq \theta_\text{C}$. This small
quantity can be motivated by Grand Unified Theories connecting quarks and leptons.  
Quark-lepton complementarity can be interpreted as a phenomenological implementation of this approach, obtaining the solar mixing angle as $\theta_{12} + \theta_\text{C} \simeq \pi/4$.
We have introduced  an extended quark-lepton complementary approach as an extension
of this relationship. Since $U_{\mathrm{PMNS}}$
arises as a product of the charged lepton and neutrino mixing matrices, \ie, $U_{\mathrm{PMNS}} = U_\ell^\dagger U_\nu$, we have postulated that
all mixing angles in $U_\ell$ and $U_\nu$ be given by either maximal mixing or powers of $\epsilon$. 
In this way, the observed solar mixing angle can only
result from taking the product in $U_{\mathrm{PMNS}} = U_\ell^\dagger U_\nu$.  In addition,
deviations from maximal atmospheric mixing, $\theta_{13}$, \etc, emerge as predictions
in this minimal unified approach as being given by powers of $\epsilon$.
Note that we
have not assumed specific forms for $U_\ell$ and $U_\nu$, such as bimaximal mixing. However,
we have obtained configurations involving bimaximal mixing as special cases.

Using this assumption for the mixing angles in $U_\ell$ and $U_\nu$ up to the order $\epsilon^2$,
\ie, the mixing angles be $\pi/4$, $\epsilon$, $\epsilon^2$, or $0$, we have systematically
tested all possible CP conserving configurations, a total of $262 \, 144$ generated models.
Naturally, we had to introduce an automated selector for the search for models compatible
with current experimental data, leading to a best sample of $2 \, 468$ models.
As a first analysis, we have scanned for particularly interesting textures, such as with $U_\ell \simeq V_{\mathrm{CKM}}$ and $U_\nu \simeq U_{\mathrm{bimax}}$, or models being perfectly consistent with
current data. We have found that these models imply that $\theta_{13}$ be large, which means that 
many of these models could be ruled out very soon due to stronger bounds on $\theta_{13}$. However, if $\theta_{13}$ turns out to be large, we have identified examples with $(\theta_{12},\theta_{13},\theta_{23}) = (33.4^\circ,7.5^\circ,43.5^\circ)$, providing a perfect fit. As the next
step, we have particularly focused on models with small $\theta_{13}$ and increased pressure on the
other oscillation parameters in order to further reduce
our sample. These models are interesting because they could survive the next ten years even if $\theta_{13}$ was not discovered. We have classified the corresponding textures systematically by their leading order
entries, and we have found 20 distinctive sets of textures. These
textures can be divided into different classes, such as lopsided models, anarchic models, \etc~As a very interesting class,
we have identified a number of models with maximal mixings $\theta_{12}^\ell$, $\theta_{23}^\ell$, and $\theta_{13}^\nu$, which we have called ``diamond models'' (because of the diamond-like structure of
the small entries in $M^{\mathrm{Maj}}_\nu$). We have shown that the sum rules for this and the anarchic-like classes of models are qualitatively different from the ones in standard quark-lepton complementarity.

In an independent approach, we have investigated the mixing angle distributions for our best sample compatible with current data. These distributions can be used to study the impact of future experiments on
the model parameter space, and should be interpreted as predictions in that way.
As the main result, we have found rather large values of $\stheta$ preferred, and $\stheta > 3.3 \cdot 10^{-5}$ for all of our models. In addition, $\sin^2 \theta_{23}$ peaks around $0.5$ and at $0.5 \pm \epsilon$, whereas $\sin^2 \theta_{12}$ peaks above the current best-fit value.
Compared to the GUT model literature (surveyed in \Ref~\cite{Albright:2006cw}), we have not found a characteristic peak of models at around $\stheta \simeq 0.04$. Since our matrix generation and selection process has been quite un-biased, this could point towards a bias in the building of specific GUT models. Compared to anarchic models, we find an excess for small $\stheta$. 
In the future, especially $\stheta$ experiments (such as superbeams or short-baseline reactor experiments) and potential $\sin^2 \theta_{12}$ experiments (such as a long-baseline reactor experiments) will put the model parameter space under pressure. Because we have used maximal mixing as an input and have generated deviations from that by powers of $\epsilon \simeq \theta_\text{C}$, $\sin^2 \theta_{12}$ is an important discriminator for this class of models. However, in order to exclude all models, very strong bounds for $\stheta$ are needed, such as those coming from a neutrino factory.

In a more specific predictability part, we have generated $\delta^\ell$ (in $U_\ell$) and $\delta^\nu$ (in $U_\nu$) with uniform distributions in order to obtain predictions for $\delta$ and the Majorana phases in $U_{\mathrm{PMNS}}$. This set of assumptions is more model-dependent than the rest of this work, but it allows a very powerful handle on phase predictions. For example, we have found that for $\delta$ (in $U_{\mathrm{PMNS}}$), maximal CP violation is significantly disfavored, because it requires large imaginary parts that are hard to obtain from the construction $U_{\mathrm{PMNS}}=U^\dagger_\ell U_\nu$. Furthermore, we have combined the non-trivial predictions of the Majorana phases with the mixing angle predictions for each model, and we have predicted the $0\nu\beta \beta$ decay rates. Not a single of our models predicts a vanishing $0\nu\beta\beta$ decay rate, because the necessary phase cancellation requires fine-tuning not present in the discussed model parameter space. We have found that 99\% of all  models predict $| m_{ee}| > 0.001 \, \mathrm{eV}$ independent of the mass hierarchy and lightest neutrino mass, \ie, the ``chimney'' is, in practice, not present. 

In conclusion, we have used a novel approach for studying neutrino mass matrices, which is
a mixture between basic fundamental assumptions and the systematic machinized parameter space scan
of a very high dimensional parameter space. This approach has turned out to be extremely powerful,
because it does not require the diagonalization of matrices. The primary objective of this work
has been to stay as far away from specific assumptions -- which may introduce a bias -- as currently possible from the computational point of view. With this approach, we have not only been able to scan a large parameter space systematically, but also
been able to make predictions based on a large sample of models compatible with current data. Each of these
 predictions for the mixing angles can be connected to a particular texture, which is different from
parameter space scans using 
particular assumptions for the input variable distributions and investigating the parameter space density.
Only in the last part, we have combined these two methods because we have used uniform distributions for the Dirac-like phases as an assumption. Naturally, we have not been able to present all of our results here, but have focused on the most interesting parts -- in our opinion, of course. The interested reader can find
a tool to view all models from our best sample compatible with current data in \Ref~\cite{psw-web}. Finally, we believe that the connection between quarks and leptons could be the key element in the motivation of future neutrino facilities, and we have demonstrated how this element can be implemented phenomenologically in a straightforward scheme. 

\subsubsection*{Acknowledgments}
The research of F.P. is supported by Research Training Group 1147 \textit{Theoretical
Astrophysics and Particle Physics} of Deutsche Forschungsgemeinschaft.
G.S. was supported by the Federal Ministry of Education and
Research (BMBF) under contract number 05HT1WWA2.
W.W. would like to acknowledge support from the Emmy Noether program of Deutsche Forschungsgemeinschaft.


\begin{thebibliography}{10}
\expandafter\ifx\csname bibnamefont\endcsname\relax
  \def\bibnamefont#1{#1}\fi
\expandafter\ifx\csname bibfnamefont\endcsname\relax
  \def\bibfnamefont#1{#1}\fi
\expandafter\ifx\csname url\endcsname\relax
  \def\url#1{\texttt{#1}}\fi
\expandafter\ifx\csname urlprefix\endcsname\relax\def\urlprefix{URL }\fi
\providecommand{\bibinfo}[2]{#2}
\providecommand{\eprint}[2][]{\url{#2}}

\bibitem{Yao:2006px}
\bibinfo{author}{\bibfnamefont{W.~M.} \bibnamefont{Yao}} \emph{et~al.}
  (\bibinfo{collaboration}{Particle Data Group}), \bibinfo{journal}{J. Phys.}
  \textbf{\bibinfo{volume}{G33}}, \bibinfo{pages}{1} (\bibinfo{year}{2006}).

\bibitem{Cabibbo:1963yz}
\bibinfo{author}{\bibfnamefont{N.}~\bibnamefont{Cabibbo}},
  \bibinfo{journal}{Phys. Rev. Lett.} \textbf{\bibinfo{volume}{10}},
  \bibinfo{pages}{531} (\bibinfo{year}{1963}).

\bibitem{Kobayashi:1973fv}
\bibinfo{author}{\bibfnamefont{M.}~\bibnamefont{Kobayashi}} \bibnamefont{and}
  \bibinfo{author}{\bibfnamefont{T.}~\bibnamefont{Maskawa}},
  \bibinfo{journal}{Prog. Theor. Phys.} \textbf{\bibinfo{volume}{49}},
  \bibinfo{pages}{652} (\bibinfo{year}{1973}).

\bibitem{Fukuda:2002pe}
\bibinfo{author}{\bibfnamefont{S.}~\bibnamefont{Fukuda}} \emph{et~al.}
  (\bibinfo{collaboration}{Super-Kamiokande}), \bibinfo{journal}{Phys. Lett.}
  \textbf{\bibinfo{volume}{B539}}, \bibinfo{pages}{179} (\bibinfo{year}{2002}),
  \eprint{hep-ex/0205075}.

\bibitem{Ahmad:2002ka}
\bibinfo{author}{\bibfnamefont{Q.~R.} \bibnamefont{Ahmad}} \emph{et~al.}
  (\bibinfo{collaboration}{SNO}), \bibinfo{journal}{Phys. Rev. Lett.}
  \textbf{\bibinfo{volume}{89}}, \bibinfo{pages}{011302}
  (\bibinfo{year}{2002}), \eprint[http://arXiv.org/abs]{nucl-ex/0204009}.

\bibitem{Fukuda:1998mi}
\bibinfo{author}{\bibfnamefont{Y.}~\bibnamefont{Fukuda}} \emph{et~al.}
  (\bibinfo{collaboration}{Super-Kamiokande}), \bibinfo{journal}{Phys. Rev.
  Lett.} \textbf{\bibinfo{volume}{81}}, \bibinfo{pages}{1562}
  (\bibinfo{year}{1998}), \eprint{hep-ex/9807003}.

\bibitem{Araki:2004mb}
\bibinfo{author}{\bibfnamefont{T.}~\bibnamefont{Araki}} \emph{et~al.}
  (\bibinfo{collaboration}{KamLAND}), \bibinfo{journal}{Phys. Rev. Lett.}
  \textbf{\bibinfo{volume}{94}}, \bibinfo{pages}{081801}
  (\bibinfo{year}{2005}), \eprint{hep-ex/0406035}.

\bibitem{Apollonio:2002gd}
\bibinfo{author}{\bibfnamefont{M.}~\bibnamefont{Apollonio}} \emph{et~al.}
  (\bibinfo{collaboration}{CHOOZ}), \bibinfo{journal}{Eur. Phys. J.}
  \textbf{\bibinfo{volume}{C27}}, \bibinfo{pages}{331} (\bibinfo{year}{2003}),
  \eprint{hep-ex/0301017}.

\bibitem{Aliu:2004sq}
\bibinfo{author}{\bibfnamefont{E.}~\bibnamefont{Aliu}} \emph{et~al.}
  (\bibinfo{collaboration}{K2K}), \bibinfo{journal}{Phys. Rev. Lett.}
  \textbf{\bibinfo{volume}{94}}, \bibinfo{pages}{081802}
  (\bibinfo{year}{2005}), \eprint{hep-ex/0411038}.

\bibitem{Tegmark:2003ud}
\bibinfo{author}{\bibfnamefont{M.}~\bibnamefont{Tegmark}} \emph{et~al.}
  (\bibinfo{collaboration}{SDSS}), \bibinfo{journal}{Phys. Rev.}
  \textbf{\bibinfo{volume}{D69}}, \bibinfo{pages}{103501}
  (\bibinfo{year}{2004}), \eprint{astro-ph/0310723}.

\bibitem{typeIseesaw}
\bibinfo{journal}{P. Minkowski, Phys. Lett. {\bf B67}, 421 (1977); T. Yanagida,
  in {\it Proceedings of the Workshop on the Unified Theory and Baryon Number
  in the Universe}, KEK, Tsukuba, 1979; M. Gell-Mann, P. Ramond, and R.
  Slansky, in {\it Proceedings of the Workshop on Supergravity}, Stony Brook,
  New York, 1979.}

\bibitem{typeIIseesaw}
\bibinfo{journal}{R.N. Mohapatra and G. Senjanovi\'c, Phys. Rev. Lett. {\bf
  44}, 912 (1980); Phys. Rev. {\bf D23}, 165 (1981); J. Schechter and J.W.F.
  Valle, Phys. Rev. {\bf D22}, 2227 (1980); G. Lazarides, Q. Shafi, and C.
  Wetterich, Nucl. Phys. {B181}, 287 (1981).}

\bibitem{Schwetz:2006dh}
\bibinfo{author}{\bibfnamefont{T.}~\bibnamefont{Schwetz}},
  \bibinfo{journal}{Phys. Scripta} \textbf{\bibinfo{volume}{T127}},
  \bibinfo{pages}{1} (\bibinfo{year}{2006}), \eprint{hep-ph/0606060}.

\bibitem{Huber:2006vr}
\bibinfo{author}{\bibfnamefont{P.}~\bibnamefont{Huber}},
  \bibinfo{author}{\bibfnamefont{J.}~\bibnamefont{Kopp}},
  \bibinfo{author}{\bibfnamefont{M.}~\bibnamefont{Lindner}},
  \bibinfo{author}{\bibfnamefont{M.}~\bibnamefont{Rolinec}}, \bibnamefont{and}
  \bibinfo{author}{\bibfnamefont{W.}~\bibnamefont{Winter}},
  \bibinfo{journal}{JHEP} \textbf{\bibinfo{volume}{05}}, \bibinfo{pages}{072}
  (\bibinfo{year}{2006}), \eprint{hep-ph/0601266}.

\bibitem{Antusch:2004yx}
\bibinfo{author}{\bibfnamefont{S.}~\bibnamefont{Antusch}},
  \bibinfo{author}{\bibfnamefont{P.}~\bibnamefont{Huber}},
  \bibinfo{author}{\bibfnamefont{J.}~\bibnamefont{Kersten}},
  \bibinfo{author}{\bibfnamefont{T.}~\bibnamefont{Schwetz}}, \bibnamefont{and}
  \bibinfo{author}{\bibfnamefont{W.}~\bibnamefont{Winter}},
  \bibinfo{journal}{Phys. Rev.} \textbf{\bibinfo{volume}{D70}},
  \bibinfo{pages}{097302} (\bibinfo{year}{2004}), \eprint{hep-ph/0404268}.

\bibitem{Minakata:2004jt}
\bibinfo{author}{\bibfnamefont{H.}~\bibnamefont{Minakata}},
  \bibinfo{author}{\bibfnamefont{H.}~\bibnamefont{Nunokawa}},
  \bibinfo{author}{\bibfnamefont{W.~J.~C.} \bibnamefont{Teves}},
  \bibnamefont{and}
  \bibinfo{author}{\bibfnamefont{R.}~\bibnamefont{Zukanovich~Funchal}},
  \bibinfo{journal}{Phys. Rev.} \textbf{\bibinfo{volume}{D71}},
  \bibinfo{pages}{013005} (\bibinfo{year}{2005}), \eprint{hep-ph/0407326}.

\bibitem{Smirnov:2004ju}
\bibinfo{author}{\bibfnamefont{A.~Y.} \bibnamefont{Smirnov}}
  (\bibinfo{year}{2004}), \eprint{hep-ph/0402264}.

\bibitem{Raidal:2004iw}
\bibinfo{author}{\bibfnamefont{M.}~\bibnamefont{Raidal}},
  \bibinfo{journal}{Phys. Rev. Lett.} \textbf{\bibinfo{volume}{93}},
  \bibinfo{pages}{161801} (\bibinfo{year}{2004}), \eprint{hep-ph/0404046}.

\bibitem{Minakata:2004xt}
\bibinfo{author}{\bibfnamefont{H.}~\bibnamefont{Minakata}} \bibnamefont{and}
  \bibinfo{author}{\bibfnamefont{A.~Y.} \bibnamefont{Smirnov}},
  \bibinfo{journal}{Phys. Rev.} \textbf{\bibinfo{volume}{D70}},
  \bibinfo{pages}{073009} (\bibinfo{year}{2004}), \eprint{hep-ph/0405088}.

\bibitem{Petcov:1993rk}
\bibinfo{author}{\bibfnamefont{S.~T.} \bibnamefont{Petcov}} \bibnamefont{and}
  \bibinfo{author}{\bibfnamefont{A.~Y.} \bibnamefont{Smirnov}},
  \bibinfo{journal}{Phys. Lett.} \textbf{\bibinfo{volume}{B322}},
  \bibinfo{pages}{109} (\bibinfo{year}{1994}), \eprint{hep-ph/9311204}.

\bibitem{Pontecorvo:1957cp}
\bibinfo{author}{\bibfnamefont{B.}~\bibnamefont{Pontecorvo}},
  \bibinfo{journal}{Sov. Phys. JETP} \textbf{\bibinfo{volume}{6}},
  \bibinfo{pages}{429} (\bibinfo{year}{1957}).

\bibitem{Maki:1962mu}
\bibinfo{author}{\bibfnamefont{Z.}~\bibnamefont{Maki}},
  \bibinfo{author}{\bibfnamefont{M.}~\bibnamefont{Nakagawa}}, \bibnamefont{and}
  \bibinfo{author}{\bibfnamefont{S.}~\bibnamefont{Sakata}},
  \bibinfo{journal}{Prog. Theor. Phys.} \textbf{\bibinfo{volume}{28}},
  \bibinfo{pages}{870} (\bibinfo{year}{1962}).

\bibitem{Altarelli:2006ri}
\bibinfo{author}{\bibfnamefont{G.}~\bibnamefont{Altarelli}}
  (\bibinfo{year}{2006}), \eprint{hep-ph/0611117}.

\bibitem{bimaximal}
\bibinfo{journal}{F. Vissani, {\tt hep-ph/9708483}; V.D. Barger, S. Pakvasa,
  T.J. Weiler, and K. Whisnant, Phys. Lett. {\bf B437}, 107 (1998), {\tt
  hep-ph/9806387}; A.J. Baltz, A.S. Goldhaber, and M. Goldhaber, Phys. Rev.
  Lett. {\bf 81}, 5730 (1998), {\tt hep-ph/9806540}; G. Altarelli and F.
  Feruglio, Phys. Lett. {\bf B439}, 112 (1998), {\tt hep-ph/9807353}; M.
  Jezabek and Y. Sumino, Phys. Lett. {\ bf B440}, 327 (1998), {\tt
  hep-ph/9807310}; D.V. Ahluwalia, Mod. Phys. Lett. {\bf A13}, 2249 (1998),
  {\tt hep-ph/9807267}}.

\bibitem{Jezabek:1999ta}
\bibinfo{author}{\bibfnamefont{M.}~\bibnamefont{Jezabek}} \bibnamefont{and}
  \bibinfo{author}{\bibfnamefont{Y.}~\bibnamefont{Sumino}},
  \bibinfo{journal}{Phys. Lett.} \textbf{\bibinfo{volume}{B457}},
  \bibinfo{pages}{139} (\bibinfo{year}{1999}), \eprint{hep-ph/9904382}.

\bibitem{Giunti:2002pp}
\bibinfo{author}{\bibfnamefont{C.}~\bibnamefont{Giunti}} \bibnamefont{and}
  \bibinfo{author}{\bibfnamefont{M.}~\bibnamefont{Tanimoto}},
  \bibinfo{journal}{Phys. Rev.} \textbf{\bibinfo{volume}{D66}},
  \bibinfo{pages}{113006} (\bibinfo{year}{2002}), \eprint{hep-ph/0209169}.

\bibitem{Frampton:2004ud}
\bibinfo{author}{\bibfnamefont{P.~H.} \bibnamefont{Frampton}},
  \bibinfo{author}{\bibfnamefont{S.~T.} \bibnamefont{Petcov}},
  \bibnamefont{and}
  \bibinfo{author}{\bibfnamefont{W.}~\bibnamefont{Rodejohann}},
  \bibinfo{journal}{Nucl. Phys.} \textbf{\bibinfo{volume}{B687}},
  \bibinfo{pages}{31} (\bibinfo{year}{2004}), \eprint{hep-ph/0401206}.

\bibitem{Ohlsson:2005js}
\bibinfo{author}{\bibfnamefont{T.}~\bibnamefont{Ohlsson}},
  \bibinfo{journal}{Phys. Lett.} \textbf{\bibinfo{volume}{B622}},
  \bibinfo{pages}{159} (\bibinfo{year}{2005}), \eprint{hep-ph/0506094}.

\bibitem{Antusch:2005kw}
\bibinfo{author}{\bibfnamefont{S.}~\bibnamefont{Antusch}} \bibnamefont{and}
  \bibinfo{author}{\bibfnamefont{S.~F.} \bibnamefont{King}},
  \bibinfo{journal}{Phys. Lett.} \textbf{\bibinfo{volume}{B631}},
  \bibinfo{pages}{42} (\bibinfo{year}{2005}), \eprint{hep-ph/0508044}.

\bibitem{Rodejohann:2003sc}
\bibinfo{author}{\bibfnamefont{W.}~\bibnamefont{Rodejohann}},
  \bibinfo{journal}{Phys. Rev.} \textbf{\bibinfo{volume}{D69}},
  \bibinfo{pages}{033005} (\bibinfo{year}{2004}), \eprint{hep-ph/0309249}.

\bibitem{Li:2005ir}
\bibinfo{author}{\bibfnamefont{N.}~\bibnamefont{Li}} \bibnamefont{and}
  \bibinfo{author}{\bibfnamefont{B.-Q.} \bibnamefont{Ma}},
  \bibinfo{journal}{Phys. Rev.} \textbf{\bibinfo{volume}{D71}},
  \bibinfo{pages}{097301} (\bibinfo{year}{2005}), \eprint{hep-ph/0501226}.

\bibitem{Xing:2005ur}
\bibinfo{author}{\bibfnamefont{Z.-z.} \bibnamefont{Xing}},
  \bibinfo{journal}{Phys. Lett.} \textbf{\bibinfo{volume}{B618}},
  \bibinfo{pages}{141} (\bibinfo{year}{2005}), \eprint{hep-ph/0503200}.

\bibitem{Datta:2005ci}
\bibinfo{author}{\bibfnamefont{A.}~\bibnamefont{Datta}},
  \bibinfo{author}{\bibfnamefont{L.}~\bibnamefont{Everett}}, \bibnamefont{and}
  \bibinfo{author}{\bibfnamefont{P.}~\bibnamefont{Ramond}},
  \bibinfo{journal}{Phys. Lett.} \textbf{\bibinfo{volume}{B620}},
  \bibinfo{pages}{42} (\bibinfo{year}{2005}), \eprint{hep-ph/0503222}.

\bibitem{Everett:2005ku}
\bibinfo{author}{\bibfnamefont{L.~L.} \bibnamefont{Everett}},
  \bibinfo{journal}{Phys. Rev.} \textbf{\bibinfo{volume}{D73}},
  \bibinfo{pages}{013011} (\bibinfo{year}{2006}), \eprint{hep-ph/0510256}.

\bibitem{Schmidt:2006rb}
\bibinfo{author}{\bibfnamefont{M.~A.} \bibnamefont{Schmidt}} \bibnamefont{and}
  \bibinfo{author}{\bibfnamefont{A.~Y.} \bibnamefont{Smirnov}}
  (\bibinfo{year}{2006}), \eprint{hep-ph/0607232};
 \bibinfo{author}{\bibfnamefont{A.} \bibfnamefont{Dighe}, \bibfnamefont{S.} \bibfnamefont{Goswami} \bibfnamefont{and} \bibfnamefont{P.} \bibfnamefont{Roy},
  Phys.~Rev. {\bf D73} (2006) 071301, {\tt hep-ph/0602062}}.

\bibitem{Frampton:2004vw}
\bibinfo{author}{\bibfnamefont{P.~H.} \bibnamefont{Frampton}} \bibnamefont{and}
  \bibinfo{author}{\bibfnamefont{R.~N.} \bibnamefont{Mohapatra}},
  \bibinfo{journal}{JHEP} \textbf{\bibinfo{volume}{01}}, \bibinfo{pages}{025}
  (\bibinfo{year}{2005}), \eprint{hep-ph/0407139}.

\bibitem{Antusch:2005ca}
\bibinfo{author}{\bibfnamefont{S.}~\bibnamefont{Antusch}},
  \bibinfo{author}{\bibfnamefont{S.~F.} \bibnamefont{King}}, \bibnamefont{and}
  \bibinfo{author}{\bibfnamefont{R.~N.} \bibnamefont{Mohapatra}},
  \bibinfo{journal}{Phys. Lett.} \textbf{\bibinfo{volume}{B618}},
  \bibinfo{pages}{150} (\bibinfo{year}{2005}), \eprint{hep-ph/0504007}.

\bibitem{Ohlsson:2002rb}
\bibinfo{author}{\bibfnamefont{T.}~\bibnamefont{Ohlsson}} \bibnamefont{and}
  \bibinfo{author}{\bibfnamefont{G.}~\bibnamefont{Seidl}},
  \bibinfo{journal}{Nucl. Phys.} \textbf{\bibinfo{volume}{B643}},
  \bibinfo{pages}{247} (\bibinfo{year}{2002}), \eprint{hep-ph/0206087}.

\bibitem{Albright:1998vf}
\bibinfo{author}{\bibfnamefont{C.~H.} \bibnamefont{Albright}},
  \bibinfo{author}{\bibfnamefont{K.~S.} \bibnamefont{Babu}}, \bibnamefont{and}
  \bibinfo{author}{\bibfnamefont{S.~M.} \bibnamefont{Barr}},
  \bibinfo{journal}{Phys. Rev. Lett.} \textbf{\bibinfo{volume}{81}},
  \bibinfo{pages}{1167} (\bibinfo{year}{1998}), \eprint{hep-ph/9802314}.

\bibitem{Nir:1999xp}
\bibinfo{author}{\bibfnamefont{Y.}~\bibnamefont{Nir}} \bibnamefont{and}
  \bibinfo{author}{\bibfnamefont{Y.}~\bibnamefont{Shadmi}},
  \bibinfo{journal}{JHEP} \textbf{\bibinfo{volume}{05}}, \bibinfo{pages}{023}
  (\bibinfo{year}{1999}), \eprint{hep-ph/9902293}.

\bibitem{Nomura:1999ty}
\bibinfo{author}{\bibfnamefont{Y.}~\bibnamefont{Nomura}} \bibnamefont{and}
  \bibinfo{author}{\bibfnamefont{T.}~\bibnamefont{Sugimoto}},
  \bibinfo{journal}{Phys. Rev.} \textbf{\bibinfo{volume}{D61}},
  \bibinfo{pages}{093003} (\bibinfo{year}{2000}), \eprint{hep-ph/9903334}.

\bibitem{Albright:2001uh}
\bibinfo{author}{\bibfnamefont{C.~H.} \bibnamefont{Albright}} \bibnamefont{and}
  \bibinfo{author}{\bibfnamefont{S.~M.} \bibnamefont{Barr}},
  \bibinfo{journal}{Phys. Rev.} \textbf{\bibinfo{volume}{D64}},
  \bibinfo{pages}{073010} (\bibinfo{year}{2001}), \eprint{hep-ph/0104294}.

\bibitem{tHooft:1979}
\bibinfo{journal}{G.~'t Hooft, Lecture at the Cargese Summer Institute (1979)}.

\bibitem{Froggatt:1978nt}
\bibinfo{author}{\bibfnamefont{C.~D.} \bibnamefont{Froggatt}} \bibnamefont{and}
  \bibinfo{author}{\bibfnamefont{H.~B.} \bibnamefont{Nielsen}},
  \bibinfo{journal}{Nucl. Phys.} \textbf{\bibinfo{volume}{B147}},
  \bibinfo{pages}{277} (\bibinfo{year}{1979}).

\bibitem{Leurer:1992wg}
  \bibinfo{journal}{M.~Leurer, Y.~Nir, and N.~Seiberg, Nucl. Phys. {\bf B398}, 319 (1993), {\tt hep-ph/9212278}; Nucl. Phys.
  {\bf B420}, 468 (1994), {\tt hep-ph/9310320}}.

\bibitem{Green:1984sg}
     \bibinfo{journal}{M.~B.~Green and J.~H.~Schwarz, Phys. Lett. {\bf
     B149}, 117 (1984); M.~Dine, N.~Seiberg, and E.~Witten,
     Nucl. Phys. {\bf B289}, 589 (1987); J.J.~Atick, L.J.~Dixon, and
     A.~Sen, Nucl. Phys. {\bf B292}, 109 (1987)}.

\bibitem{Ibanez:1994ig}
     \bibinfo{journal}{L.~E.~Ibanez and G.~G.~Ross, Phys. Lett. {\bf
     B332}, 100 (1994), {\tt hep-ph/9403338}; P.~Binetruy and
     P.~Ramond, Phys. Lett. {\bf B350}, 49 (1995), {\tt
     hep-ph/9412385}; P.~Binetruy, S.~Lavignac, and P.~Ramond,
     Nucl. Phys. {\bf B477}, 353 (1996), {\tt hep-ph/9601243};
     K.S.~Babu, T.~Enkhbat, and I.~Gogoladze, Nucl. Phys. {\bf B678},
     233 (2004), {\tt hep-ph/0308093}; H.K.~Dreiner, H.~Murayama, and
     M.~Thormeier, Nucl. Phys. {\bf B729}, 278, {\tt hep-ph/0312012}}.

\bibitem{Enkhbat:2005xb}
  \bibinfo{journal}{T.~Enkhbat and G.~Seidl,
  Nucl. Phys. {\bf B730}, 223 (2005),
  {\tt hep-ph/0504104}.}

\bibitem{Babu:2002dz}
  \bibinfo{journal}{See, {\it e.g.}, K.~S.~Babu, E.~Ma, and J.~W.~F.~Valle,
  Phys. Lett. {\bf B552} (2003) 207, {\tt hep-ph/0206292};
  G.~Seidl, {\tt hep-ph/0301044};
  G.~Altarelli and F.~Feruglio,
  Nucl. Phys. {\bf B720}, 64 (2005), {\tt hep-ph/0504165};
  J.~Kubo, Phys. Lett. {\bf B622}, 303 (2005),
  {\tt hep-ph/0506043};
  K.~S.~Babu and X.~G.~He, {\tt hep-ph/0507217};
  C.~Hagedorn, M.~Lindner, and R.~N.~Mohapatra,
  JHEP {\bf 0606}, 042 (2006), {hep-ph/0602244};
  C.~Hagedorn, M.~Lindner, and F.~Plentinger,
  Phys.~Rev. {\bf D74} (2006) 025007, {\tt hep-ph/0604265};
  I.~de Medeiros Varzielas, S.~F.~King, and G.~G.~Ross,
  {\tt hep-ph/0607045};
  Y.~Cai and H.~B.~Yu, {\tt hep-ph/0608022};
  Y.~Kajiyama, J.~Kubo, and H.~Okada,
  {\tt hep-ph/0610072};
  S.F.~King and M.~Malinsky,
  {\tt hep-ph/0610250};
  E.~Ma, {\tt hep-ph/0612013}}.

\bibitem{Albright:2006cw}
\bibinfo{author}{\bibfnamefont{C.~H.} \bibnamefont{Albright}} \bibnamefont{and}
  \bibinfo{author}{\bibfnamefont{M.-C.} \bibnamefont{Chen}}
  (\bibinfo{year}{2006}), \eprint{hep-ph/0608137}.
  
\bibitem{tribimaximal}
\bibinfo{journal}{P.F. Harrison, D.H. Perkins, and W.G. Scott, Phys. Lett. {\bf
  B458}, 79 (1999), {\tt hep-ph/9904297}; Phys. Lett. {\bf B530}, 167 (2002),
  {\tt hep-ph/0202074}}.

\bibitem{Plentinger:2005kx}
\bibinfo{author}{\bibfnamefont{F.}~\bibnamefont{Plentinger}} \bibnamefont{and}
  \bibinfo{author}{\bibfnamefont{W.}~\bibnamefont{Rodejohann}},
  \bibinfo{journal}{Phys. Lett.} \textbf{\bibinfo{volume}{B625}},
  \bibinfo{pages}{264} (\bibinfo{year}{2005}), \eprint{hep-ph/0507143}.

\bibitem{Huber:2005jk}
\bibinfo{author}{\bibfnamefont{P.}~\bibnamefont{Huber}},
  \bibinfo{author}{\bibfnamefont{M.}~\bibnamefont{Lindner}},
  \bibinfo{author}{\bibfnamefont{M.}~\bibnamefont{Rolinec}}, \bibnamefont{and}
  \bibinfo{author}{\bibfnamefont{W.}~\bibnamefont{Winter}},
  \bibinfo{journal}{Phys. Rev.} \textbf{\bibinfo{volume}{D73}},
  \bibinfo{pages}{053002} (\bibinfo{year}{2006}), \eprint{hep-ph/0506237}.

\bibitem{Huber:2003ak}
\bibinfo{author}{\bibfnamefont{P.}~\bibnamefont{Huber}} \bibnamefont{and}
  \bibinfo{author}{\bibfnamefont{W.}~\bibnamefont{Winter}},
  \bibinfo{journal}{Phys. Rev.} \textbf{\bibinfo{volume}{D68}},
  \bibinfo{pages}{037301} (\bibinfo{year}{2003}), \eprint{hep-ph/0301257}.

\bibitem{Bandyopadhyay:2004cp}
\bibinfo{author}{\bibfnamefont{A.}~\bibnamefont{Bandyopadhyay}},
  \bibinfo{author}{\bibfnamefont{S.}~\bibnamefont{Choubey}},
  \bibinfo{author}{\bibfnamefont{S.}~\bibnamefont{Goswami}}, \bibnamefont{and}
  \bibinfo{author}{\bibfnamefont{S.~T.} \bibnamefont{Petcov}},
  \bibinfo{journal}{Phys. Rev.} \textbf{\bibinfo{volume}{D72}},
  \bibinfo{pages}{033013} (\bibinfo{year}{2005}), \eprint{hep-ph/0410283}.

\bibitem{deGouvea:2003xe}
\bibinfo{author}{\bibfnamefont{A.}~\bibnamefont{de~Gouvea}} \bibnamefont{and}
  \bibinfo{author}{\bibfnamefont{H.}~\bibnamefont{Murayama}},
  \bibinfo{journal}{Phys. Lett.} \textbf{\bibinfo{volume}{B573}},
  \bibinfo{pages}{94} (\bibinfo{year}{2003}), \eprint{hep-ph/0301050}.

\bibitem{Lindner:2005kr}
\bibinfo{author}{\bibfnamefont{M.}~\bibnamefont{Lindner}},
  \bibinfo{author}{\bibfnamefont{A.}~\bibnamefont{Merle}}, \bibnamefont{and}
  \bibinfo{author}{\bibfnamefont{W.}~\bibnamefont{Rodejohann}},
  \bibinfo{journal}{Phys. Rev.} \textbf{\bibinfo{volume}{D73}},
  \bibinfo{pages}{053005} (\bibinfo{year}{2006}), \eprint{hep-ph/0512143}.

\bibitem{Petcov:2004wz}
\bibinfo{author}{\bibfnamefont{S.~T.} \bibnamefont{Petcov}},
  \bibinfo{journal}{New J. Phys.} \textbf{\bibinfo{volume}{6}},
  \bibinfo{pages}{109} (\bibinfo{year}{2004}).

\bibitem{Klapdor-Kleingrothaus:2000sn}
\bibinfo{author}{\bibfnamefont{H.~V.} \bibnamefont{Klapdor-Kleingrothaus}}
  \emph{et~al.}, \bibinfo{journal}{Eur. Phys. J.}
  \textbf{\bibinfo{volume}{A12}}, \bibinfo{pages}{147} (\bibinfo{year}{2001}),
  \eprint{hep-ph/0103062}.

\bibitem{psw-web}
\bibinfo{author}{\bibfnamefont{F.}~\bibnamefont{Plentinger}},
  \bibinfo{author}{\bibfnamefont{G.}~\bibnamefont{Seidl}}, \bibnamefont{and}
  \bibinfo{author}{\bibfnamefont{W.}~\bibnamefont{Winter}}, 
\bibinfo{note}{{\tt http://theorie.physik.uni-wuerzburg.de/ $\sim$winter/Resources/Textures/index.html}}.

\end{thebibliography}

\end{document}